\documentclass{aa} 
\def\mbh{$M_{\rm BH}$\/}
\def\nh{$n_{\mathrm{H}}$\/}
\def\lledd{$L/L_{\rm Edd}$}

\def\rfe{$R_{\rm FeII}$}

\def\feiiq{\rm Fe{\sc ii}$\lambda$4570\/}
\def\msol{M$_\odot$\/}

\def\ltsima{$\; \buildrel < \over \sim \;$}
\def\ltsim{\lower.5ex\hbox{\ltsima}}  
\def\gtsima{$\; \buildrel > \over \sim \;$}

\def\gtsim{\lower.5ex\hbox{\gtsima}} 

\def\lya{{ Ly}$\alpha$}
\def\civ{{\sc{Civ}}$\lambda$1549\/}
\def\civonly{{\sc Civ\/}}
\def\civnc{{\sc{Civ}}$\lambda$1549$_{\rm NC}$\/}
\def\civbc{{\sc{Civ}}$\lambda$1549$_{\rm BC}$\/}

\def\cm3{cm$^{-3}$\/}
\def\hb{{\sc{H}}$\beta$}

\def\hbbc{{\sc{H}}$\beta_{\rm BC}$\/}

\def\hbnc{{\sc{H}}$\beta_{\rm NC}$\/}
\def\mgii{{Mg\sc{ii}}$\lambda$2800\/}

\def\oiiiopt{{\sc{[Oiii]}}$\lambda\lambda$4959,5007\/}

\def\caii{{Ca{\sc ii}}}

\def\oiiiuv{{\sc{Oiii]}}$\lambda$1663\/}

\def\siiii{Si{\sc iii]}$\lambda$1892\/}
\def\aliii{Al{\sc iii}$\lambda$1860\/}
\def\heiiuv{He{\sc{ii}}$\lambda$1640}

\def\feii{{Fe\sc{ii}}\/}

\def\fe{{\sc{Fe}}\/}

\def\fe76087{{\sc [Fe vii]}$\lambda$6087\/}

\def\kms{km~s$^{-1}$}

\def\ergss{erg s$^{-1}$\/}

\def\mb{$m_{\mathrm B}$\/}
\def\siiv{Si{\sc iv}$\lambda$1397\/}
\def\oiv{O{\sc iv]}$\lambda$1402\/}

\def\cmp{$c(\frac{1}{2})$}
\def\cqp{$c(\frac{1}{4})$}
\usepackage{enumerate}

\usepackage{natbib}
\usepackage{graphicx}
\usepackage{txfonts}
\usepackage{xcolor}
\usepackage{pdflscape}
\usepackage{enumerate}
\def\civonly{{C\sc{iv}\,}}
\def\civbc{{\sc{Civ}}$_{\rm BC}$\ }
\def\hb{{\sc{H}}$\beta$\,}
\def\mgii{{Mg\sc{ii}}$\lambda$2800\/}

\def\aliii{Al{\sc iii}$\lambda$1860\/}
\def\lledd{$\rm{L/L_{\rm Edd}}$}
\def\hbbc{{\sc{H}}$\beta_{\rm BC}$\/}
\def\kms{\,km\,s$^{-1}\,$}
\def\k2{$\rm{km/s}$}
\def\feii{{Fe\sc{ii}}\/}

\def\feiiq{\rm Fe{\sc ii}$\lambda$4570\/}
\def\rfe{$R_{\rm FeII}$}
\def\feiiq{\rm Fe{\sc ii}$\lambda$4570\/}
\def\cmp{$c(\frac{1}{2})$}

\def\rfe{$R_{\rm FeII}$}
\def\mb{$m_{\mathrm B}$\/}

\begin{document} 
\citeindextrue

\title{What does \civ\ tell us about the physical driver of the Eigenvector Quasar Sequence?\thanks{Based on 
observations made with ESO Telescopes at the La Silla Paranal Observatory under 
programme ID082.B-0572(A) and with the Italian Telescopio Nazionale Galileo (TNG) operated by the Fundaci\'on Galileo Galilei (INAF) at the Observatorio del Roque de los Muchachos.}
}
 
\titlerunning{What drives \civ\ blueshifts?}

\author{J. W. Sulentic\inst{1} \and  A. del Olmo\inst{1} \and  P. Marziani\inst{2} \and M. A. Mart\'\i nez-Carballo\inst{1}\thanks{Present address: Departamento de Matemática Aplicada and IUMA, Universidad de Zaragoza, E-50009 Zaragoza, Spain}  \and M. D'Onofrio\inst{3}  \and D. Dultzin\inst{4} \and J. Perea\inst{1} \and 
\ \ \ \  M.~L. Mart\'\i nez-Aldama\inst{1} \and C.A. Negrete\inst{4} \and G.M. Stirpe\inst{5} \and S. Zamfir\inst{6} }
\institute{{Instituto de Astrofis\'{\i}ca de Andaluc\'{\i}a, IAA-CSIC, E-18008 Granada, Spain.}
\and{INAF, Osservatorio Astronomico di Padova, IT 35122, Padova, Italy}
\and {Dipartimento di Fisica \& Astronomia ``Galileo Galilei'', Univer. Padova, Padova,  Italia}
\and {Instituto de Astronom\'{\i}a, UNAM, Mexico D.F. 04510, Mexico}
\and {INAF, Osservatorio Astronomico di Bologna, Italy }
\and {Dep. of Physics \& Astronomy,University of Wisconsin, Stevens Points, USA}
}
   \date{}

%
%

\abstract{ Broad emission lines in quasars enable us to ``resolve'' structure and kinematics 
of the broad line emitting region (BLR) thought to involve an accretion disk feeding a supermassive 
black hole. Interpretation of broad line measures within the 4DE1 formalism simplifies 
the apparent confusion among such data by contrasting and unifying properties of 
so-called high and low accreting Population A and B sources. \hb\ serves as an estimator of
black hole mass, Eddington ratio and source rest frame, the latter a valuable input 
for \civ\ studies which allow us to isolate the blueshifted wind component. 
Optical and HST-UV spectra yield \hb and \civ\ spectra for low-luminosity sources while 
VLT-ISAAC  and FORS and TNG-LRS provide spectra for high Luminosity sources. New high S/N data for \civonly\ in 
high-luminosity quasars are presented here for comparison with the other previously published 
data. Comparison of \hb and \civ\ profile widths/shifts indicates that much of the 
emission from the two lines arise in regions with different structure and kinematics. 
Covering a wide range of luminosity and redshift shows evidence for a correlation between \civ\ blueshift and source Eddington ratio, with a weaker trend with source luminosity (similar amplitude outflows are seen over 4 of the 5 dex luminosity range  in our combined samples). At low luminosity ($z \lesssim 0.7$)\  only Population A sources show evidence for a significant outflow while at high luminosity the outflow signature begins to appear in Population B quasars as well.}

\keywords{quasars: general -- quasars: emission lines -- quasars: supermassive black holes }
\maketitle
   
%
\defcitealias{marzianietal09}{M09}
\defcitealias{sulenticetal14}{S14}
\defcitealias{sulenticetal07}{S07}

 
\section{Introduction}

Slightly more than fifty years after their discovery we are beginning to see progress in both defining and contextualizing 
the properties of Type 1 AGN/quasars which can be argued to be the ``parent population'' of the AGN phenomenon. We now know that Type 1 sources show considerable diversity in most observed  properties. They all show broad optical/UV emission lines from ionic species covering a wide range of ionization potential. Balmer lines (especially \hb) and \civ\AA\  (hereafter \civonly) have been used as  representative of low- and high-ionization lines. They are almost always accompanied by broad permitted \feii\ emission extending from the FUV to the NIR. The Balmer lines have played a prominent role for characterizing the broad line region (BLR) in lower redshift (z < 0.8) quasars, with \hb providing information for the largest number of sources  \citep{osterbrockshuder82,willsetal85,sulentic89,marzianietal03a,zamfiretal10,huetal12,shen13}. Most quasars lie 
beyond z $\sim$ 0.8 where we lose \hb\ as a characteristic low ionization broad line, in the absence of expensive IR spectra. 
Other higher ionization broad lines appear in their optical spectra. Can any of these lines be used as useful substitute 
of \hb? The higher ionization \civonly\ line has long been considered the best candidate for a BLR diagnostic beyond z 
$\sim$  1.4. 

\civonly\ and other high ionization lines -- unlike \hb\ and other low-ionization lines -- frequently show blueshifted profiles  
over a broad range in z and Luminosity \citep{gaskell82,espeyetal89,tytlerfan92,marzianietal96,richardsetal02,bachevetal04,baskinlaor05b,sulenticetal07,netzeretal07,richardsetal11,coatmanetal16}. The \civonly\ emission line has always played a major role in the study of quasars at $z \gtrsim 1.4$. Over the years, we have come to consider outflows as almost ubiquitous in quasars right because of the blueshifts of \civonly. Early reports at low-$z$\ encompassed  IUE observations of individual Seyfert 1 galaxies \citep[e.g.,][]{wamstekeretal90,zhengetal95}, or of high-$z$\ quasars with \civonly\ redshifted into the optical band \citep[e.g.][]{osmersmith76}. Since the 1990s the mainstream interpretation of the blueshift has involved emission by outflowing gas, possibly in a wind \citep{collinsouffrinetal88,emmeringetal92,murrayetal95} but alternate interpretations are still being discussed \citep{gaskellgoosmann13}. A less incomplete view came from the first analysis of HST observations \citep{willsetal93,corbinboroson96,marzianietal96}. For the first time, it was possible to take advantage of moderate resolution in the UV domain, and a reliable rest frame set by optical narrow lines. For intermediate to high-$z$ quasars a turning point came with the first data releases of the SDSS \citep{richardsetal02}, where shifts were found to be common, especially in the amplitude range from few hundreds \kms\ to $\sim 1000$ \kms. It was realized since the 1990s and later confirmed that the behavior of \civonly\ in powerful radio loud (RL) sources is different from the behavior in radio quiet (RQ), both in terms of shifts amplitude and dependence on luminosity \citep{marzianietal96,richardsetal11,marzianietal16}. The previous works suggest that \hb\ and \civonly\ are not surrogates of one another -- both lines arising e.g. in the same stratified medium \citep{marzianietal10}\ and that RQ and RL sources should be considered separately when dealing with a high-ionization line such as \civonly. The next major development  came with the availability of IR spectrometers that allow the observations of narrow emission lines redshifted into the NIR and therefore to set \civonly\ shifts measurements on a firm quantitative basis. Shifts larger than $\sim 1000 $ \kms\ were found to be frequent at high luminosity \citep[e.g.,][]{coatmanetal16,marzianietal16,vietri17,bisognietal17}.  

Current interpretation of the broad line region (BLR) in quasars sees the broad lines arising in a region that is  physically and dynamically composite \citep[][]{collinsouffrinetal88,elvis00,ferlandetal09,marzianietal10,kollatschnyzetzl13,grieretal13,duetal16}.  It appears that we need both \civonly\ and \hb\  to effectively characterize the low and high ionization gas in the BLR of quasars. For the past 20+ years our group has focused on spectroscopy of Type 1 AGN/quasars, with one of our goals the comparison of \hb and \civonly properties in the same sources. The work has involved in two main steps.

1) LOWZ: Using a large low z sample  ($\sim$ 160+ sources) we introduced the 4DE1  contextualization to distinguish and unify Type 1 quasar diversity. Measures of the \hb\  profile width and \civonly profile shift were adopted as two of the principal 4DE1 parameters  in that formalism \citep{sulenticetal00a,sulenticetal00b,sulenticetal02}.  Comparison of \hb\ measures with all high s/n \civonly HST-UV spectra (130 sources by 2007) completed the low redshift study \citep{sulenticetal07}. 
The 4DE1 formalism serves as a multidimensional set of HR diagrams  for type-1 sources \citep{sulenticetal00a}. Broad line measures, FWHM \hbbc\   and \feii\  strength as parameterized by the intensity ratio involving the \feii\  blue blend at 4570\AA\ and broad \hb (i.e., \rfe  =  I(\feiiq)/ I(\hb)), coupled with measures of the \civonly\ profile velocity displacement provide three dimensions of the 4DE1 parameter space that are observationally independent  and measure different physical aspects \citep[see][and references therein]{marzianietal10}. Earlier evidence for a correlation between the strength of a soft X-ray excess and some of the above measures  \citep{wangetal96,bolleretal96} also motivated the inclusion $\Gamma_\mathrm{soft}$ as a 4DE1 parameter.  

This first phase yielded a systematic view of the main observational properties along the 4DE1 sequence, which included parameters ranging from FIR colors to prevalence of radio-loud (RL) sources, to \oiiiopt\ blueshifts.  {Low z SDSS samples \citep{zamfiretal10} corroborates a 4DE1 main sequence extending from Population A quasars with 
FWHM \hb < 4000 \kms, strong optical FeII emission, a frequent \civonly\ blueshift and soft X-ray excess and, at 
the other extreme, Population B sources usually showing broader lines, weaker optical FeII and absence of a \civonly\ blueshift or soft X-ray excess (\citealt{sulenticetal11} and \citealt{fraix-burnetetal17} summarize several parameter that change among the 4DE1  sequence and are therefore systematically different between Pop. A and B). 

We suggested that the distinction between NLSy1 and rest of type-1 AGN was not meaningful, and that, at low $z$, a break in properties occurs at a larger FWHM(\hb), $\approx$ 4000 \kms\, distinguishing two Populations A and B of narrower and broader (or high and low Eddington ratio) sources respectively \citep{sulenticetal11}. Eventually, the break was associated with a critical Eddington ratio  value \citep{marzianietal03b},  and possibly with a change in  accretion disk structure \citep[][and references therein]{sulenticetal14a}. Eddington ratio (\lledd) was suggested by several authors  as the key parameter governing the 4DE1 sequence \citep[e.g.,][]{borosongreen92,marzianietal01,kuraszkiewiczetal00,dongetal09,ferlandetal09,sunshen15}. The roles of orientation, black hole mass, spin, and line emitting gas chemical composition in shaping the observed spectra remain not convincingly understood to this date \citep[][and references therein]{sulenticetal12a}, especially for RQ sources. Nonetheless, the 4DE1 has offered a consistent organization of quasars observational properties whose physical interpretation is still lacunose.}

2) High luminosity, HE sample: the advent of ESO VLT-ISAAC NIR spectrograph opened the possibility to 
obtain high S/N spectra for \hb\ in much higher luminosity quasars at intermediate-to-high redshift.
Our extreme $L$\ quasar sample was chosen from the Hamburg ESO Survey \citep{wisotzkietal00} in the  range z=0.9--3.1 and bolometric luminosity $\log L = {\rm 47.4-48.4}$. We obtained spectra of the \hb region 
in 53 sources \citep{sulenticetal04,sulenticetal06,marzianietal09}. 
{{Some  properties of the emission lines in the \hb\ region at  high luminosity $\log L \gtrsim 47$ [\ergss] are relevant to the interpretation of the results of the present paper. The \hb\ profile still supported the dichotomy between narrower and broader sources, with the broader sources showing a  redward asymmetry, even more prominent that at low-$L$.  The precise definition of Pop. A and Pop. B is however somewhat blurred at high $L$, since the minimum FWHM (\hb) significantly increases with source bolometric luminosity \citep[hereafter M09]{marzianietal09}. 

In this paper we add very high S/N VLT-FORS and TNG-LRS spectral data for \civonly\ of 28 high-luminosity quasars. Earlier high S/N \civonly spectra existed for other quasar samples, but no matter how good the \civonly data, 
measures of \civonly\ profiles remained  of uncertain interpretation without a precise determination of the quasar rest frame that we have obtained in our sample from the \hb spectral region. Accurate $z$ measurement is not easy to obtain from UV broad lines and redshift determinations from optical survey data, even at $z \gtrsim 1$\, suffer systematic biases as large as several hundred \kms \citep{hewettwild10,shenetal16}.

This paper presents {new observations and data analysis along with detailed profile decompositions (reported in \S \ref{sample})} that become possible with high S/N spectra. Results from \civonly data  complete step 2 of our \hb - \civonly study and also allow us a comparison with previous low redshift work. Our comparisons with low z sources is unprecedented because we introduce a sample of extreme luminosity sources with accurate rest frame estimates as well as spectra with high enough S/N to enable decomposition of \civonly into the blueshifted component and the unshifted component \civonly counterpart to broad component \hb.  Such comparisons are simply the best clues yet about the relative behaviour of Pop A and B quasars at low and high z and low and high L. 
 {Measurements and empirical models of spectra are reported in \S \ref{immediate}. They are discussed in \S \ref{disc} with a focus on the correlations between \civonly\ shifts and luminosity and Eddington ratio (\S \ref{corr}) which constrain the physical driver of the outflow (\S \ref{dynamics}). A follow-up paper presents an analysis of possible virial broadening estimators based on rest-frame UV observations. }

\section{Samples,  observations, and data analysis}
\subsection{Contextualizing our old and new \civonly samples}

Figure \ref{fig:locationSample} shows the locations in a redshift -- absolute magnitude plane
of our extreme luminosity Hamburg-ESO (HE) sample and the low z
comparison sample of 70 radio-quiet quasars (RQ) used in this study.
The HE sample, and quasars in the low z sample
above z=0.2 involve the bright end of the optical luminosity function at their respective redshifts. Quasars with 
luminosities similar to those in the HE sample do not exist at low redshift. {We also included in Figure 
\ref{fig:locationSample} a sample of 22 high redshift (z$\sim$2.3) quasars (GTC sample) with luminosities similar to local quasars. Spectra for these sources were obtained with the 10m GTC
telescope \citep{sulenticetal14}. We were motivated to see if low luminosity quasars 
similar to our low z sample (luminosity analogues with $\log L =46.1\pm0.4$) existed at z$\sim$2.3  
and whether they would show differences from their low z counterparts indicating a role for redshift 
evolution. The lack of \hb spectra for this sample required rest frame  estimation from UV lines.}
\begin{figure}[htp!]
\centering
\includegraphics[width=0.85\columnwidth]{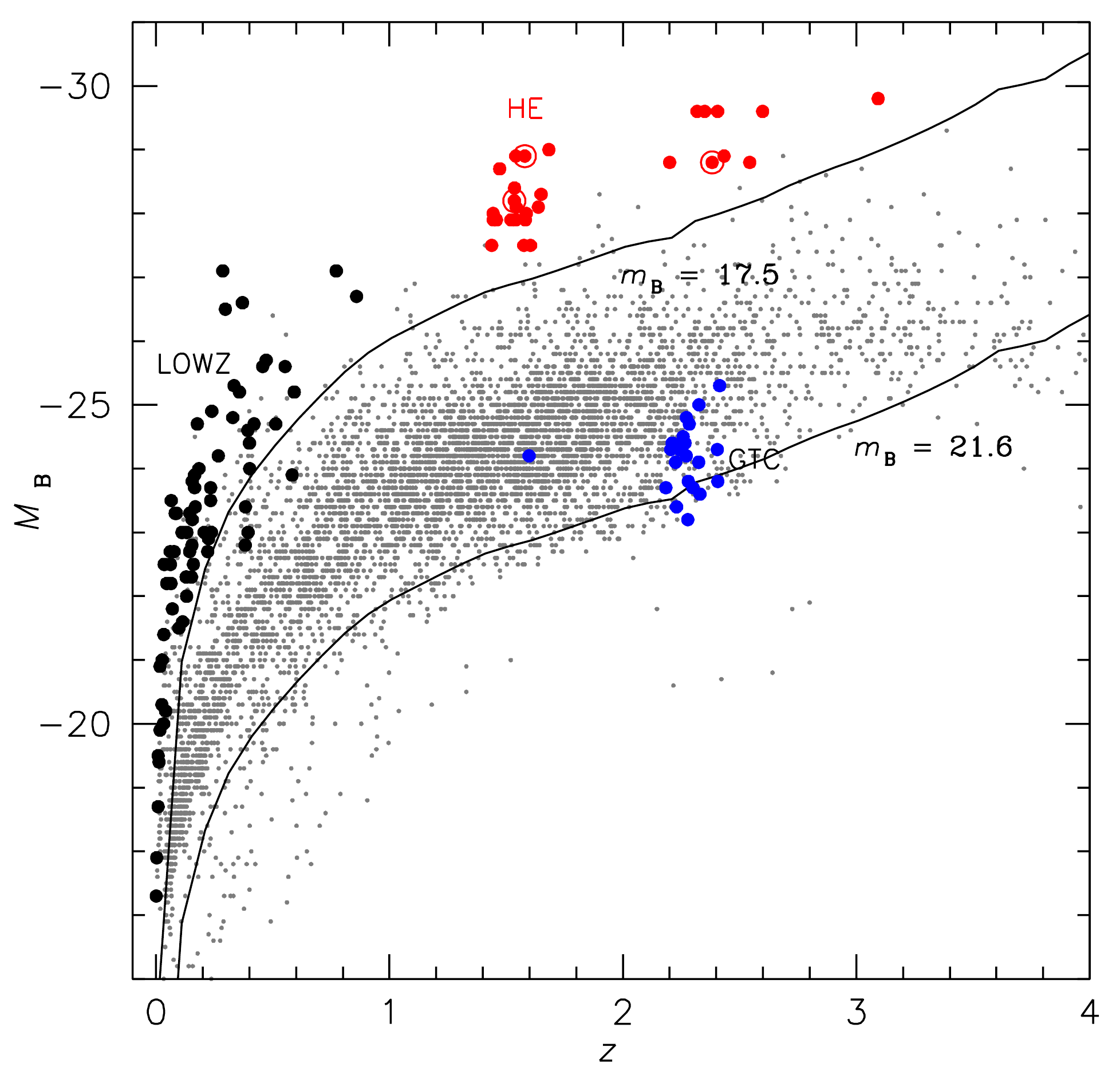}
\caption{Location of the HE sample in the plane M$_{B}$ vs. 
redshift (red spots), superimposed to a random subsample of the \citet{veroncettyveron10}
catalog (grey dots). Data points representing our low-$z$ RQ control sample and GTC sample of 
faint quasars at $z\sim 2.3$\, are also shown as black and 
blue filled circles respectively. The $\rm{M_B}$ associated with two apparent magnitude limited 
sample (17.5 appropriate for the HE survey, and 21.6 for the SDSS) are shown as function of redshift}
\label{fig:locationSample} 
\end{figure}
\label{sample}

\subsection{The Hamburg-ESO (HE)  sample }

\begin{table}
\scriptsize
\caption{Basic sample properties}             
\label{tab:he}      
\centering                         
\begin{tabular}{l c  c c c c   c c}   
\hline\hline \noalign{\vskip 0.05cm}      
Name & $m_\mathrm{B}$     & $z$  &Ref.$^\mathrm{a}$ &  $M_\mathrm{B}^\mathrm{b}$&  $\log L$ & Pop. &   Notes \\[0.05cm]
\hline\noalign{\vskip 0.07cm}
\object{HE0035-2853}	&	17.03	&	1.6377	&	1	&	-28.1	&	47.7	&	B	&		\\
\object{HE0043-2300}	&	17.06	&	1.5402	&	1	&	-27.9	&	47.6	&	A	&	RL 	\\
\object{HE0058-3231}	&	17.14	&	1.5821	&	1	&	-27.9	&	47.7	&	B	&		\\
\object{HE0109-3518}	&	16.44	&	2.4057	&	1	&	-29.6	&	48.3	&	A	&		\\
\object{HE0122-3759}	&	16.94	&	2.2004	&	2	&	-28.8	&	48.0	&	A	&		\\
\object{HE0203-4627}	&	17.34	&	1.4381	&	1	&	-27.5	&	47.5	&	B	&	RL	\\
\object{HE0205-3756}	&	17.17	&	2.4335	&	1	&	-28.9	&	48.0	&	A	&		\\
\object{HE0248-3628}	&	16.58	&	1.5355	&	1	&	-28.4	&	47.9	&	A	&		\\
\object{HE0251-5550}	&	16.59	&	2.3505	&	1	&	-29.6	&	48.3	&	A	&		\\
\object{HE0349-5249}	&	16.13	&	1.5409	&	1	&	-28.9	&	48.1	&	B	&		\\
\object{HE0359-3959}	&	17.09	&	1.5209	&	1	&	-27.9	&	47.6	&	A	&		\\
\object{HE0436-3709}	&	16.84	&	1.4447	&	1	&	-27.9	&	47.7	&	B	&		\\
\object{HE0507-3236} 	&	17.36	&	1.5759	&	3	&	-27.6	&	47.5	&	A	&		\\
\object{HE0512-3329} 	&	17.03	&	1.5862	&	3	&	-28.0	&	47.7	&	A	&		\\
\object{HE0926-0201} 	&	16.23	&	1.6828	&	3	&	-29.0	&	48.1	&	B	&		\\
\object{HE0940-1050}	&	16.70	&	3.0932	&	1	&	-30.1	&	48.4	&	A	&		\\
\object{HE1039-0724}	&	16.90	&	1.4584	&	1	&	-28.1	&	47.6	&	B	&		\\
\object{HE1104-1805} 	&	16.45	&	2.3180	&	3	&	-29.6	&	48.3	&	A	&		\\
\object{HE1120+0154}	&	16.31	&	1.4720	&	1	&	-28.7	&	48.0	&	B	&		\\
\object{HE1347-2457}	&	16.83	&	2.5986	&	1	&	-29.6	&	48.3	&	A	&		\\
\object{HE1349+0007}	&	16.83	&	1.4442	&	1	&	-28.0	&	47.7	&	B	&		\\
\object{HE1409+0101}	&	16.92	&	1.6497	&	1	&	-28.3	&	47.8	&	B	&		\\
\object{HE2147-3212}	&	16.84	&	1.5432	&	1	&	-28.1	&	47.7	&	A	&		\\
\object{HE2156-4020}	&	17.39	&	2.5431	&	1	&	-28.8	&	48.0	&	B	&		\\
\object{HE2202-2557} 	&	16.71	&	1.5347	&	2	&	-28.2	&	47.8	&	B	&		\\
\object{HE2349-3800} 	&	17.46	&	1.6040	&	2	&	-27.5	&	47.5	&	B	&	RL	\\
\object{HE2352-4010} 	&	16.05	&	1.5799	&	2	&	-28.9	&	48.1	&	A	&		\\
\object{HE2355-4621} 	&	17.13	&	2.3825	&	2	&	-28.8	&	48.0	&	B	&		\\[0.05cm]\hline
\noalign{\vskip 0.07cm}
\multicolumn{8}{l}{$^\mathrm{a}${References for redshift: 1: \citet{marzianietal09}; 2: \citet{sulenticetal04};}}\\
\multicolumn{8}{l}{~~3: \citet{sulenticetal06}}\\
\multicolumn{8}{l}{$^\mathrm{b}${Absolute B Magnitude, obtained from references in column Ref., corrected of galactic}}\\
\multicolumn{8}{l}{~~absorption and k-correction, and computed for H{$_0\ ={\rm 70\ km\ s^{-1}\ Mpc^{-1}}$}, $\Omega_\mathrm{M}=0.3$,}\\
\multicolumn{8}{l}{~~$\Omega_\Lambda=0.7$}\\
\end{tabular}
\end{table}

The quasars considered in the present study are selected from ISAAC \hb\  observations of 52 quasars identified in the Hamburg ESO survey \citep{wisotzkietal00}, in the redshift range $0.9 \lesssim z \lesssim 3$. Previous observations of \hb\   were discussed in a series of papers \citep[][hereafter M09]{sulenticetal02,sulenticetal06,marzianietal09}. Out of the 52 \hb\ sources 32 have $z \gtrsim 1.4$\ that allows for coverage of redshifted \civonly, and of them 28 were actually observed.

Table \ref{tab:he} summarizes the main properties of the sources including: identification in the HE catalog, apparent B magnitude taken from \citet{wisotzkietal00}, redshift, reference to redshift, absolute K-corrected B magnitude also corrected for Galactic extinction, logarithm of the bolometric luminosity estimated from the tabulated \mb, assuming $10 \lambda L_{\lambda}(5100)$\AA\ = $L$,   population assignment (A or B, discussed in \S \ref{pop}) in the 4DE1 scheme. In the last column, we identify RL sources. The redshifts  were estimated  from the narrow component of \hb\ and from \oiiiopt\ (only if   consistent with \hbnc) and are as reported in \citetalias{marzianietal09} and in the previous papers of the series dealing with \hb\ VLT/ISAAC observations. They are adopted as systemic redshifts for both the \hb\ and \civonly\ spectral range. 

\subsubsection{Objects belonging to particular classes}

\object{HE0512-3329} and \object{HE1104-1805} are known to be gravitationally lensed quasars. They were discussed with the presentation of the \hb\ ISAAC spectra \citep{sulenticetal06}. Here we just note that the FORS1 slit PA was  $-112$\ deg for  HE1104, and that the  PA values exclude contamination by the fainter lensed image. The HE 0512 PA   $\approx -41$\ deg implies an angle with the lens axis of $\approx 58$ deg  on average. The small separation between the two lensed images $\approx 0.64$\ arcsec (which are of comparable brightness in the $B$-band) implies a separation within the slit 0.5 arcsec, and that the two images should be well within  the slit if the slit was centered on their midpoint. This is however not to be given for granted since the A component is $\approx 0.44$ mag brighter than B in the R band  \citep{greggetal00}. The possibility of lensing or of the occurrence of a close quasar pair was also raised  for \object{HE1120+0154} ($\equiv$ UM 425). In this case, the separation  of the FORS observations ($\approx -171 $\ deg) excludes   contamination by fainter lensed quasar images. In addition, the second-brightest component of the lens  is   very faint, $V \approx 20$    \citep[][]{meylandjorgovski89}.   

An interesting aspect of HE1120 is the presence of variable \civonly\ absorption lines \citep{smalletal97}.  Given the limited radial velocity extent of the absorptions, the source would qualify as a mini-BAL, rather than a BAL QSO since the BALnicity index  \citep{weymannetal91} estimated from our spectrum is 0 \kms. 

 \object{HE0359-3959} and \object{HE2352-4010} belong to the class of weak-lined quasars (WLQs, \citealt[][HE 2352 as a borderline source]{diamond-stanicetal09,shemmeretal10}), for which $W$(\civonly)$\lesssim 10$\AA. WLQs in the HE sample will be discussed in \S \ref{phys}. 

 \begin{table}
\scriptsize
\caption{Log of Observations   \label{tab:obs}}      
\centering      \tabcolsep=2pt                   
\begin{tabular}{l c c r c c c}        
\hline\hline\noalign{\vskip 0.05cm}         
Name &   \multicolumn{2}{c}{Coordinates}  & Date        & Telescope/ & Exp.Time & Slit    \\
          & $\alpha$ (2000)  & $\delta$ (2000) & Observation & Instrument &   (s)    & arcsecs  \\[0.05cm]
\hline
\noalign{\vskip 0.07cm}
HE0035-2853	&00	38	06.5	&	-28	36	49	&	7-Oct-08	&	VLT-U2/FORS	&	1980	&	0.7	\\
HE0043-2300	&00	45	39.5	&	-22	43	56	&	30-Oct-09	&	VLT-U2/FORS	&	2340	&	1.0	\\
HE0058-3231	&01	00	39.2	&	-32	14	57	&	30-Oct-09	&	VLT-U2/FORS	&	2160	&	1.0	\\
HE0109-3518	&01	11	43.5	&	-35	03	01	&	12-Oct-08	&	VLT-U2/FORS	&	1800	&	1.0	\\
HE0122-3759	&01	24	17.4	&	-37	44	23	&	7-Oct-08	&	VLT-U2/FORS	&	1980	&	0.7	\\
HE0203-4627	&02	05	52.4	&	-46	13	30	&	11-Oct-08	&	VLT-U2/FORS	&	2520	&	1.0	\\
HE0205-3756	&02	07	27.2	&	-37	41	57	&	7-Oct-08	&	VLT-U2/FORS	&	1980	&	0.7	\\
HE0248-3628	&02	50	55.3	&	-36	16	35	&	12-Oct-08	&	VLT-U2/FORS	&	1800	&	1.0	\\
HE0251-5550	&02	52	40.1	&	-55	38	32	&	7-Oct-08	&	VLT-U2/FORS	&	1800	&	0.7	\\
HE0349-5249	&03	50	59.3	&	-52	40	35	&	7-Oct-08	&	VLT-U2/FORS	&	1800	&	1.0	\\
HE0359-3959	&04	01	14.0	&	-39	51	33	&	11-Oct-08	&	VLT-U2/FORS	&	1440	&	1.0	\\
HE0436-3709	&04	38	37.3	&	-37	03	41	&	11-Oct-08	&	VLT-U2/FORS	&	1440	&	1.0	\\
HE0507-3236	&05	09	17.8	&	-32	32	45	&	11-Oct-08	&	VLT-U2/FORS	&	1440	&	1.0	\\
HE0512-3329	&05	14	10.8	&	-33	26	23	&	11-Oct-08	&	VLT-U2/FORS	&	2340	&	1.0	\\
HE0926-0201	&09	29	13.5	&	-02	14	47	&	4-Mar-08	&	TNG/LRS		&	2400	&	1.0	\\
HE0940-1050	&09	42	53.5	&	-11	04	27	&	27-Jan-09	&	VLT-U2/FORS	&	1980	&	0.7	\\
HE1039-0724	&10	42	19.3	&	-07	40	37	&	6-Feb-09	&	VLT-U2/FORS	&	2340	&	1.0	\\
HE1104-1805	&11	06	33.5	&	-18	21	25	&	6-Feb-09	&	VLT-U2/FORS	&	1980	&	1.0	\\
HE1120+0154	&11	23	20.7	&	01	37	48	&	6-Feb-09	&	VLT-U2/FORS	&	1980	&	1.0	\\
HE1347-2457	&13	50	38.8	&	-25	12	16	&	4-May-08	&	TNG/LRS		&	2400	&	1.0	\\
HE1349+0007	&13	51	50.5	&	00	07	39	&	1-Apr-08	&	TNG/LRS		&	2400	&	1.0	\\
HE1409+0101	&14	12	21.7	&	00	47	19	&	3-May-08	&	TNG/LRS		&	2400	&	1.0	\\
HE2147-3212	&21	50	52.3	&	-31	58	26	&	21-Oct-09	&	VLT-U2/FORS	&	1980	&	1.0	\\
HE2156-4020	&21	59	54.7	&	-40	05	50	&	30-Oct-09	&	VLT-U2/FORS	&	2340	&	0.7	\\
HE2202-2557	&22	05	29.8	&	-25	42	23	&	2-Oct-08	&	VLT-U2/FORS	&	1980	&	1.0	\\
HE2349-3800	&23	52	10.7	&	-37	43	22	&	30-Sep-08&	VLT-U2/FORS	&	2700	&	0.7	\\
HE2352-4010	&23	55	34.5	&	-39	53	54	&	2-Oct-08	&	VLT-U2/FORS	&	1980	&	1.0	\\
HE2355-4621	&23	58	09.2	&	-46	05	00	&	30-Sep-08&	VLT-U2/FORS	&	2520	&	0.7	\\[0.05cm]
\hline                                   
\end{tabular}
\end{table}

\subsection{Observations and data reduction}

 Most (24) of the 28 HE quasar observations were obtained with FORS on the 8m VLT-U2 telescope (ESO, Paranal Observatory, Chile) with four additional obtained with the Low Resolution Spectrograph (LRS) at the 3.6m TNG telescope (Observatorio del Roque de los Muchachos, La Palma, Spain). Table \ref{tab:obs} presents an observational log with:  equatorial coordinates at J2000, date of observation, telescope and spectrograph, exposure time and slit width in arcsec. Long-slit observations with VLT/FORS1 were obtained with the 600B grism that provided a reciprocal dispersion of 1.5\AA/pixel. The LRS was equipped with  the LR-B grism which yielded  a dispersion of 2.7 \AA/px. 

Data reduction was carried out using standard IRAF procedures. Spectra were first trimmed and overscan corrected.  Bias subtraction and flat-fielding were performed each night. 1D wavelength calibration  using arc lamps were obtained with the same configuration and slit width. The rms of the wavelength calibration was in all  cases better than 0.07 \AA. The  wavelength scales of all  individual exposures  were realigned with sky lines present in the spectrum before object extraction and background subtraction, to avoid significant zero point errors. The scatter of sky line peak wavelengths after reassignment, $\approx$ 0.5  \AA\ or 30 \kms\ at 5000 \AA\ provides a realistic estimate of the uncertainty in the wavelength scale.  The spectral resolution estimated from the FWHM of  skylines is about 200 -- 300 \kms\ for VLT/FORS data. The  S/N ratio is $\sim$ 100 -- 200 in the continuum of the \civonly\ spectral region. The TNG/LRS observations yielded a spectral resolution of $\approx$\ 600 \kms\ with a typical S/N $\gtrsim$   50, and  $\sim 200$ in the case of {HE0926-0201} and {HE1409+0101}. Flux calibration was obtained each night from spectrophotometric standard
stars observed with the same configuration. Due to the narrow slit width, and the wavelength coverage extending to the NUV ($\approx$ 3300 \AA), an additional correction due to differential atmospheric refraction following \citet{filippenko82}, and an absolute flux recalibration proved to be necessary. 

We derived synthetic photometry on the spectra (computed using the \texttt{sband} task of IRAF). Specific fluxes in physical units (\ergss\ cm$^{-2}$ \AA$^{-1}$) were rescaled to the synoptic Catalina observations \citep{drakeetal09} for the wide majority of the objects,  to the SDSS (HE1120) and to the NOMAD catalogues (HE0248; \citealt{zachariasetal12}; HE 0058 and HE 0203; \citealt{zachariasetal05}). Similarly, the original \hb\ spectra synthetic magnitudes have been rescaled  to the 2MASS J,H,K magnitudes \citep{cutrietal03}. The rescaling allowed for a validation of the bolometric luminosity computed from the 1450 \AA\ and 5100 \AA\ fluxes, assuming bolometric correction factors of 3.5 and  10, respectively. After rescaling, the bolometric luminosities derived from the optical and UV are in good agreement, with average    $\delta \log L = \log  (3.5 \lambda L_{\lambda})_{1450} - \log (10 \lambda L_{\lambda})_{5100}\ \approx\  -0.072 \pm 0.310 $\ ($\log L \gtrsim 43.5$\ [\ergss]) for the combined LOWZ+HE sample, and $\delta \log  L \approx -0.086 \pm 0.264$\ for the HE sample. Only 4 sources of the HE sample (HE0205, HE0248, HE0436, and HE0512) have $|\delta \log L |\gtrsim 0.5$. {    The gravitationally lensed source HE 0512  might have been affected by a different magnification in the UV and optical.}  If they  are excluded, the HE average becomes $\delta \log L \approx -0.05 \pm 0.17$.

Systematic losses in emission line fluxes  will cancel out when computing intensity ratios, provided that flux losses are the same for both lines.   Nonetheless, great care should be used if computing intensity ratios of individual sources between the \civonly\ and \hb\ spectral range. Observations are not simultaneous  and separation in time  $\Delta t$\ between optical and IR span the range $  \Delta \tau \approx    \Delta t /(1 +z) \ltsim 1 - 3\  $\ yr\ in the rest frame.  Major optical /UV variability of highly luminous quasars is expected to occur on long timescales set by the expected emissivity-weighted radius of the BLR,  $c \Delta \tau \propto L(5100)_{48}^{\frac{1}{2}}$\ $\approx 4000$ ld \citep[if we can extrapolate the radius-luminosity relation of][]{bentzetal09}.  Previous works have also shown that, for a given time interval, higher luminosity sources  show lower variability levels \citep{vandenberketal04,simmetal16,caplaretal17}. However, moderate short-term variations  ($\approx 20$\%\ on $\Delta \tau \approx$ 15d on the rest frame)  are  possible  \citep{wooetal13,punslyetal16} and, along with intrinsic spectral energy distribution differences affecting the bolometric correction, may be responsible for part of the scatter between rest frame optical and UV-based $L$\ estimates.

\subsection{LOWZ: FOS-based comparison sample}
\label{fos}
A low-$L$\ comparison sample was selected from the 71 RQ sources of  \citet{sulenticetal07}\footnote{The FOS measurements used in this paper are available on CDS Vizier at URL \url{http://vizier.u-strasbg.fr}.}. We will consider 70 RQ FOS sources (one source at $\log L < 42$\ \ergss , with $L$\ estimated from the \hb\ spectral range, was excluded due to its outlying low luminosity) as a complementary low-$L$, FOS  sample for a comparative analysis of \civonly\ line parameters (LOWZ). Measurements  on the \hb\ line profiles that include centroids, widths and asymmetry are available  for 49 of them from \citet{marzianietal03a}. The comparison will be restricted to this sample if \hb\ parameters beyond FWHM are involved. RL sources were excluded from the LOWZ  sample because they are overrepresented with respect to their prevalence in optically selected samples ($\approx 10\%$, \citealt{zamfiretal08}) and especially because radio loudness is affecting the high-ionization outflows traced by \civonly\ \citep{marzianietal96,punsly10,richardsetal11,marzianietal16}. 

The LOWZ sample cannot be considered complete, but: (1) previous work has shown that it samples well the 4DE1 optical main sequence; (2) its bolometric luminosity $L$\  distribution is relatively uniform over the range $44 \lesssim \log L\ \lesssim 47$ [\ergss]. 

\subsection{Data analysis: \civ\ and \hb\ profile interpretation}
\label{contex}

\label{anal}
 
The broad profile of the \hb\ and \civonly\ profiles in each quasar spectrum can be modeled  by changing the relative intensity of three main components.  

\begin{itemize}
\item  The broad component (BC) has been referred to by various authors as the broad component,  intermediate  component, central broad component \citep[e.g.,][]{brothertonetal94a,popovicetal02}. Here the BC is represented by a symmetric and unshifted profile  (Lorentzian for  Pop. A or Gaussian for Pop. B) and assumed to be associated with a virialized BLR subsystem.
\item The blue shifted component (BLUE).  A strong  blue excess in Pop. A \civonly\ profiles is not in doubt.  In some \civonly\  profiles like  the extreme Population A  prototype I Zw1 the blue excess   is by far the dominant contributor to the total emission line flux \citep{marzianietal96,leighlymoore04}.  BLUE is modeled  by one or more skew-normal distributions   \citep{azzaliniregoli12}. The ``asymmetric gaussian'' use is, at present, motivated empirically by the often irregular appearance of the blueshfted excess in \civonly\ and \mgii.  
\item The very broad component (VBC). We expect a prominent \civonly\ VBC associated with high ionization gas  in the innermost BLR \citep{sneddengaskell07,marzianietal10,wangli11,goadkorista14}. The  VBC was postulated because of typical \hb\ profile of Pop. B sources, that can be (empirically) modeled with amazing fidelity using the sum of two Gaussians, one narrower and almost unshifted (the BC) and one broader showing a significant redshift $\sim 2000$ \kms\ \citep[the VBC, ][]{willsetal93,brothertonetal94,zamfiretal10}. {Past works provide evidence of a VBC in \civonly\ and other lines \citep[e.g.,][]{marzianietal96,punsly10,marzianietal10}.}
\end{itemize}

In addition, superimposed to the broad profile there is narrower emission most likely associated with larger distances from the central continuum sources  (i.e., with the narrow line region, NLR)  and unresolved in the FORS spectra. The projected linear size at $z \approx 1.5$ is 8.6\ kpc/arcsec, meaning that within the ISAAC and FORS slits were collecting light within few  kpc from the nucleus. This includes not only the innermost part of the NLR but also a significant contribution form the inner bulge of the host galaxy.  A fourth component has been considered in the analysis of the integrated \civonly\ and \hb\ profiles. 

\begin{itemize}
\item The narrow component (NC). The \hbnc\ is not in question even in the HE sample, although the ratio \hbnc\ to \hbbc\ is on average lower at high $L$.  Considering the large projected spatial apertures involved in our observations, the spatially-unresolved,  emissivity-weighted  \hbnc\ is expected to show a relatively narrow profile.   The rationale for a significant \civnc\ at low-$z$, and moderate luminosity was provided by \citet{sulenticmarziani99}.  The lack of a well-defined  critical density associated with \civonly\ (\oiiiopt\ is instead suppressed at density \nh $\gtrsim 10^{6}$\ cm$^{-3}$), and   high ionization parameter favoring \civonly\ emission supported the assumption that \civonly\ NC could be up to 3 times broader than \oiiiopt\ (and \hb) following the simple model of \citet{netzer90} based on  a density radial trend.   Hence, the \civnc\  profile may merge smoothly with \civbc.  Since \civnc\ is mainly due  to the inner  NLR  in this scenario, \civnc\ blueshifts  of a few hundreds \kms\ per second are expected.  Both  \civnc\ and \hbnc\ are modeled by a single, symmetric Gaussian, but the \hbnc\ is assumed as reference for rest-frame, while \civnc\ peak shift is allowed to vary in the \civonly\ profile fits.  
\end{itemize}

\subsection{Non-linear multicomponent fits and full profile measurements of \civ\ and \hb}

The data analysis of the present paper follows the same approach employed in several recent works, and consists in applying a non-linear multicomponent fit to continuum, a scalable \feii\ emission template, and emission line components \citep[e.g.,][]{negreteetal13,sulenticetal14}.   The minimum $\chi_{\nu}^{2}$\ multicomponent fit was carried out with the {\tt specfit} routine within IRAF \citep{kriss94} yielding FWHM, peak wavelength, and intensity for the four line components  BC, BLUE, VBC, NC. An important caveat is that the line profile models are heuristic because the actual component shapes are unknown.  

In addition to the  \civonly\ spectra, the \hb\ spectral range  covered by the VLT-ISAAC was  also measured using the IRAF task {\tt specfit}. {The {\tt specfit} measurements   were found to be generally  consistent with the earlier ones reported in \citetalias{marzianietal09}.\footnote{The measurement technique was different for the \hb\ observations in \citetalias{marzianietal09} and earlier papers by \citet{sulenticetal04} and \citet{sulenticetal06}. In those papers  \feii\ and continuum templates were sequentially subtracted before measuring \hb\ parameters, and not obtained through a multicomponent nonlinear fit.} \citetalias{marzianietal09} FWHM, \cmp, and EW of \hb\ are on average equal  with a scatter $\lesssim $20 \%. Measures toward the line base are influenced by the line profile model: especially on the red side there can be differences associated with the correction by \oiiiopt\ emission. The correction of  \citetalias{marzianietal09} was empirical, while in the present paper we assume an  intrinsic \hb\ broad line shape (\S \ref{anal}). This latter approach is preferable because of the strong and irregular \oiiiopt\ emission recently detected in several luminous quasars \citep[e.g.,][and references therein]{canodiazetal12,carnianietal15,marzianietal16a}. The \oiiiopt\ emission is modeled using a core component plus a  semi-broad component, usually blueshifted \citep[e.g., ][]{zhangetal11,craccoetal16,marzianietal16a}.

\begin{figure}[pth!]
\centering
\includegraphics[width=1.\columnwidth]{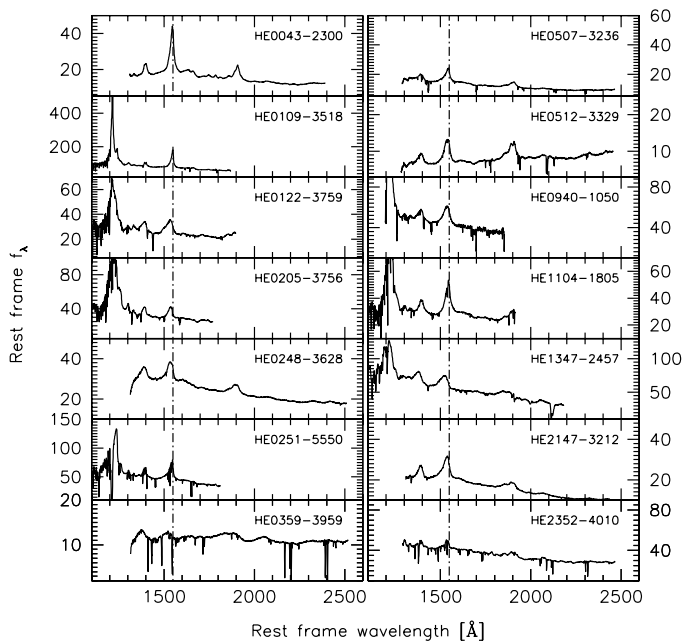}
\caption{Rest frame UV spectra of the 14 quasars belonging to Pop. A. Abscissae are rest frame wavelength in \AA, ordinates are specific flux in units of  $10^{-15}$ \ergss\ cm$^{-2}$\ \AA$^{-1}$. The dot dashed line traces the rest-frame wavelength of the \civonly\ line.}
\label{fig:speca}
\end{figure}

\begin{figure}[pth!]
\begin{center}
\includegraphics[width=1\columnwidth]{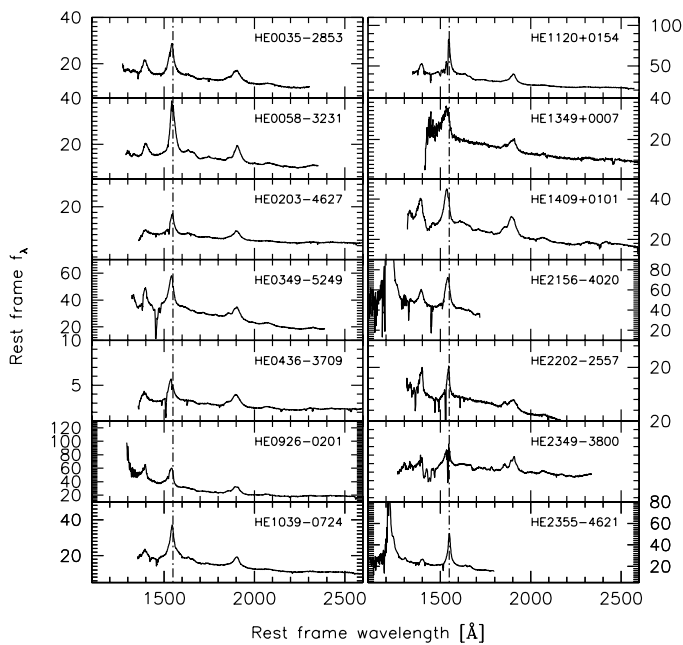} 
\end{center}
\caption{Rest frame UV spectra of the 14 Pop. B HE quasars presented in this paper. Abscissae are rest frame wavelength in \AA, ordinates are specific flux in units of  $10^{-15}$ \ergss\ cm$^{-2}$\ \AA$^{-1}$.}
\label{fig:specb}
\end{figure}

Line profile parameters (FWHM, centroids, asymmetry, and kurtosis  defined by \citealt{zamfiretal10}) were measured on the full broad \civonly\ and \hb\ profiles (excluding the NC) to provide a qualitative  description independent of the {\tt specfit }   modeling. Here we just stress that line centroids at different fractional peak intensity ($c(\frac{i}{4})$\ for $i=1,2,3$ and $c$(0.9)) are computed with respect to rest frame, while the asymmetry index (AI), computed only at $\frac{1}{4}$\ peak line intensity, is referred to the line peak. In addition $c(\frac{1}{4})$ measures the deviations from rest frame close to the line base, while $c(\frac{1}{2})$ measures the deviations from a symmetric and unshifted line at the fractional intensity where the broadening estimator (the FWHM) is measured.  The $c$(0.9) is used as a proxy for the line peak radial velocity.  The mock profiles of  Fig. 2 in \citet{marzianietal16} illustrate the quantitative centroid and width measurements extracted from the broad \civonly\ profile (the narrow component is supposed to have been removed beforehand, if significant). The mock profiles  show the main components expected in the \civonly\ profile of Pop. A   (BLUE and BC), and Pop. B  sources (BLUE, BC, VBC). 

\subsection{Population classification}
\label{pop}

The A/B classification of the quasars in HE sample (Table \ref{tab:he}) has been assigned by considering a limit in  FWHM(\hb) following the luminosity-dependent criterion of \citet{sulenticetal04}:  FWHM$_\mathrm{AB} \approx 3500 + 500 \cdot 10^{-0.06 \cdot (M_\mathrm{B} + 20.24)} $ \kms, where the $M_\mathrm{B}$ are reported in Table \ref{tab:he}    (see \citealt{sulenticetal04} for the derivation of the minimum FWHM as a function of the absolute magnitude). 

Given the very weak dependence on $L$, the luminosity-dependent FWHM separation is consistent with an approximately constant limit at 4000 \kms\ if $\log L \lesssim 47 $ [\ergss]. {\bf  The same population assignments are obtained if  the luminosity-dependent limit FWHM$_\mathrm{AB}$\ mentioned in the previous paragraph is applied to FOS sample in place of the constant limit at 4000 \kms.} If virial broadening dominates the width of \hb, the line emitting region size scales roughly as $r \propto L^{0.5}$, and \lledd $\lesssim$1, then there is a minimum FWHM corresponding to the maximum \lledd. At very high luminosity ($\log L \gtrsim 47$\ [\ergss]), if all sources radiate sub-Eddington or close to the Eddington limit, we do not expect to observe sources with very narrow lines, such as the one falling in the domain of narrow-line Seyfert 1s (NLSy1s), as further discussed in \S \ref{lum}. For instance, at $\log L \gtrsim 47$\ [\ergss], sources with FWHM$\lesssim$ 2500 \kms\ are not possible, if the conditions mentioned earlier in this section are satisfied. At the extreme $L \approx 10^{15} L_{\odot}$, no source may show FWHM$\lesssim$ 4000 \kms. This underlines the inadequacy of assuming FWHM$=$ 4000 \kms  at the typical luminosity of the HE sources. The FWHM$_\mathrm{AB}$\ limit is empirical (as discussed in \S 4.5.1 of \citealt{marzianietal09}).  It is found a posteriori that Pop. A and B have systematically different \lledd\ values (\citealt{marzianietal03b}; \S \ref{helowz}).   

With respect to \citetalias{marzianietal09} the classification was changed for a few border-line sources close to the luminosity-dependent Pop. A / B boundary: {HE1347-2457} and {HE2147-3212}  $\rightarrow$ Pop. A (they were previously assigned to spectral type B2, $4000 < $ FWHM(\hb) $\le 8000$ \kms, 1.0 $ \le $ \rfe $\le 1.5$\ following \citealt{sulenticetal02}), {HE2156-4020}  $\rightarrow$ Pop. B (was spectral type AM, which includes sources above the canonical limit at FWHM = 4000 \kms, and below the luminosity-dependent limit for Pop. A as defined in \citealt{sulenticetal04}).

\begin{table}[htp!]
\setlength{\tabcolsep}{1pt}\scriptsize
\caption{Spectrophotometric measurements  on \civ\ and \hb\  \label{tab:civhb}}
\begin{tabular}{lccccccccc}\hline\hline\noalign{\vskip 0.05cm}
& \multicolumn{4}{c}{\civ\ spectral range} &&  \multicolumn{4}{c}{\hb\ spectral range} \\[0.05cm]  \cline{2-5} \cline{7-10}
Identification  &{$f_{\lambda,1450}^{a}$}& {$f$({\sc Civ})$^{b}$} &{W({\sc Civ})} &   {$f$(Si {\sc iv}+O{\sc iv}])$^{b}$} & & {$f_{\lambda,5100}^{a}$} & {$f$(\hb)$^{b}$} &{W(\hb)}   & {\rfe}   \\  
&   & & {[\AA]} &   & &    &  & {[\AA]} \\ 
\hline
\noalign{\vskip 0.07cm}
&& &&\multicolumn{1}{c}{Population A}& &\\ [0.07cm]
\hline
\noalign{\vskip 0.07cm}
HE0043-2300	&	17.6	$\pm$1.3	&	8.2	&	48	&	1.7	&	&	3.2$\pm$	0.2	&	2.4	&	68	&	0.37	\\
HE0109-3518	&	80.4	$\pm$7.8	&	19.5	&	26	&	3.4	&	&	3.4$\pm$	0.3	&	4.8	&	137	&	0.16	\\
HE0122-3759 	&	24.8	$\pm$2.7	&	5.3	&	22	&	2.4	&	&	2.2$\pm$	0.3	&	1.2	&	50	&	1.12	\\
HE0205-3756	&	24.5	$\pm$3.2	&	4.9	&	18	&	2.4	&	&	9.7$\pm$	0.6	&	5.9	&	56	&	0.40	\\
HE0248-3628	&	29.9	$\pm$0.9$^\mathrm{c}$&	4.3	&	15	&	2.7	&	&	0.8$\pm$	0.0	&	0.4	&	44	&	0.53	\\
HE0251-5550	&	46.7	$\pm$4.7	&	12.8	&	30	&	3.7	&	&	5.9$\pm$	0.3	&	4.0	&	64	&	0.39	\\
HE0359-3959	&	11.2	$\pm$1.3	&	0.6	&	5.3	&	1.2	&	&	1.8$\pm$	0.1	&	0.9	&	50	&	1.10	\\
HE0507-3236 	&	15.1	$\pm$2.3	&	4.1	&	29	&	1.5	&	&	2.1$\pm$	0.2	&	1.2	&	67	&	0.31	\\
HE0512-3329 	&	6.4	$\pm$1.2	&	3.0	&	45	&	1.2	&	&	2.7$\pm$	0.1	&	1.9	&	86	&	0.63	\\
HE0940-1050	&	45.3	$\pm$3.6	&	9.5	&	23	&	4.6	&	&	5.2$\pm$	0.4	&	2.7	&	45	&	0.55	\\
HE1104-1805 	&	27.8	$\pm$3.3	&	10.4	&	41	&	3.3	&	&	3.0$\pm$	0.3	&	3.7	&	121	&	0.56	\\
HE1347-2457	&	59.9	$\pm$4.8	&	8.8	&	15	&	7.4	&	&	3.9$\pm$	0.2	&	1.7	&	38	&	1.22	\\
HE2147-3212	&	20.7	$\pm$2.7	&	5.8	&	29	&	1.8	&	&	1.7$\pm$	0.1	&	1.3	&	70	&	0.77	\\
HE2352-4010 	&	42.1	$\pm$3.4	&	4.1	&	10.0	&	1.8	&	&	6.3$\pm$	0.2	&	3.5	&	50	&	0.50	\\
[0.1cm] \hline
\noalign{\vskip 0.07cm}
&& &&\multicolumn{1}{c}{Population B}& &\\[0.07cm]
\hline
\noalign{\vskip 0.1cm}
HE0035-2853	&	15.5	$\pm$2.3	&	5.6	&	39	&	1.8	&	&	2.0$\pm$	0.1	&	1.5	&	77	&	0.72	\\
HE0058-3231	&	15.3$^\mathrm{d}$&	9.9	&	72	&	1.9	&	&	1.9$\pm$	0.1	&	1.9	&	92	&	0.42	\\
HE0203-4627	&	10.0$^\mathrm{d}$&	2.7	&	27	&	0.4	&	&	1.9$\pm$	0.1	&	0.8	&	60	&	0.36	\\
HE0349-5249	&	36.1	$\pm$3.3	&	9.7	&	29	&	4.1	&	&	2.8$\pm$	0.1	&	1.8	&	64	&	0.74	\\
HE0436-3709	&	3.2	$\pm$0.3	&	1.1	&	36	&	0.3	&	&	1.4$\pm$	0.1	&	1.2	&	80	&	0.54	\\
HE0926-0201 	&	38.4	$\pm$2.7	&	10.6	&	32	&	5.5	&	&	5.1$\pm$	0.2	&	3.0	&	70	&	0.40	\\
HE1039-0724	&	18.3	$\pm$1.5	&	7.9	&	47	&	1.3	&	&	1.7$\pm$	0.1	&	1.5	&	82	&	0.24	\\
HE1120+0154	&	38.0	$\pm$2.3	&	13.8	&	37	&	3.2	&	&	5.7$\pm$	0.2	&	3.9	&	65	&	0.18	\\
HE1349+0007	&	22.9	$\pm$1.8	&	6.1	&	28	&	\ldots&	&	1.9$\pm$	0.1	&	1.6	&	73	&	0.34	\\
HE1409+0101	&	26.8	$\pm$1.9	&	7.7	&	29	&	3.8	&	&	4.2$\pm$	0.2	&	4.1	&	91	&	0.49	\\
HE2156-4020	&	43.3	$\pm$5.6	&	13.0	&	32	&	4.0	&	&	2.2$\pm$	0.2	&	2.3	&	99	&	0.43	\\
HE2202-2557 	&	14.4	$\pm$1.9	&	1.7	&	12	&	1.2	&	&	1.9$\pm$	0.2	&	1.0	&	51	&	0.40	\\
HE2349-3800 	&	8.4	$\pm$1.1	&	2.4	&	27	&	0.4	&	&	1.5$\pm$	0.2	&	0.9	&	57	&	0.47	\\
HE2355-4621 	&	21.1	$\pm$2.3	&	7.7	&	40	&	0.9	&	&	2.5$\pm$	0.2	&	2.6	&	99	&	0.18	\\
[0.07cm]
 \hline\noalign{\vskip 0.07cm}
\multicolumn{10}{l}{$^\mathrm{a}${In units of { $10^{-15}$\ergss\ cm$^{-2}$ \AA$^{-1}$}}}\\
\multicolumn{10}{l}{$^\mathrm{b}${In units of {$10^{-13}$\ergss\ cm$^{-2}$ }}}\\
\multicolumn{10}{l}{$^\mathrm{c}${Rescaled to the $V$ magnitude reported in the UCAC4 \citep{zachariasetal12}.}}\\
\multicolumn{10}{l}{$^\mathrm{d}${Rescaled to the $V$ magnitude reported in the NOMAD1 catalog \citep{zachariasetal05}.}}\\
 \end{tabular}
\end{table}

\section{Results  }
\label{immediate}
\begin{table}[htp!]
\setlength{\tabcolsep}{3pt}
\begin{center}
\scriptsize
\caption{Measurements on the \civ\ full line profile   \label{tab:civprof}}
\begin{tabular}{lcccrrr}\hline\hline\noalign{\vskip 0.05cm}
\multicolumn{1}{l}{Identification}  &\multicolumn{1}{c}{FWHM}& \multicolumn{1}{c}{A.I.} &\multicolumn{1}{c}{Kurt.}     &\multicolumn{1}{c}{$c(\frac{1}{4})$} & \multicolumn{1}{c}{$c(\frac{1}{2})$} \\  
&   [\kms]    & &  & [\kms]   & [\kms]\\ 
\hline\noalign{\vskip 0.06cm}
\multicolumn{6}{c}{Population A}&\\[0.05cm] \hline\noalign{\vskip 0.1cm}
HE0043-2300	&	5140$\pm$\ 	380	&	-0.04	$\pm$\ 	0.11	&	0.31	$\pm$\ 	0.06	&	-1110	$\pm$\ 	720	&	-980	$\pm$\ 	380	\\
HE0109-3518	&	3440$\pm$\ 	260	&	-0.16	$\pm$\ 	0.12	&	0.34	$\pm$\ 	0.08	&	-1050	$\pm$\ 	530	&	-820	$\pm$\ 	260	\\
HE0122-3759 	&	8510$\pm$\ 	870	&	-0.30	$\pm$\ 	0.07	&	0.38	$\pm$\ 	0.04	&	-4800	$\pm$\ 	540	&	-3930	$\pm$\ 	870	\\
HE0205-3756	&	4930$\pm$\ 	330	&	-0.22	$\pm$\ 	0.18	&	0.36	$\pm$\ 	0.09	&	-3320	$\pm$\ 	1080	&	-2920	$\pm$\ 	330	\\
HE0248-3628	&	7360$\pm$\ 	400	&	-0.12	$\pm$\ 	0.10	&	0.38	$\pm$\ 	0.05	&	-3950	$\pm$\ 	810	&	-2790	$\pm$\ 	400	\\
HE0251-5550	&	4220$\pm$\ 	320	&	-0.30	$\pm$\ 	0.25	&	0.32	$\pm$\ 	0.13	&	-2170	$\pm$\ 	1860	&	-1640	$\pm$\ 	320	\\
HE0359-3959	&	8590$\pm$\ 	490	&	-0.13	$\pm$\ 	0.08	&	0.42	$\pm$\ 	0.04	&	-5990	$\pm$\ 	610	&	-5880	$\pm$\ 	490	\\
HE0507-3236 	&	5410$\pm$\ 	430	&	-0.31	$\pm$\ 	0.10	&	0.27	$\pm$\ 	0.06	&	-3100	$\pm$\ 	970	&	-1670	$\pm$\ 	430	\\
HE0512-3329 	&	7130$\pm$\ 	340	&	-0.10	$\pm$\ 	0.13	&	0.48	$\pm$\ 	0.07	&	-2960	$\pm$\ 	920	&	-2470	$\pm$\ 	340	\\
HE0940-1050	&	8060$\pm$\ 	600	&	-0.15	$\pm$\ 	0.08	&	0.41	$\pm$\ 	0.05	&	-3290	$\pm$\ 	660	&	-2800	$\pm$\ 	600	\\
HE1104-1805	&	6710$\pm$\ 	370	&	-0.07	$\pm$\ 	0.10	&	0.42	$\pm$\ 	0.07	&	-1630	$\pm$\ 	830	&	-1500	$\pm$\ 	370	\\
HE1347-2457	&	8260$\pm$\ 	440	&	-0.16	$\pm$\ 	0.09	&	0.45	$\pm$\ 	0.05	&	-5850	$\pm$\ 	660	&	-5490	$\pm$\ 	440	\\
HE2147-3212	&	7770$\pm$\ 	620	&	-0.27	$\pm$\ 	0.10	&	0.34	$\pm$\ 	0.05	&	-4440	$\pm$\ 	1090	&	-3110	$\pm$\ 	620	\\
HE2352-4010 	&	5180$\pm$\ 	390	&	-0.31	$\pm$\ 	0.08	&	0.24	$\pm$\ 	0.04	&	-4370	$\pm$\ 	720	&	-2390	$\pm$\ 	390	\\[0.08cm]
 \hline\noalign{\vskip 0.05cm}
\multicolumn{6}{c}{Population B}&\\[0.06cm] \hline
\noalign{\vskip 0.1cm}
HE0035-2853	&	6730$\pm$\ 	550	&	0.08	$\pm$\ 	0.10	&	0.32	$\pm$\ 	0.06	&	-1080	$\pm$\ 	970	&	-1530	$\pm$\ 	550	\\
HE0058-3231	&	6290$\pm$\ 	400	&	0.15	$\pm$\ 	0.13	&	0.37	$\pm$\ 	0.08	&	190	$\pm$\ 	1190	&	-290	$\pm$\ 	400	\\
HE0203-4627	&	6850$\pm$\ 	460	&	0.04	$\pm$\ 	0.10	&	0.34	$\pm$\ 	0.06	&	-1020	$\pm$\ 	880	&	-910	$\pm$\ 	460	\\
HE0349-5249	&	6050$\pm$\ 	460	&	-0.13	$\pm$\ 	0.11	&	0.33	$\pm$\ 	0.06	&	-1960	$\pm$\ 	900	&	-1750	$\pm$\ 	460	\\
HE0436-3709	&	7040$\pm$\ 	580	&	0.03	$\pm$\ 	0.10	&	0.30	$\pm$\ 	0.06	&	-1680	$\pm$\ 	1070	&	-1880	$\pm$\ 	580	\\
HE0926-0201 	&	5980$\pm$\ 	410	&	-0.28	$\pm$\ 	0.12	&	0.36	$\pm$\ 	0.07	&	-3280	$\pm$\ 	1030	&	-2280	$\pm$\ 	410	\\
HE1039-0724	&	7890$\pm$\ 	740	&	0.12	$\pm$\ 	0.10	&	0.28	$\pm$\ 	0.05	&	-490	$\pm$\ 	1250	&	-700	$\pm$\ 	740	\\
HE1120+0154	&	5930$\pm$\ 	420	&	-0.06	$\pm$\ 	0.17	&	0.31	$\pm$\ 	0.09	&	-1110	$\pm$\ 	1730	&	-670	$\pm$\ 	420	\\
HE1349+0007	&	6630$\pm$\ 	460	&	-0.22	$\pm$\ 	0.10	&	0.35	$\pm$\ 	0.05	&	-4090	$\pm$\ 	840	&	-3040	$\pm$\ 	460	\\
HE1409+0101	&	6370$\pm$\ 	420	&	-0.03	$\pm$\ 	0.11	&	0.36	$\pm$\ 	0.07	&	-2900	$\pm$\ 	1040	&	-2790	$\pm$\ 	420	\\
HE2156-4020	&	6420$\pm$\ 	370	&	0.01	$\pm$\ 	0.12	&	0.41	$\pm$\ 	0.07	&	-1870	$\pm$\ 	900	&	-1850	$\pm$\ 	370	\\
HE2202-2557 	&	5300$\pm$\ 	420	&	0.08	$\pm$\ 	0.10	&	0.30	$\pm$\ 	0.06	&	-1280	$\pm$\ 	1000	&	-2080	$\pm$\ 	420	\\
HE2349-3800 	&	7140$\pm$\ 	490	&	-0.20	$\pm$\ 	0.11	&	0.34	$\pm$\ 	0.07	&	-2810	$\pm$\ 	1120	&	-2070	$\pm$\ 	490	\\
HE2355-4621 	&	5740$\pm$\ 	350	&	0.04	$\pm$\ 	0.18	&	0.37	$\pm$\ 	0.09	&	190	$\pm$\ 	1360	&	10	$\pm$\ 	350	 		\\[0.08cm] \hline
\end{tabular}
\end{center}
\end{table}

\begin{table}
\setlength{\tabcolsep}{0.5pt}
\begin{center}\scriptsize
\caption{Results of {\tt specfit} analysis  on \civ\   \label{tab:specfitciv}}
\begin{tabular}{lcccccccccccrrrr}\hline\hline\noalign{\vskip 0.05cm}
\multicolumn{1}{l}{Identification}  &\multicolumn{3}{c}{BLUE}&& \multicolumn{3}{c}{BC} &&\multicolumn{3}{c}{VBC}  &&\multicolumn{3}{c}{NC} \\ \cline{2-4}\cline{6-8}\cline{10-12}\cline{14-16}\noalign{\vskip 0.05cm}
& $I/I_\mathrm{tot}$ & FWHM & Shift  && $I/I_\mathrm{tot}$ & FWHM & Shift && $I/I_\mathrm{tot}$ & FWHM & Shift  && $I/I_\mathrm{tot}$ & FWHM & Shift \\ 
& & \multicolumn{2}{c}{$\rm [km~s^{-1}]$} &&   &\multicolumn{2}{c}{$\rm [km~s^{-1}]$} & && \multicolumn{2}{c}{$\rm [km~s^{-1}]$}& && \multicolumn{2}{c}{$\rm [km~s^{-1}]$}\\ 
\hline\noalign{\vskip 0.05cm}
&\multicolumn{14}{c}{Population A}&\\[0.05cm] \hline\noalign{\vskip 0.05cm}
HE0043-2300	&	0.28	&	3840	&	-1600	&	&	0.65	&	4420	&	120	&	&	\ldots	&	\ldots	&	\ldots	&	&	0.07	&	1560	&	30	\\
HE0109-3518	&	0.27	&	3020	&	-1490	&	&	0.64	&	2660	&	-90	&	&	\ldots	&	\ldots	&	\ldots	&	&	0.09	&	970	&	110	\\
HE0122-3759 	&	0.78	&	7520	&	-3350	&	&	0.22	&	3260	&	-90	&	&	\ldots	&	\ldots	&	\ldots	&	&	0.00	&	\ldots	&	\ldots	\\
HE0205-3756	&	0.73	&	4640	&	-2550	&	&	0.26	&	4430	&	80	&	&	\ldots	&	\ldots	&	\ldots	&	&	0.00	&	1300	&	-60	\\
HE0248-3628	&	0.63	&	5140	&	-3570	&	&	0.34	&	4070	&	80	&	&	\ldots	&	\ldots	&	\ldots	&	&	0.03	&	1410	&	-600	\\
HE0251-5550	&	0.58	&	4170	&	-1240	&	&	0.42	&	5050	&	80	&	&	\ldots	&	\ldots	&	\ldots	&	&	0.00	&	1300	&	-160	\\
HE0359-3959	&	0.93	&	7940	&	-5170	&	&	0.07	&	4050	&	80	&	&	\ldots	&	\ldots	&	\ldots	&	&	0.00	&	\ldots	&	\ldots	\\
HE0507-3236 	&	0.48	&	4720	&	-2050	&	&	0.52	&	4210	&	80	&	&	\ldots	&	\ldots	&	\ldots	&	&	0.00	&	\ldots	&	\ldots	\\
HE0512-3329 	&	0.58	&	5000	&	-3200	&	&	0.42	&	3700	&	80	&	&	\ldots	&	\ldots	&	\ldots	&	&	0.00	&	\ldots	&	\ldots	\\
HE0940-1050	&	0.53	&	6270	&	-3250	&	&	0.47	&	4360	&	80	&	&	\ldots	&	\ldots	&	\ldots	&	&	0.00	&	\ldots	&	\ldots	\\
HE1104-1805 	&	0.31	&	4610	&	-2660	&	&	0.67	&	4650	&	80	&	&	\ldots	&	\ldots	&	\ldots	&	&	0.02	&	1300	&	-350	\\
HE1347-2457	&	0.95	&	7720	&	-4870	&	&	0.05	&	5880	&	80	&	&	\ldots	&	\ldots	&	\ldots	&	&	0.00	&	\ldots	&	\ldots	\\
HE2147-3212	&	0.67	&	6650	&	-3100	&	&	0.32	&	5550	&	80	&	&	\ldots	&	\ldots	&	\ldots	&	&	0.02	&	1470	&	-1230	\\
HE2352-4010 	&	0.66	&	3760	&	-2680	&	&	0.34	&	3720	&	80	&	&	\ldots	&	\ldots	&	\ldots	&	&	0.00	&	\ldots	&	\ldots	\\[0.08cm]
\hline\noalign{\vskip 0.05cm}
&\multicolumn{14}{c}{Population B}&\\[0.05cm] \hline
\noalign{\vskip 0.05cm}
HE0035-2853	&	0.43	&	5530	&	-2180	&	&	0.35	&	6630	&	40	&	&	0.17	&	11450	&	5480	&	&	0.04	&	1530	&	-260	\\
HE0058-3231	&	0.18	&	5200	&	-2120	&	&	0.43	&	4960	&	-60	&	&	0.36	&	10620	&	3120	&	&	0.03	&	1130	&	-170	\\
HE0203-4627	&	0.31	&	5520	&	-2210	&	&	0.43	&	5870	&	90	&	&	0.19	&	12390	&	2260	&	&	0.06	&	1730	&	70	\\
HE0349-5249	&	0.52	&	5690	&	-2090	&	&	0.21	&	4560	&	-10	&	&	0.25	&	11360	&	2440	&	&	0.01	&	1050	&	10	\\
HE0436-3709	&	0.48	&	5670	&	-2650	&	&	0.24	&	5430	&	80	&	&	0.28	&	11290	&	2620	&	&	0.00	&	\ldots	&	\ldots	\\
HE0926-0201 	&	0.82	&	5560	&	-2000	&	&	0.10	&	5090	&	80	&	&	0.07	&	13230	&	810	&	&	0.01	&	1030	&	50	\\
HE1039-0724	&	0.29	&	4350	&	-1760	&	&	0.33	&	9190	&	80	&	&	0.30	&	12170	&	2500	&	&	0.07	&	2100	&	-250	\\
HE1120+0154	&	0.39	&	4400	&	-1140	&	&	0.27	&	6200	&	360	&	&	0.25	&	14080	&	3020	&	&	0.09	&	1600	&	-180	\\
HE1349+0007	&	0.87	&	5980	&	-3020	&	&	0.11	&	5440	&	80	&	&	0.00	&	13080	&	3190	&	&	0.02	&	2000	&	0	\\
HE1409+0101	&	0.65	&	5390	&	-2920	&	&	0.13	&	7020	&	80	&	&	0.20	&	14630	&	2310	&	&	0.02	&	1400	&	-140	\\
HE2156-4020	&	0.54	&	5000	&	-2520	&	&	0.17	&	4740	&	-30	&	&	0.28	&	13090	&	1420	&	&	0.01	&	860	&	-480	\\
HE2202-2557 	&	0.42	&	4490	&	-1940	&	&	0.27	&	7050	&	80	&	&	0.21	&	12420	&	3180	&	&	0.11	&	1210	&	-170	\\
HE2349-3800 	&	0.51	&	5710	&	-2760	&	&	0.19	&	3640	&	30	&	&	0.26	&	11160	&	1340	&	&	0.03	&	1090	&	80	\\
HE2355-4621 	&	0.04	&	5900	&	-3830	&	&	0.41	&	4660	&	0	&	&	0.42	&	14910	&	2150	&	&	0.13	&	1680	&	230	\\[0.05cm]
\hline
\end{tabular}
\end{center}
\end{table}

\subsection{HE sample \civ}

We identified 14 Pop. A and 14 Pop. B sources in the HE sample.  Their rest frame spectra are shown in Fig. \ref{fig:speca}  and Fig. \ref{fig:specb}, respectively.  The first columns of Table \ref{tab:civhb}  report  the rest frame specific continuum flux $f_{\lambda}$\  at 1450 \AA\ as well as the flux and equivalent width of \civonly. The total flux of the 1400 feature (interpreted as a blend of \oiv\ and \siiv\ emission) is reported in the last column. It will not be further considered in the present paper but will be used in a follow-up paper in preparation.  Uncertainties on $f_{\lambda}$\ are derived from the magnitude uncertainties used in the rescaling, and are at 1$\sigma$\ confidence level.  Measurements of the integrated line profile for \civonly\ are reported in Table \ref{tab:civprof} which yields, in the following order, FWHM, asymmetry index,   kurtosis, and centroids at $\frac{1}{4}$\  and $\frac{1}{2}$\  fractional intensity, along with the uncertainty at   {2$\sigma$}\ confidence level. The results of the {\tt specfit} analysis  are shown in Fig. \ref{fig:specfita} and \ref{fig:specfitb} for Pop. A and B sources respectively. The \oiiiopt\ emission is modeled with a core and a blueshifted component \citep{zhangetal11,marzianietal16a}.   The \heiiuv\ emission on the red side of \civonly\ has been modeled  with BLUE and BC FWHM and shifts as for \civonly, leaving only relative intensities as free parameters. The   parameter values for the individual components are listed in Table \ref{tab:specfitciv}. Flux normalized by the total line flux (including NC), FWHM, and centroid at $\frac{9}{10}$ fractional intensity  are reported for BLUE, BC, NC, and VBC (the VBC only in the case of Pop. B). 

The amplitude of the \civonly\ blueshifts is remarkable. Population A \cmp\ blueshifts exceed 2000 \kms\ in almost all cases, which is a shift amplitude extremely rare at $\log L \lesssim 47$\ [\ergss] \citep{richardsetal11}. In some extreme cases, the \civonly\ profile appears to be fully blueshifted with respect to the rest frame.  Shifts of comparable amplitude are observed only at high luminosity  $\log L \gtrsim 47$\ [\ergss]  \citep[e.g.,][ and references therein]{richardsetal11,coatmanetal16,coatmanetal16a}.
Large shifts above 1000 \kms\ in amplitude are also observed in the \civonly\ profile of most Pop. B sources.

\subsection{HE sample \hb\ measurements paired to \civ}

\begin{figure*}[htp!]
\centering
\includegraphics[width=0.34\columnwidth]{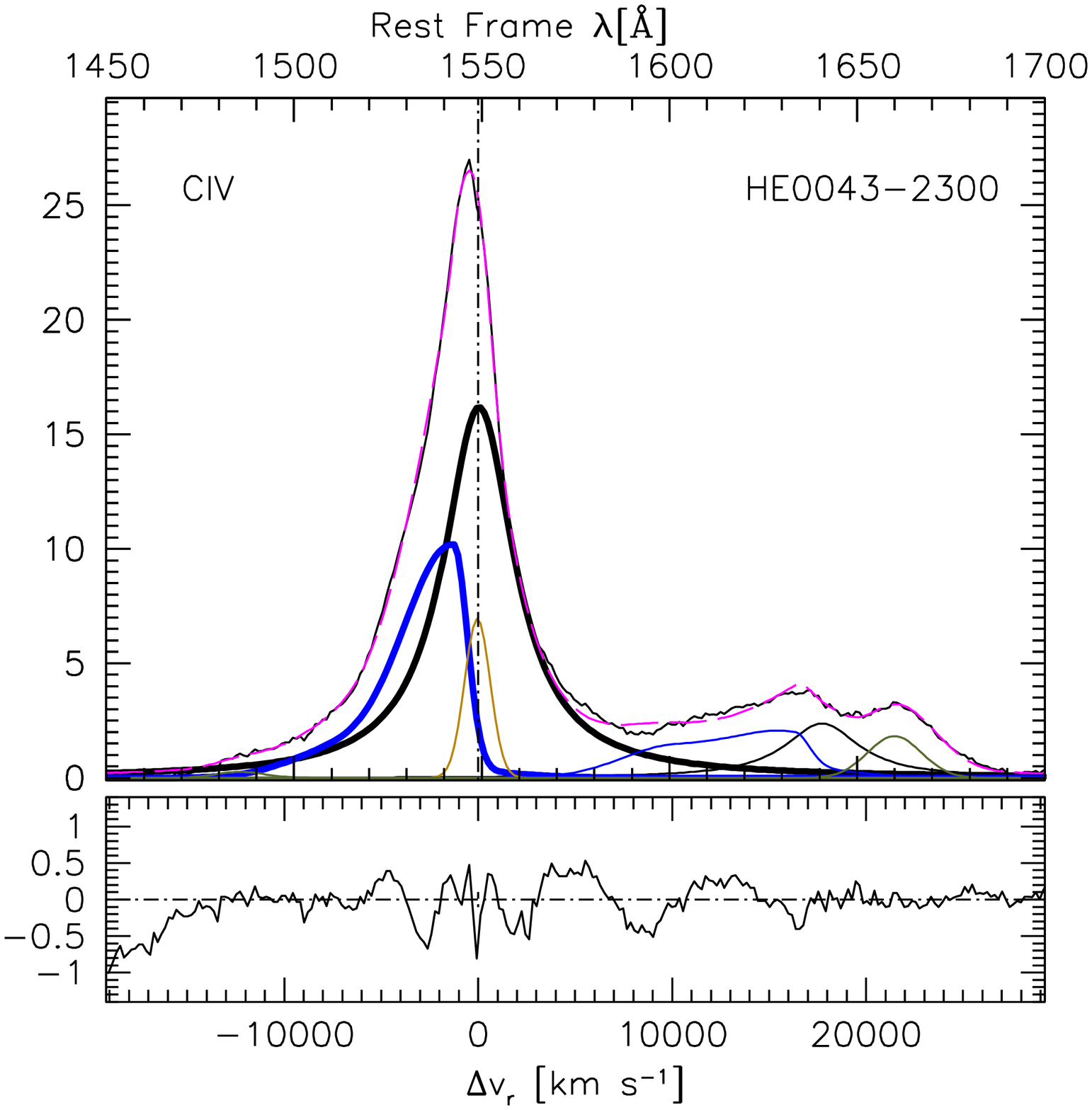}
\includegraphics[width=0.34\columnwidth]{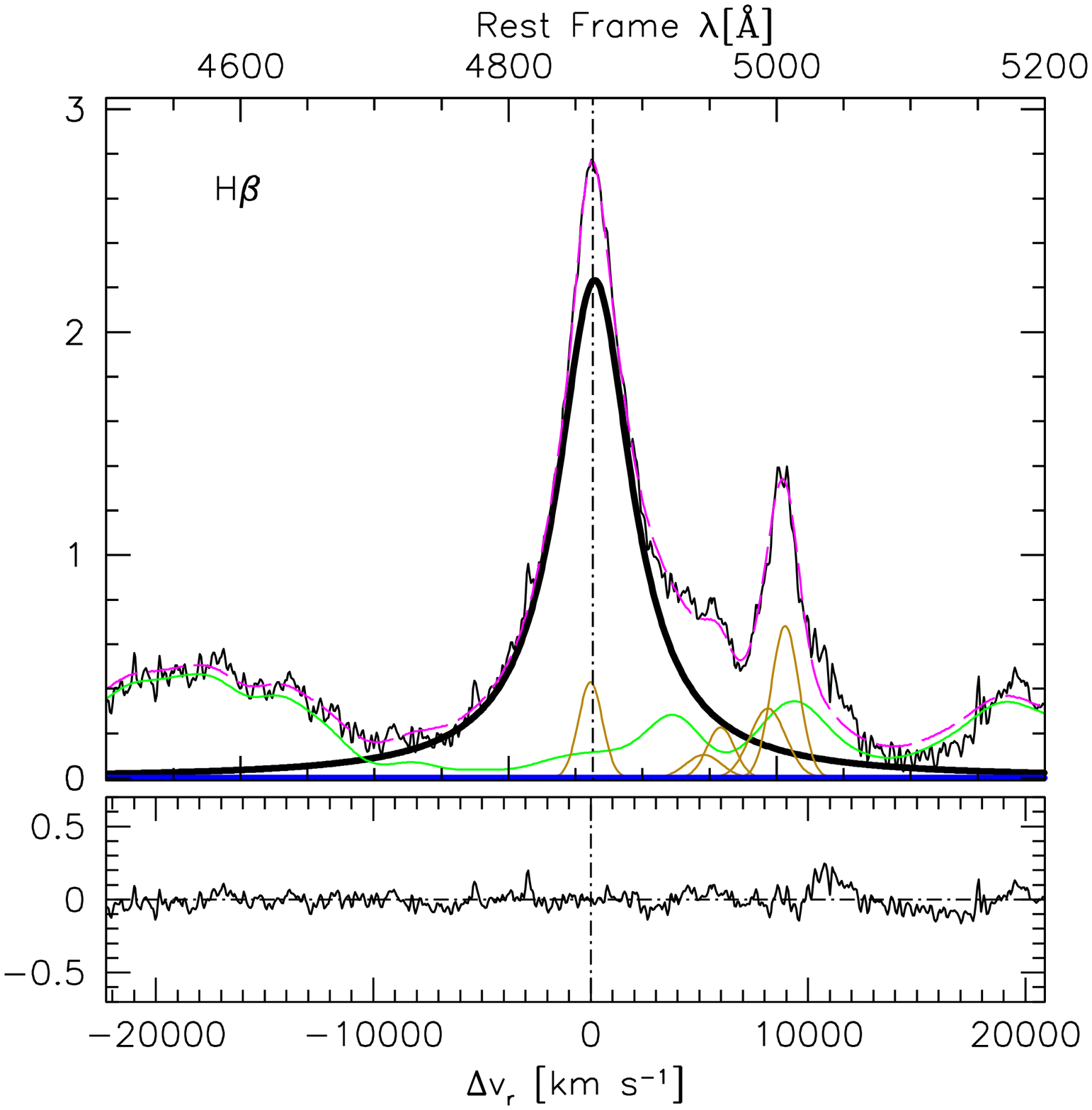}
\includegraphics[width=0.34\columnwidth]{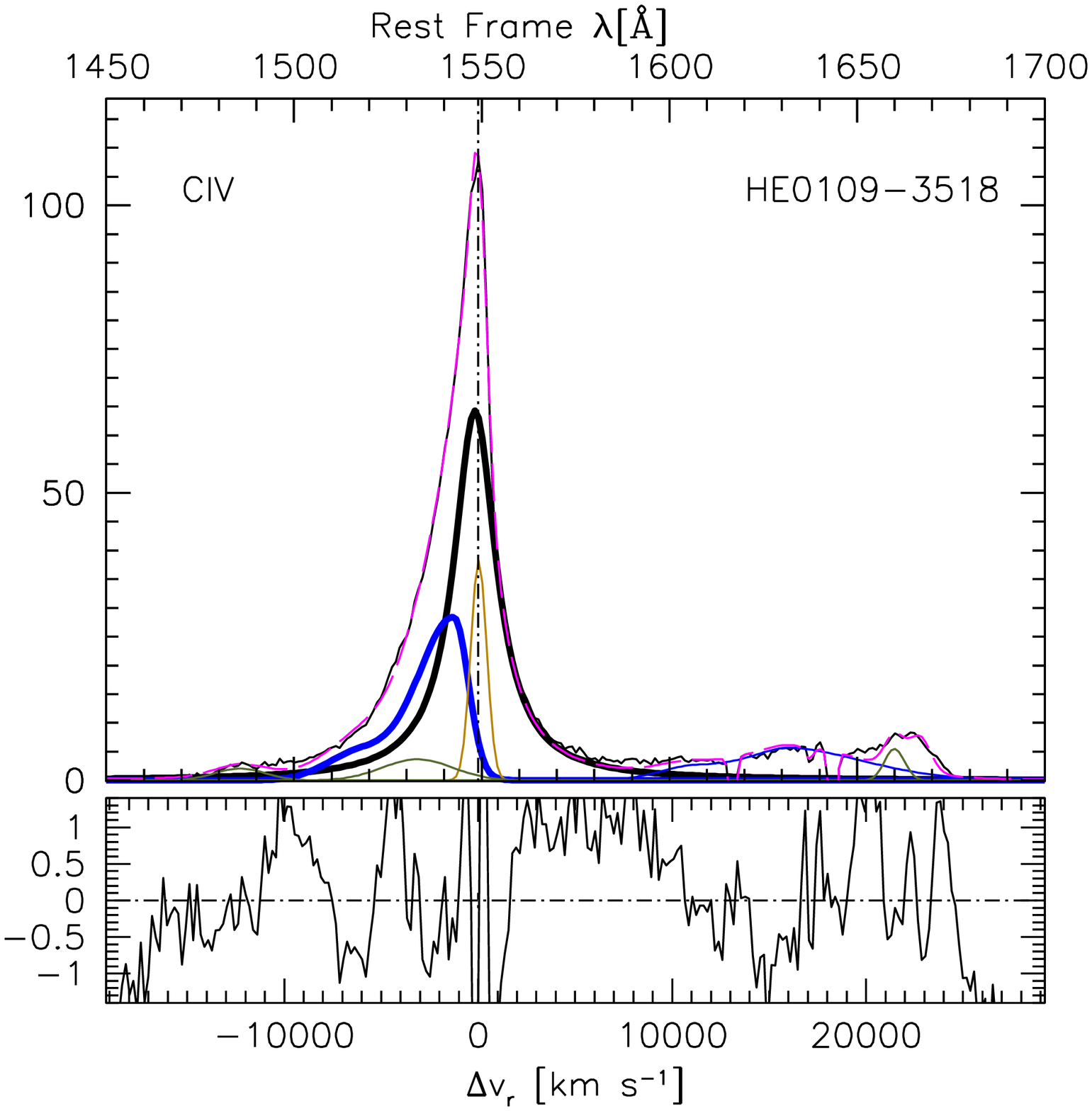}
\includegraphics[width=0.34\columnwidth]{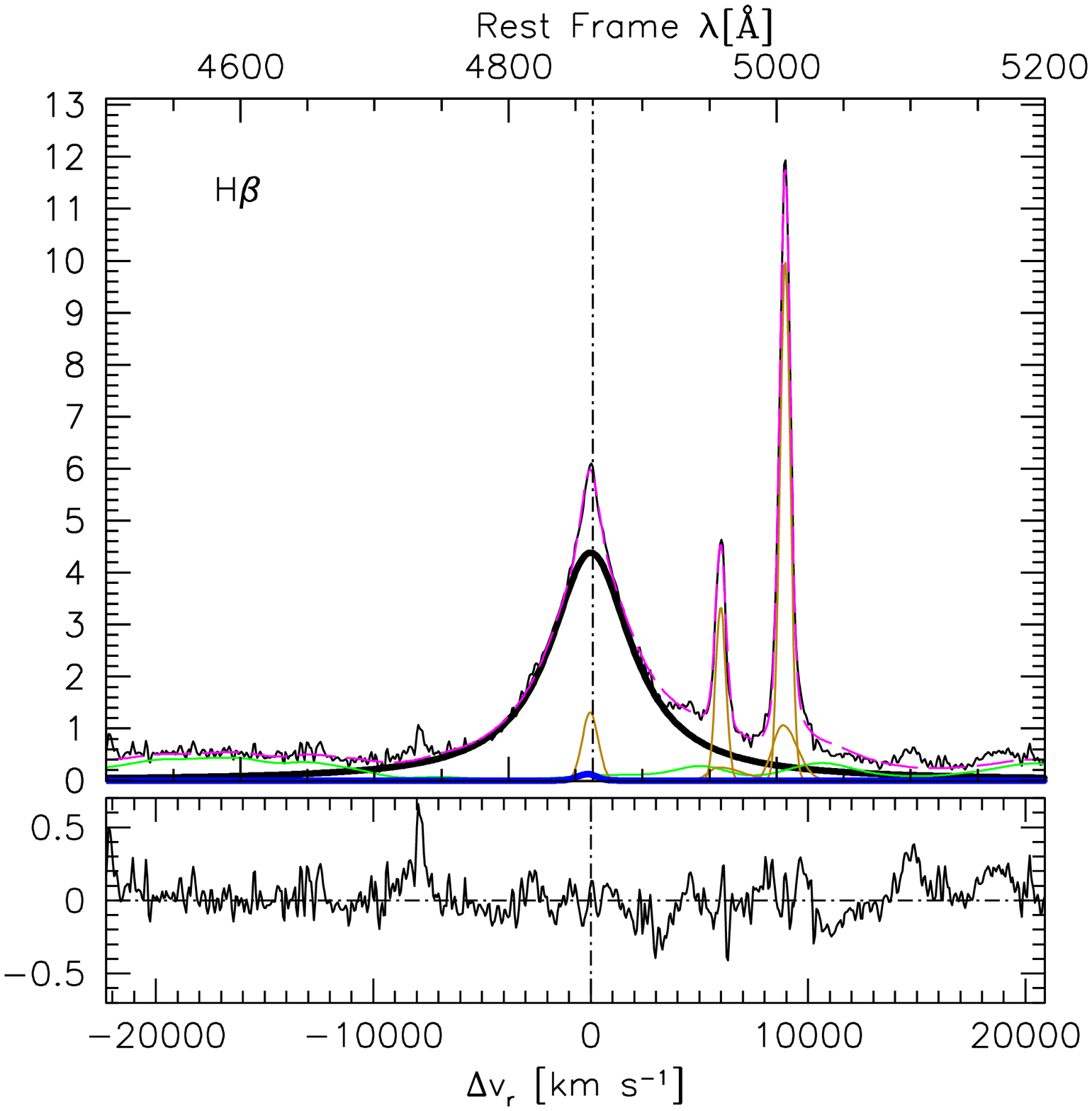}\\
\includegraphics[width=0.34\columnwidth]{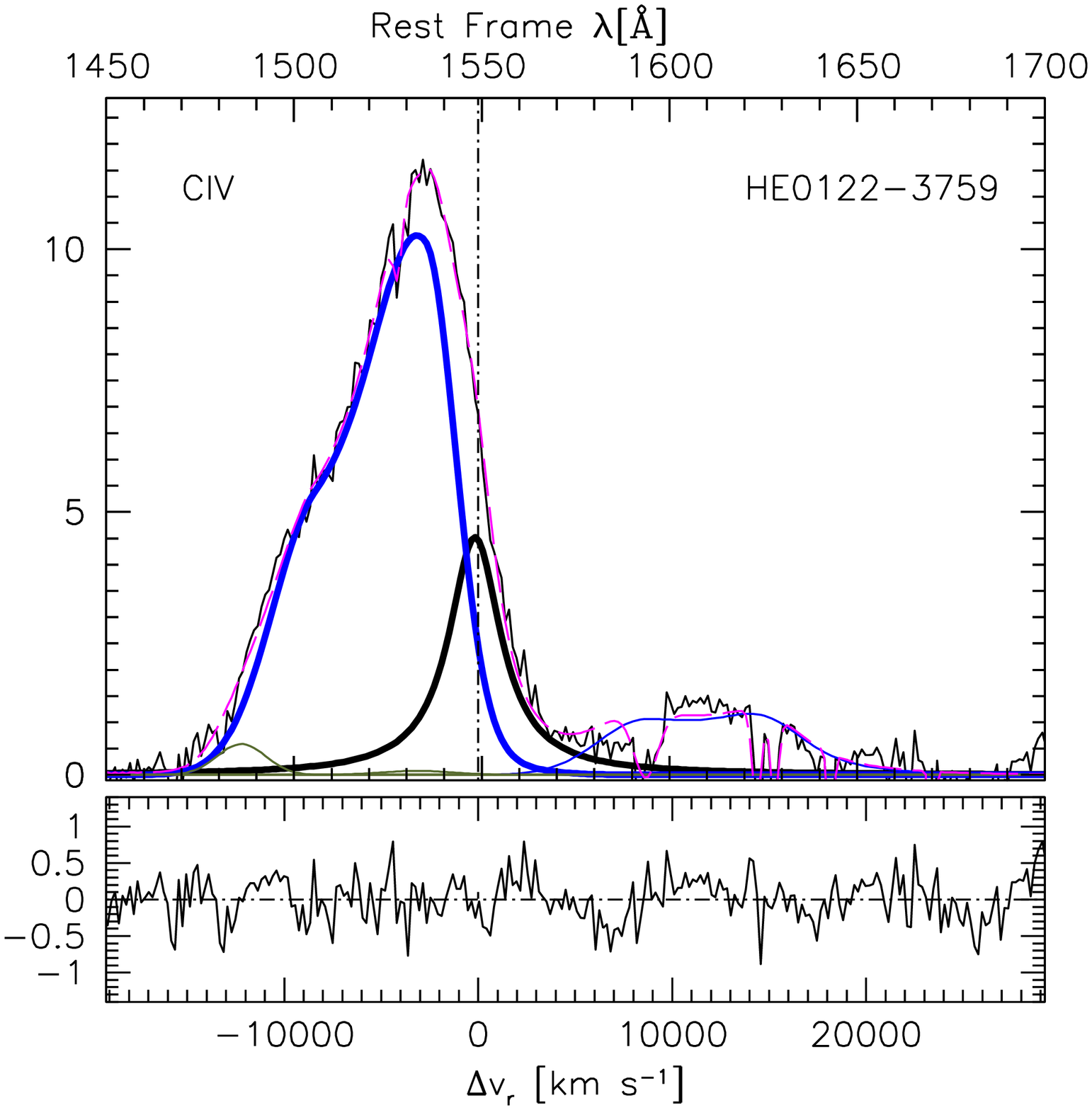}
\includegraphics[width=0.34\columnwidth]{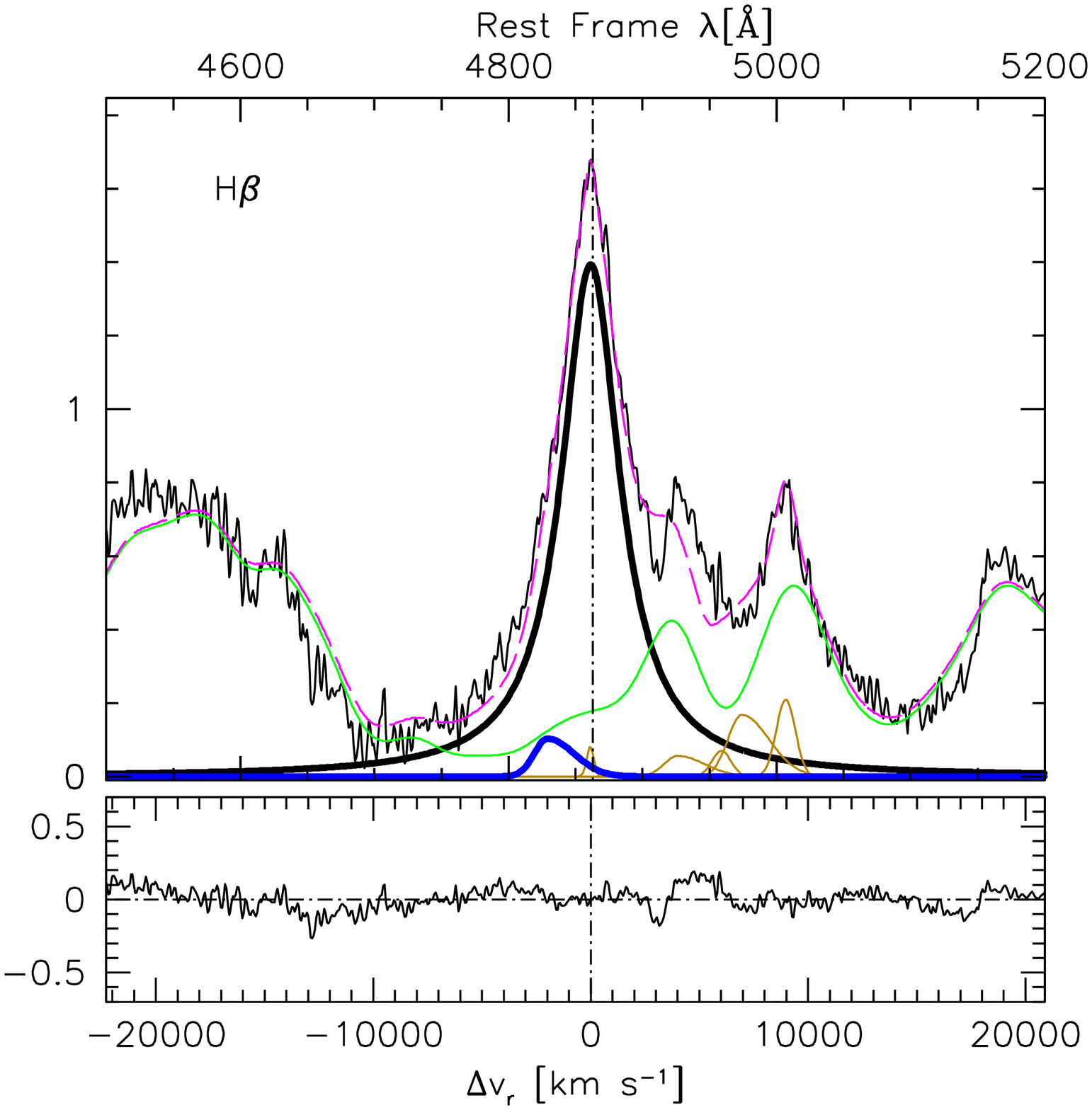} 
\includegraphics[width=0.34\columnwidth]{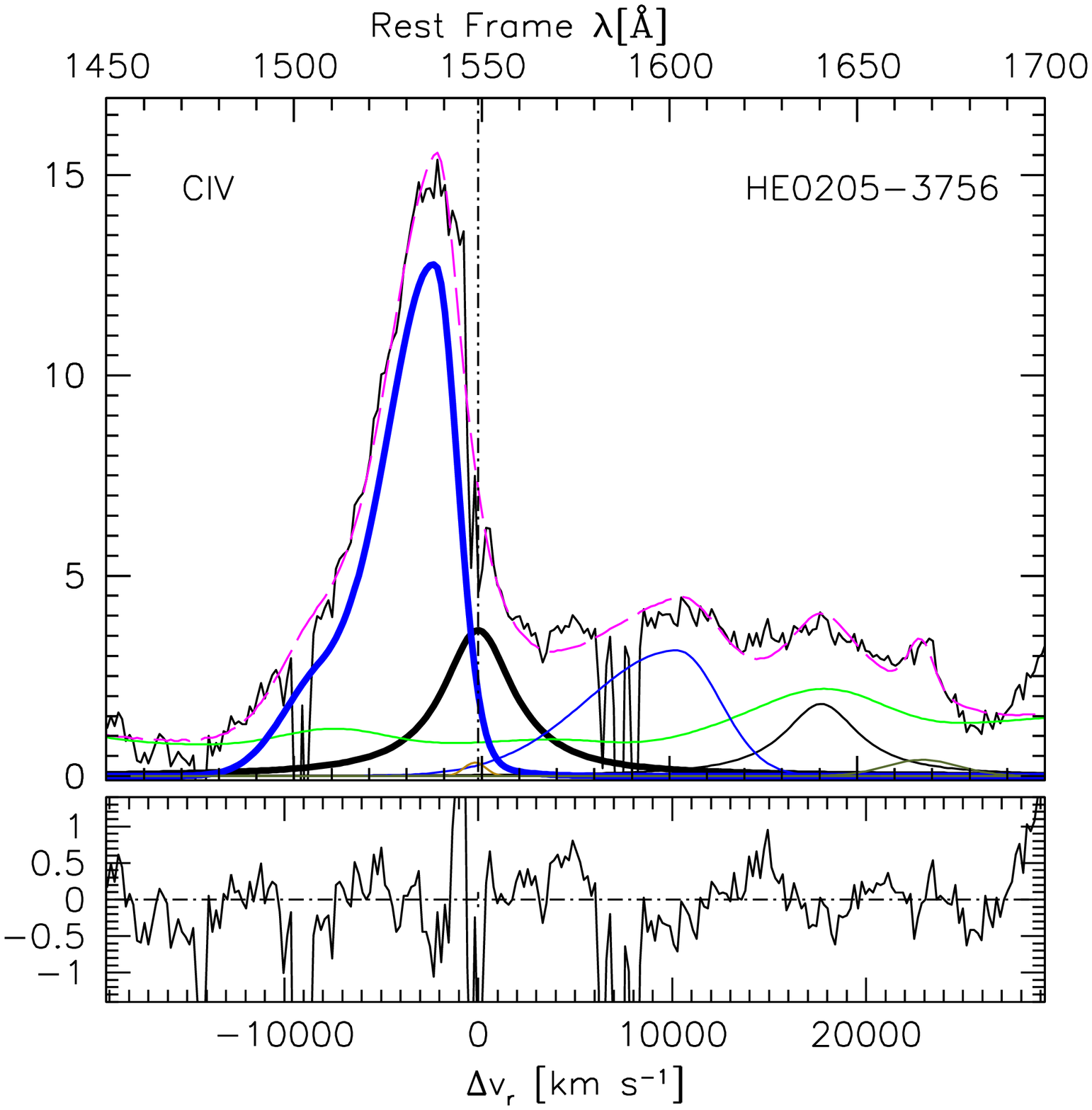}
\includegraphics[width=0.34\columnwidth]{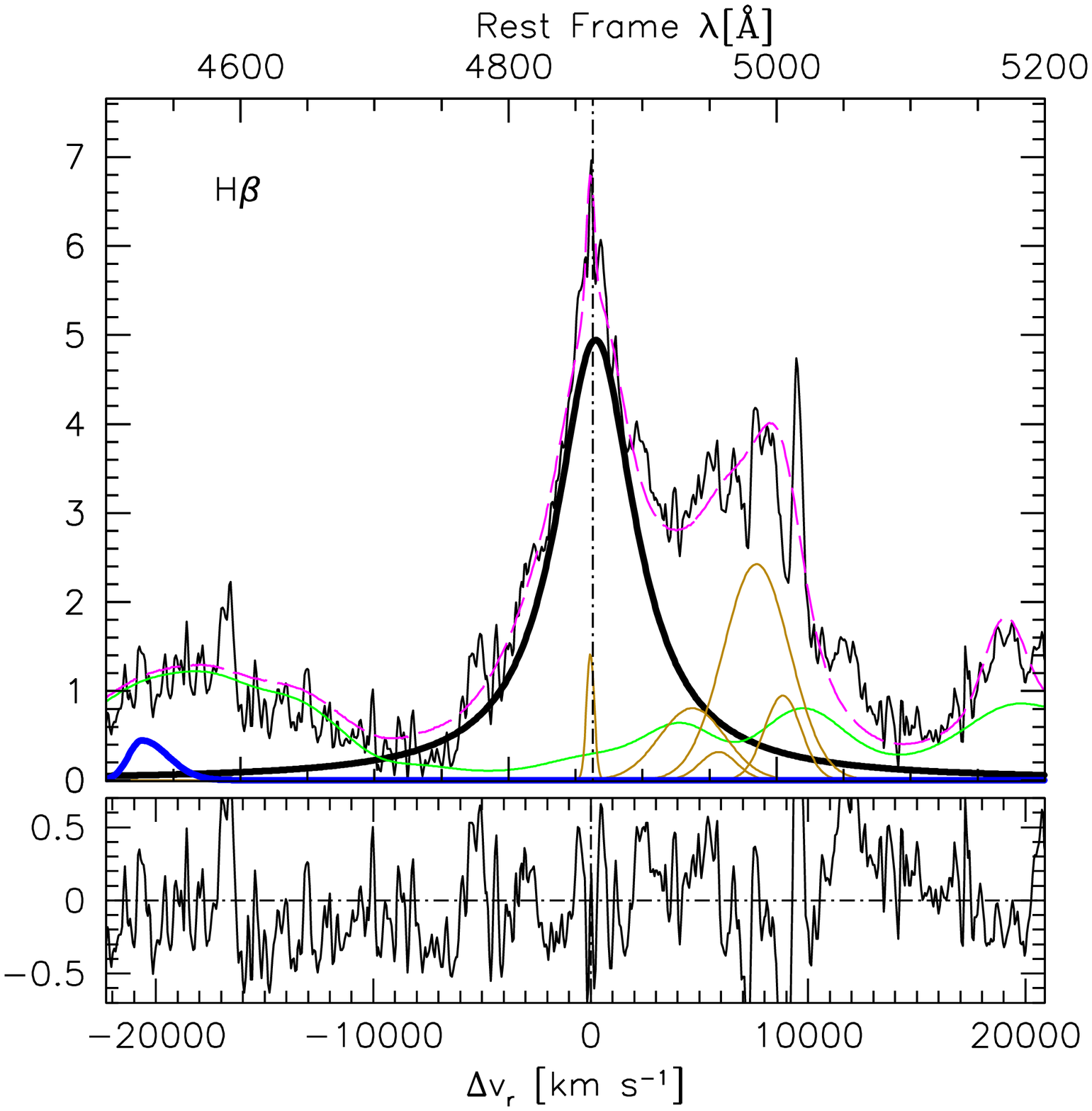}\\
\includegraphics[width=0.34\columnwidth]{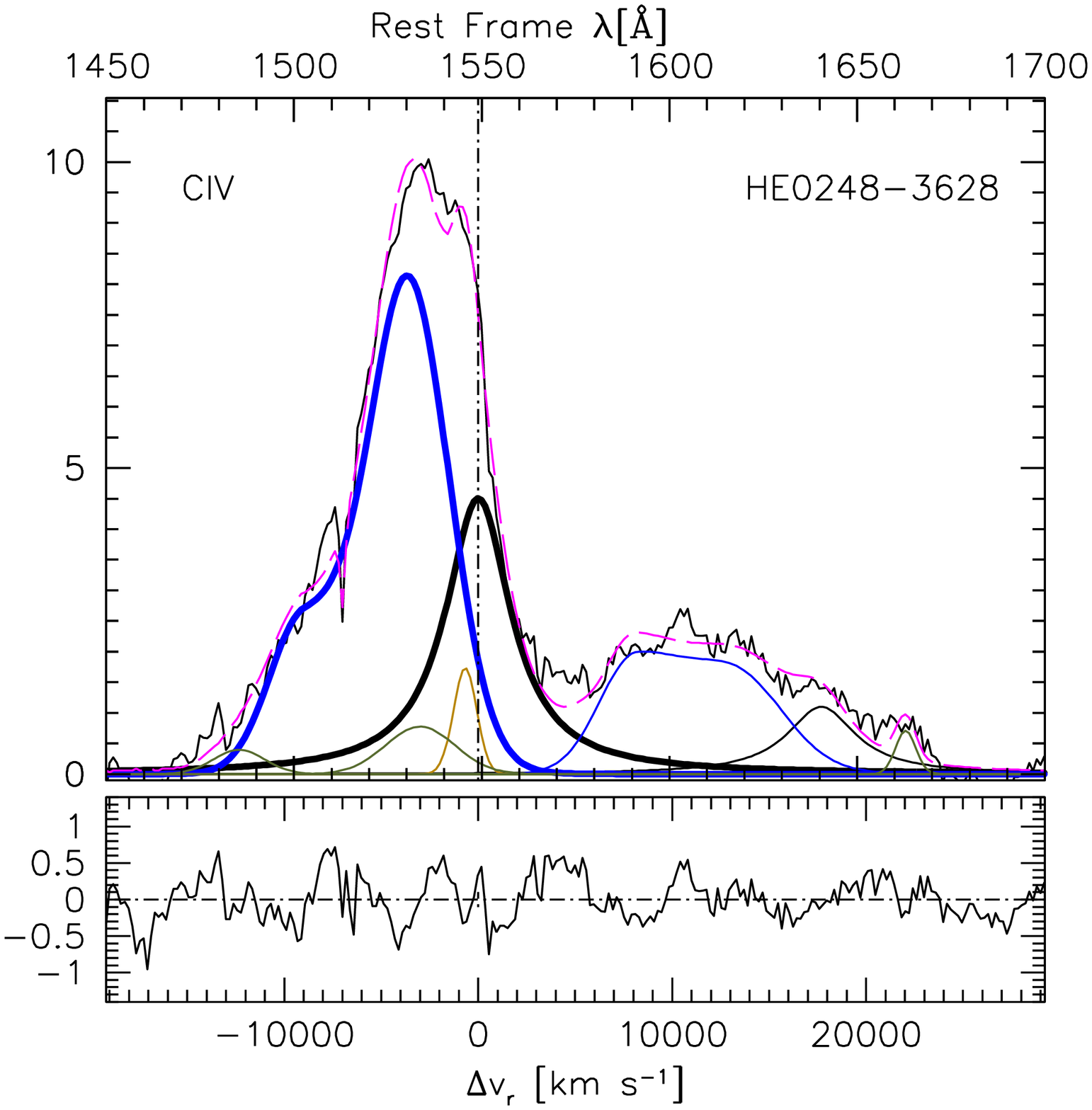}
\includegraphics[width=0.34\columnwidth]{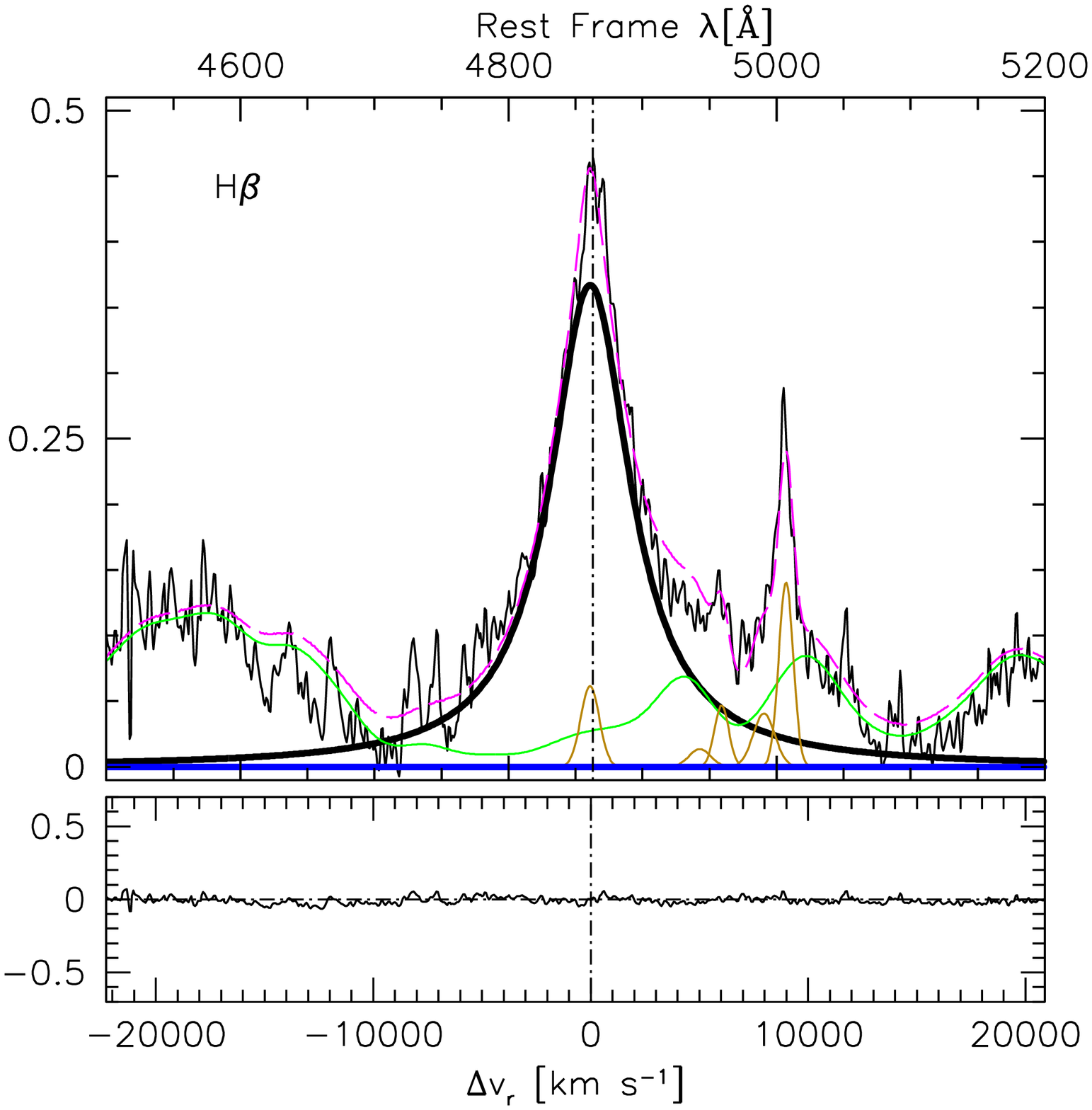}
\includegraphics[width=0.34\columnwidth]{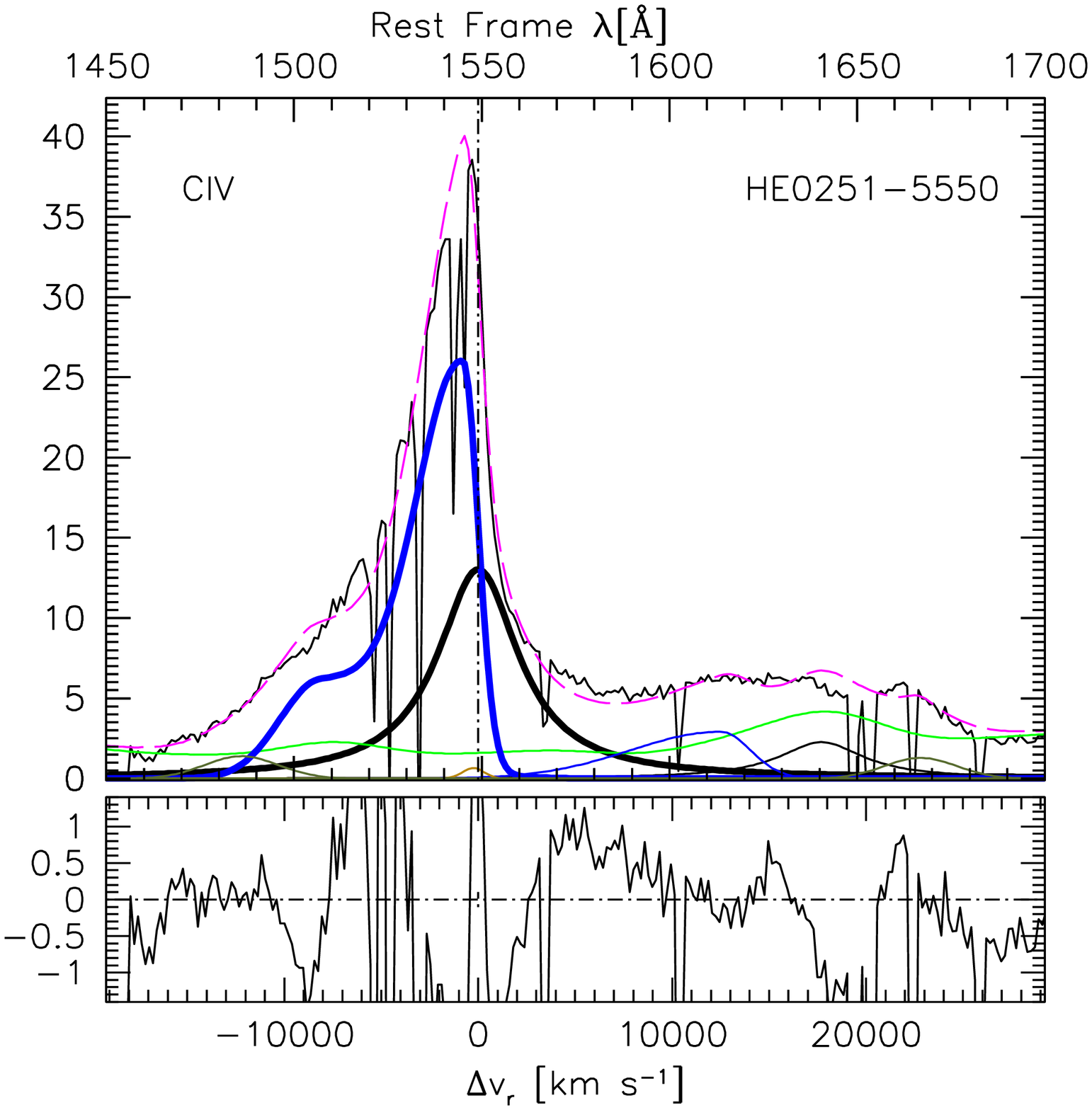}
\includegraphics[width=0.34\columnwidth]{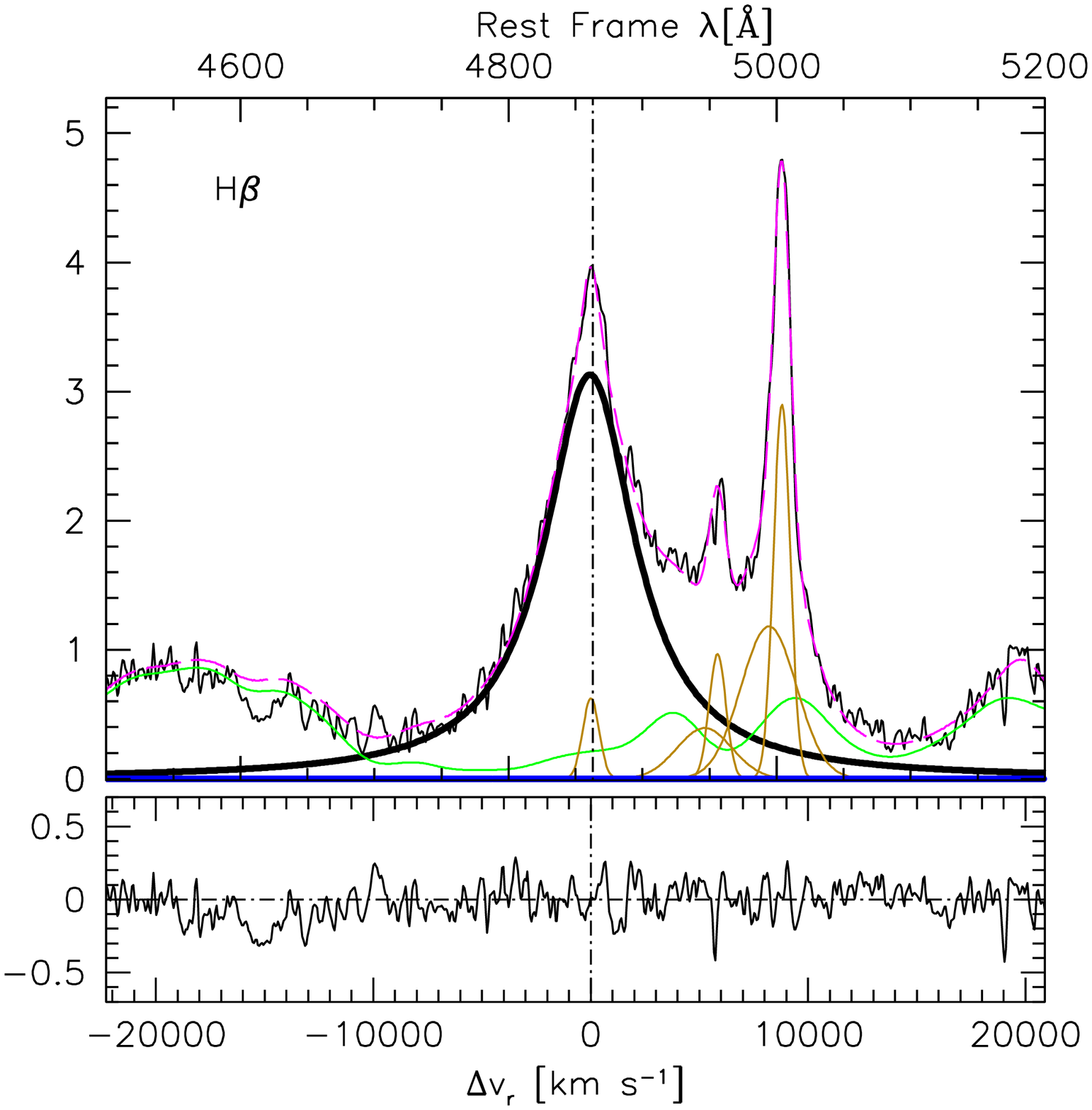}\\
\includegraphics[width=0.34\columnwidth]{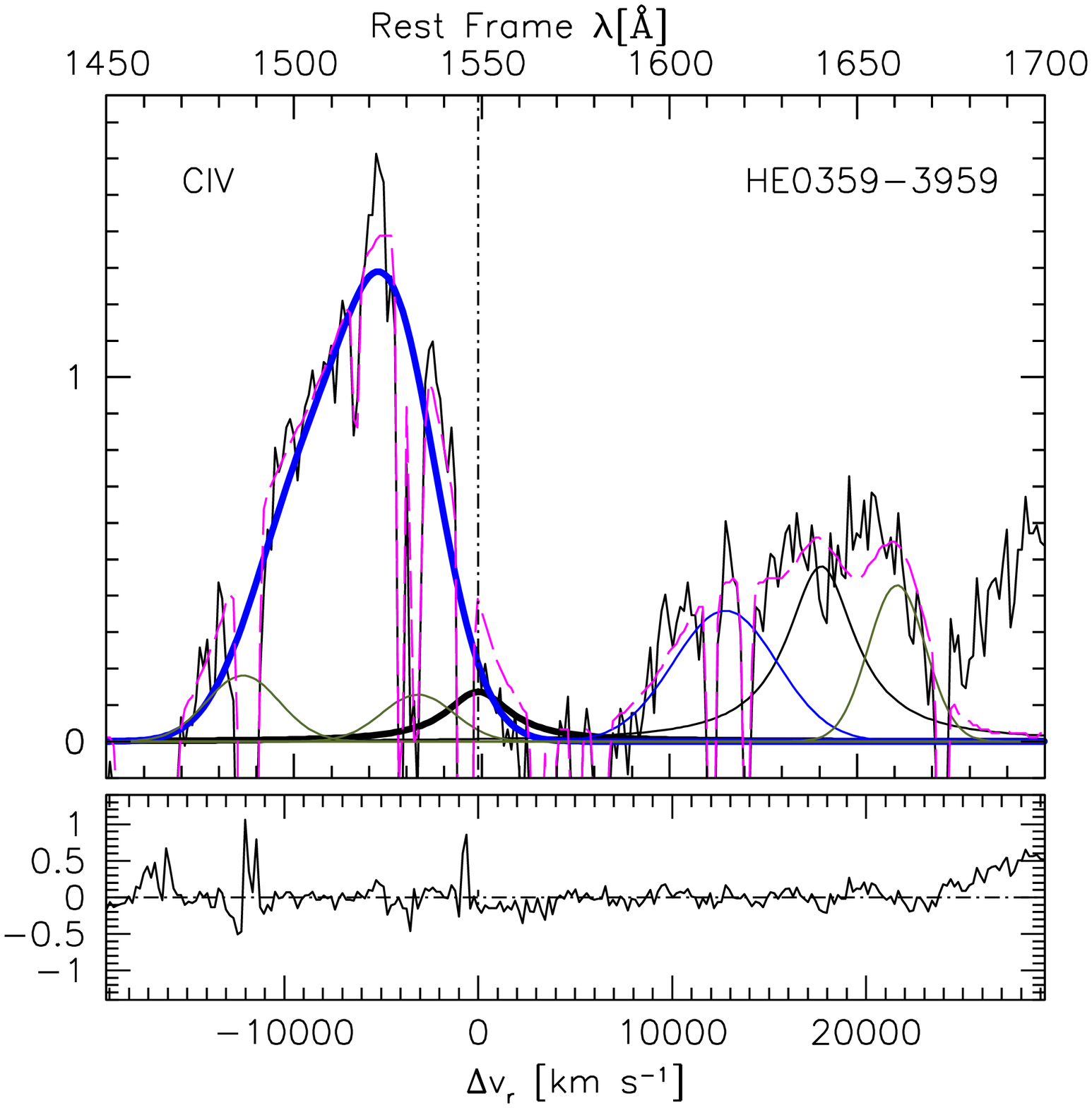}
\includegraphics[width=0.34\columnwidth]{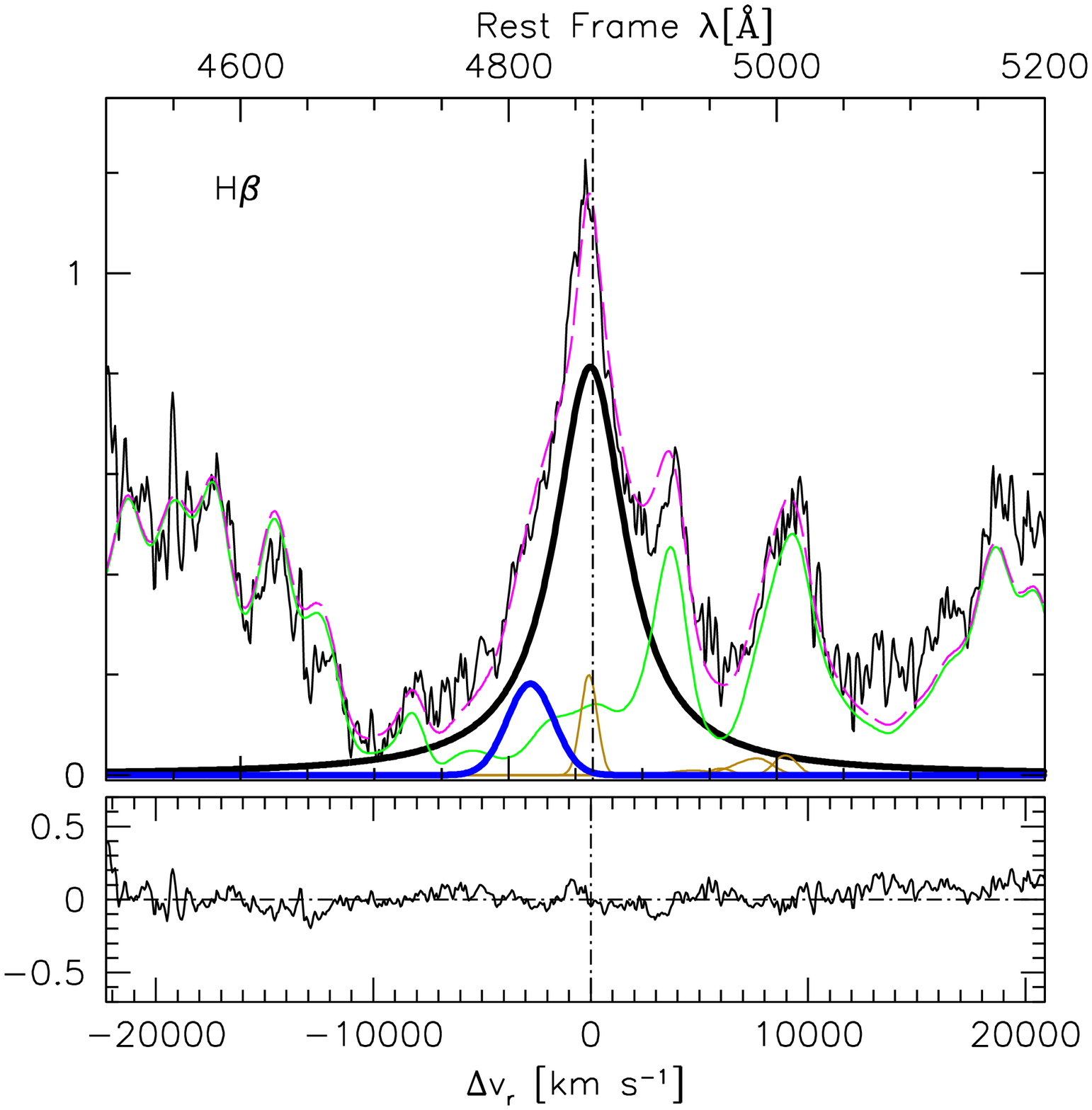}
\includegraphics[width=0.34\columnwidth]{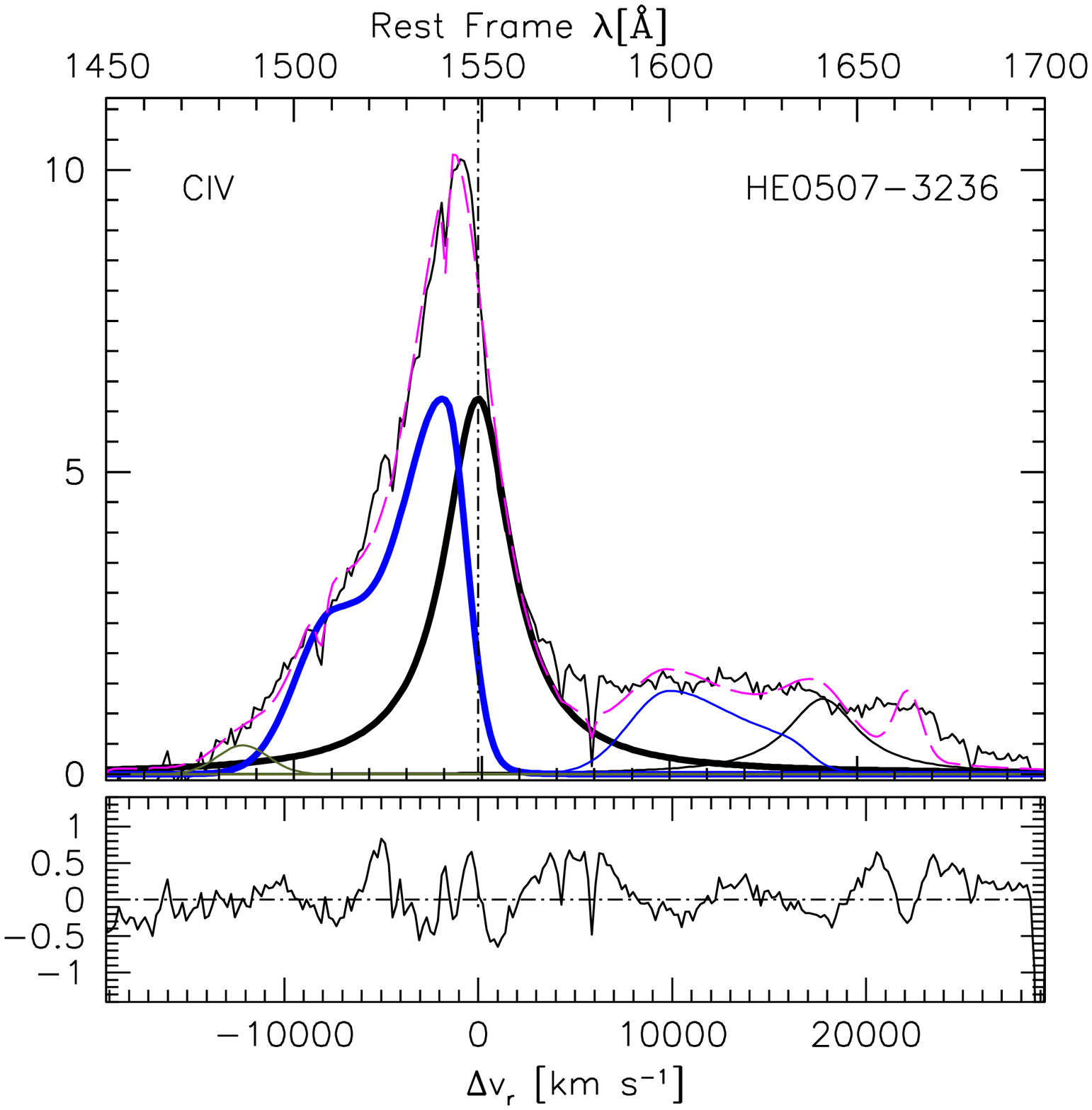}
\includegraphics[width=0.34\columnwidth]{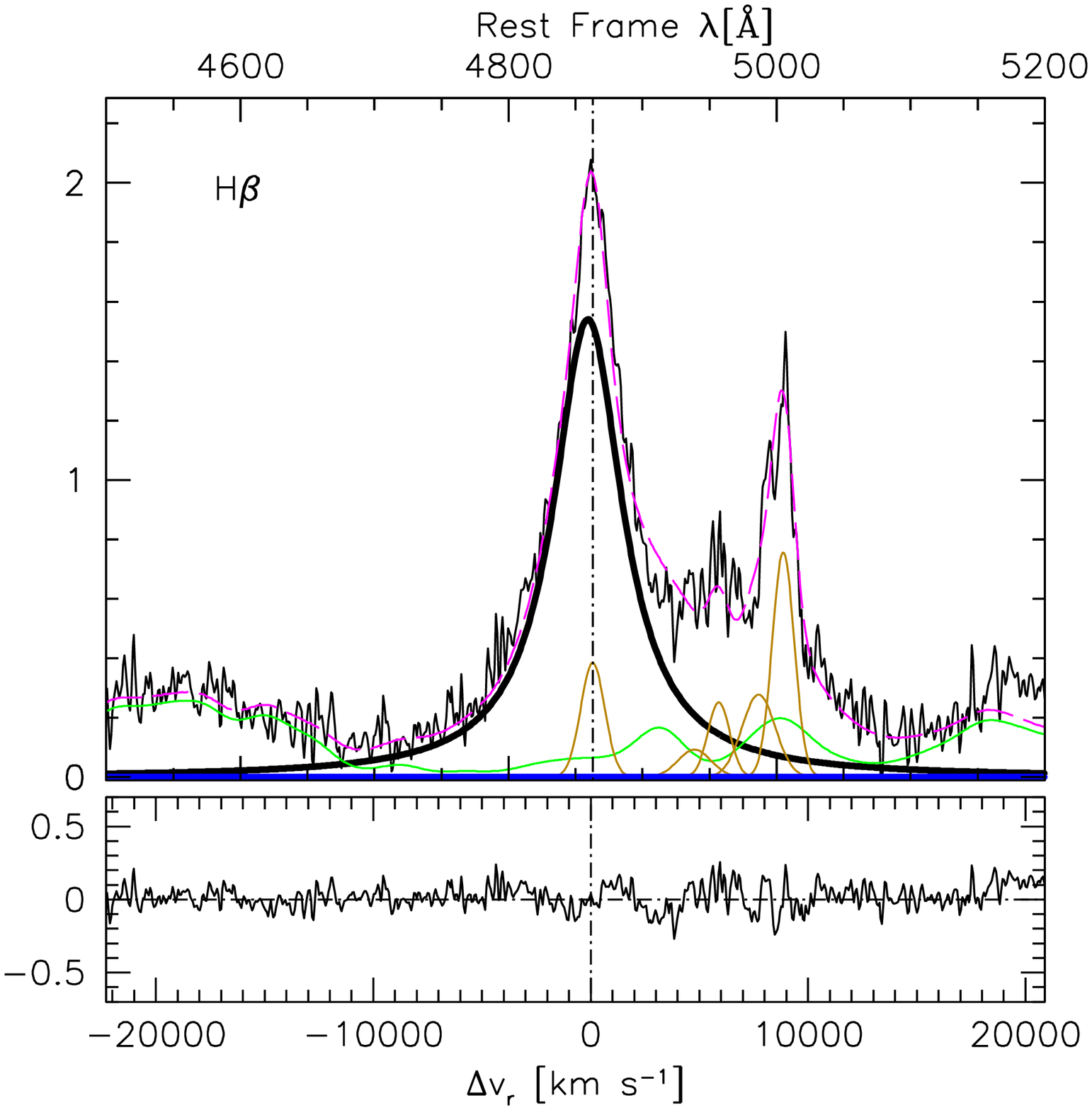}\\
\includegraphics[width=0.34\columnwidth]{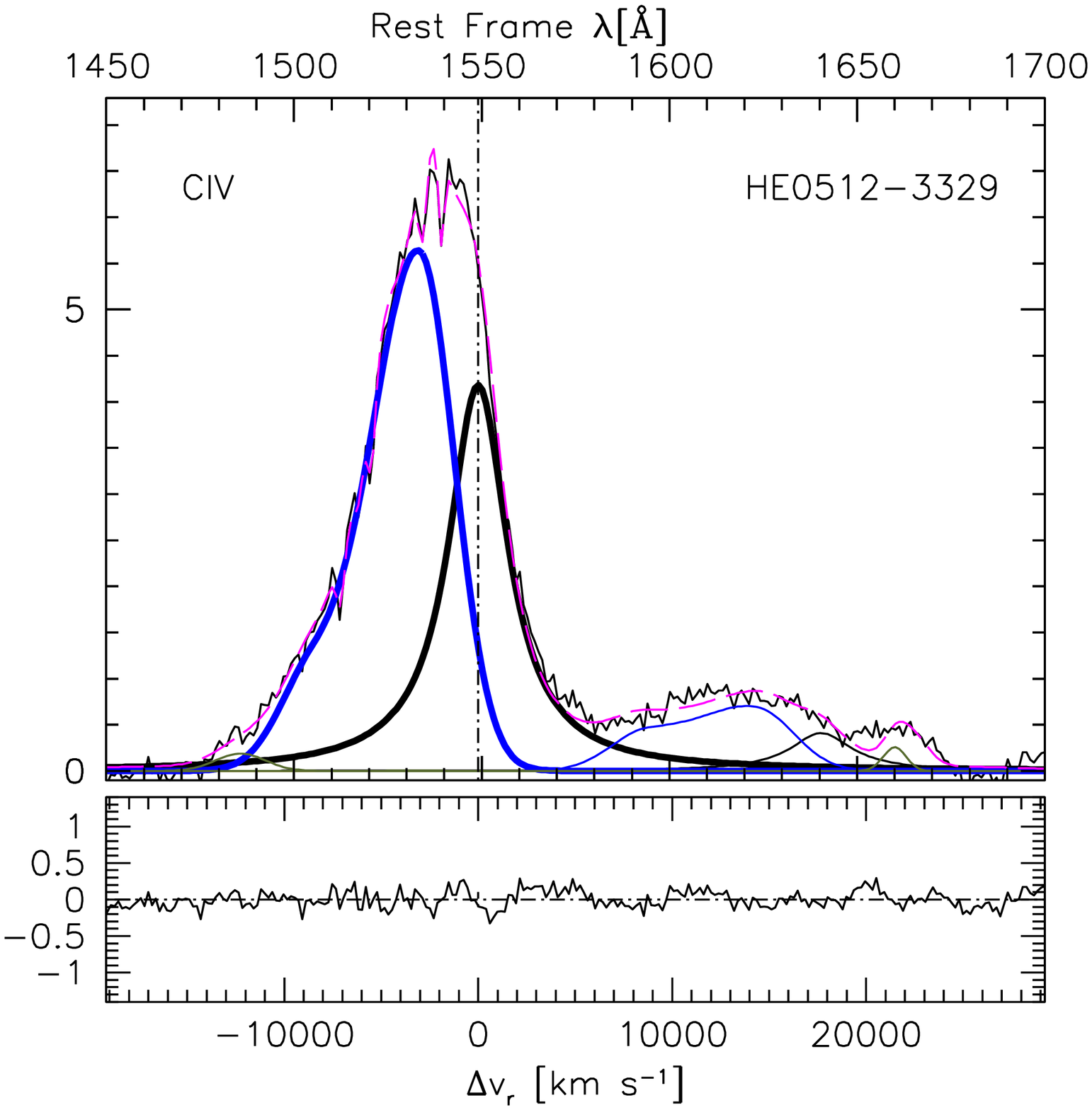}
\includegraphics[width=0.34\columnwidth]{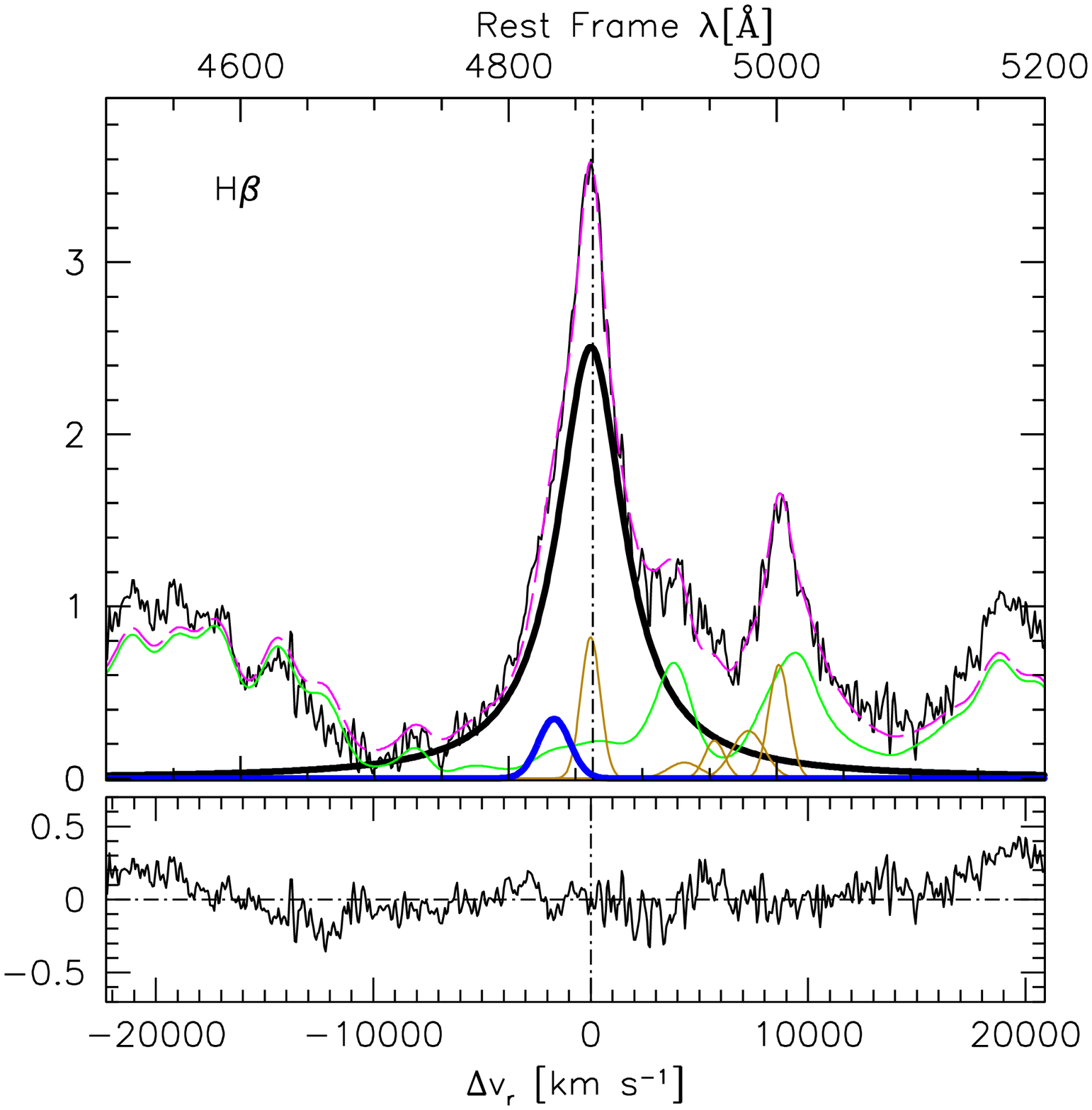}
\includegraphics[width=0.34\columnwidth]{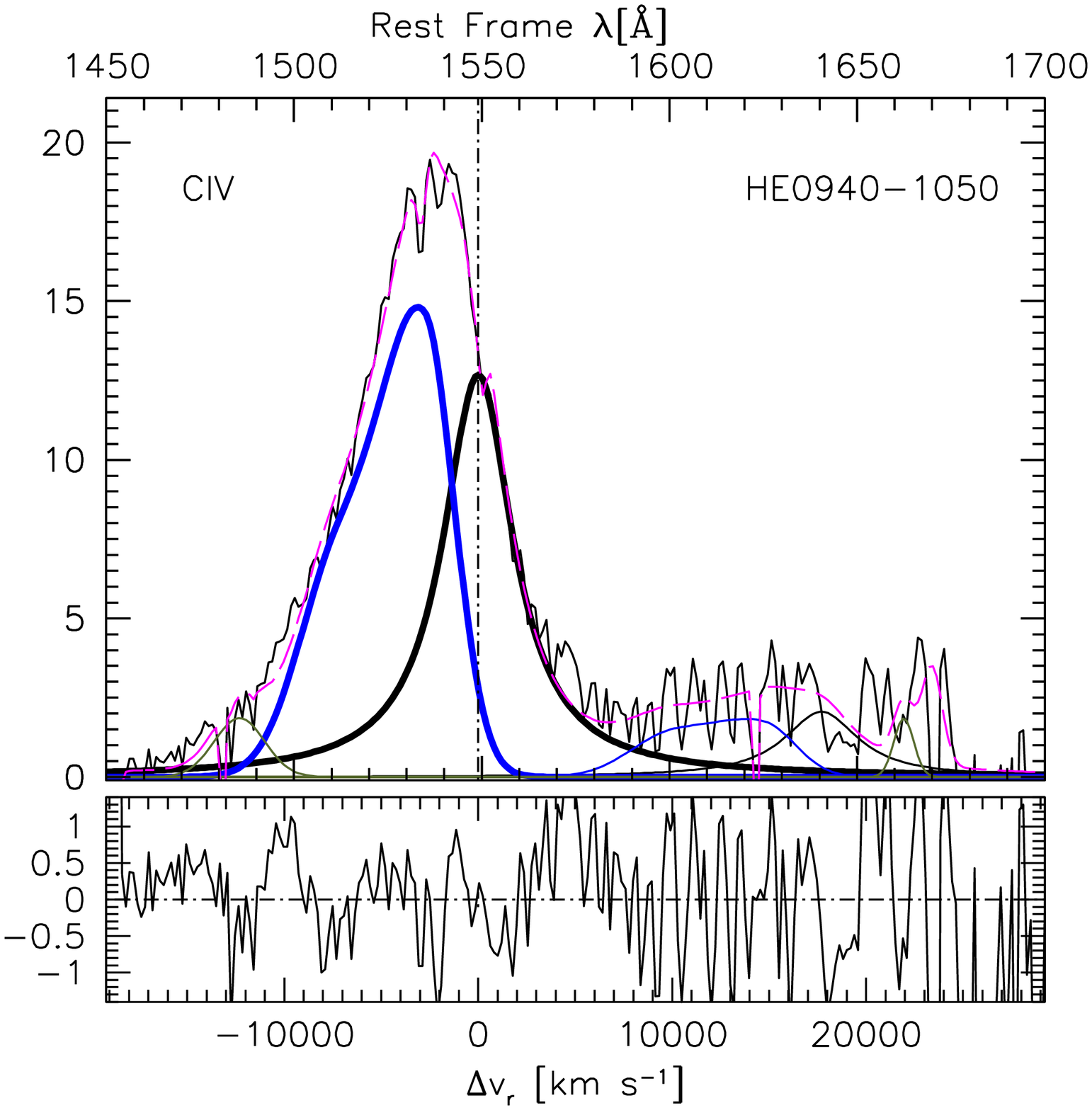} 
\includegraphics[width=0.34\columnwidth]{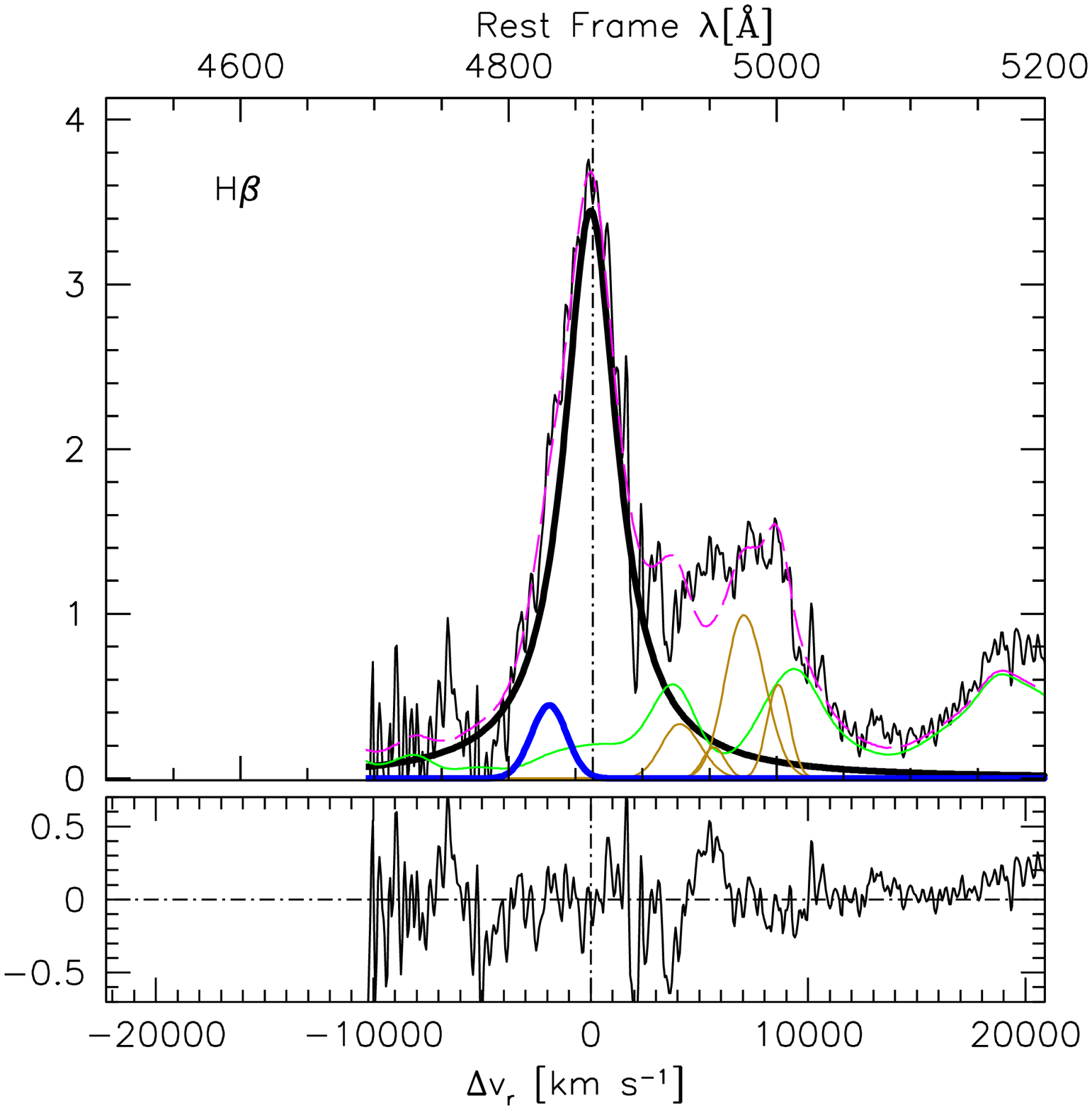}\\
\includegraphics[width=0.34\columnwidth]{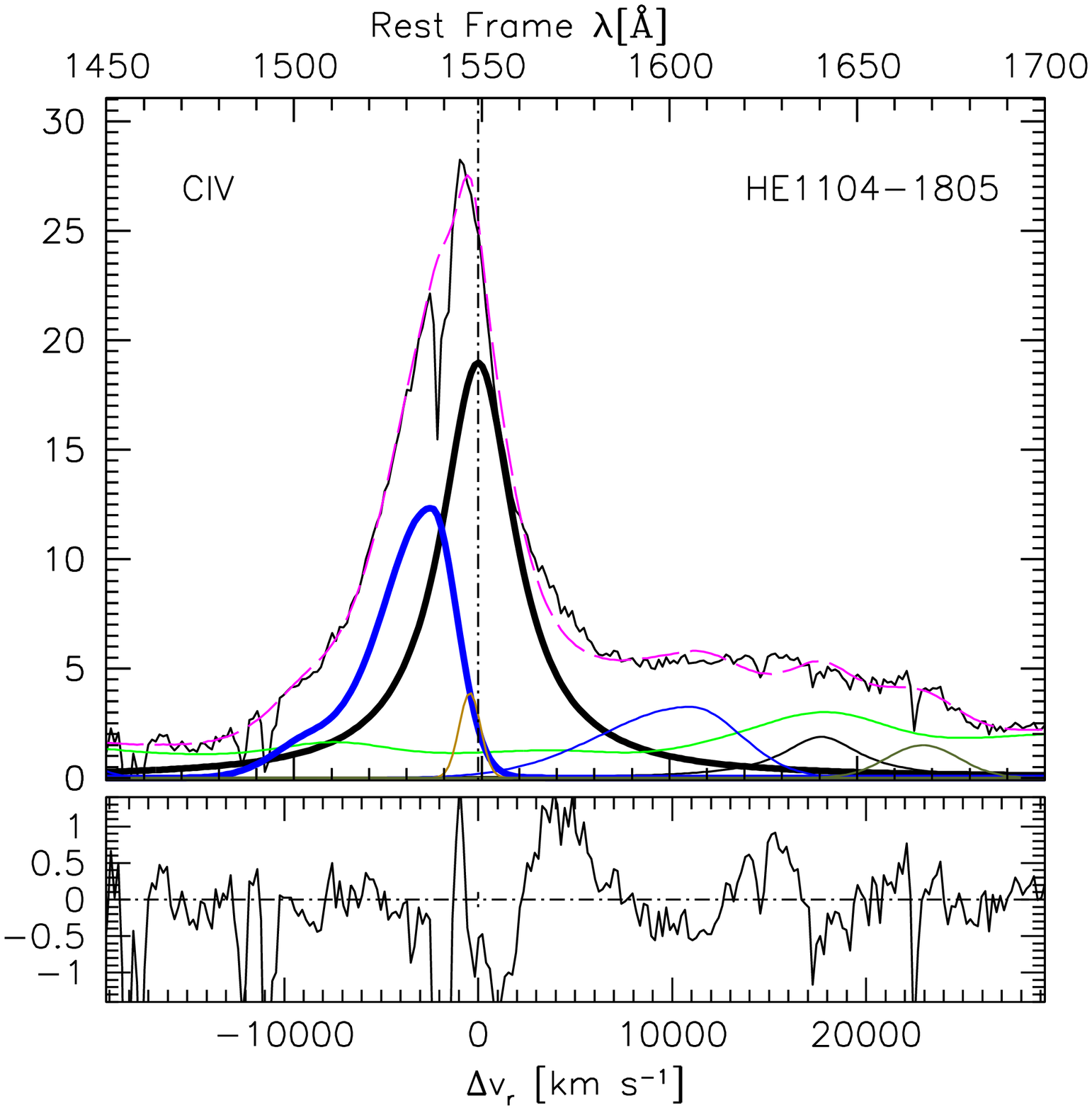}
\includegraphics[width=0.34\columnwidth]{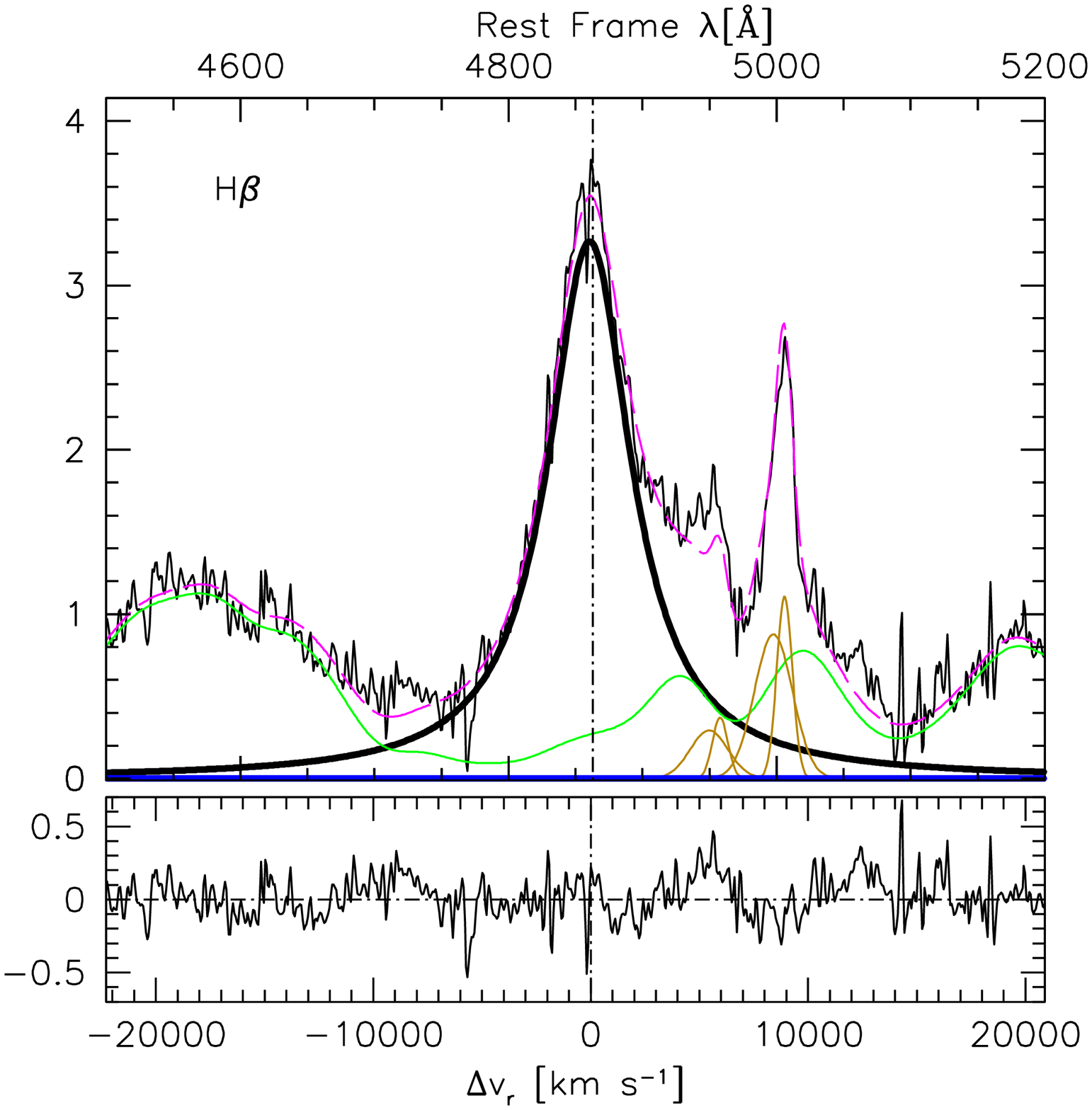}
\includegraphics[width=0.34\columnwidth]{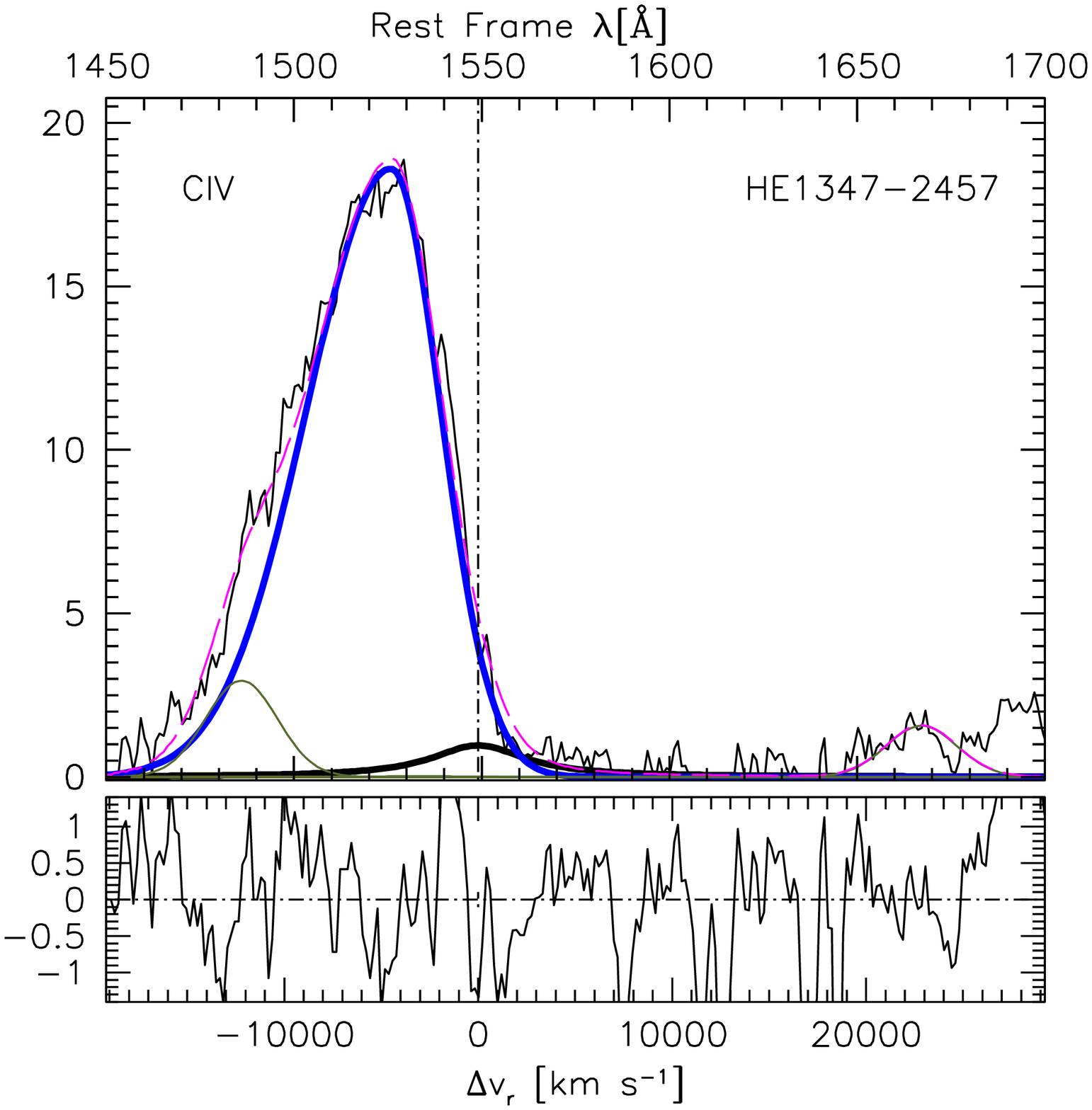}
\includegraphics[width=0.34\columnwidth]{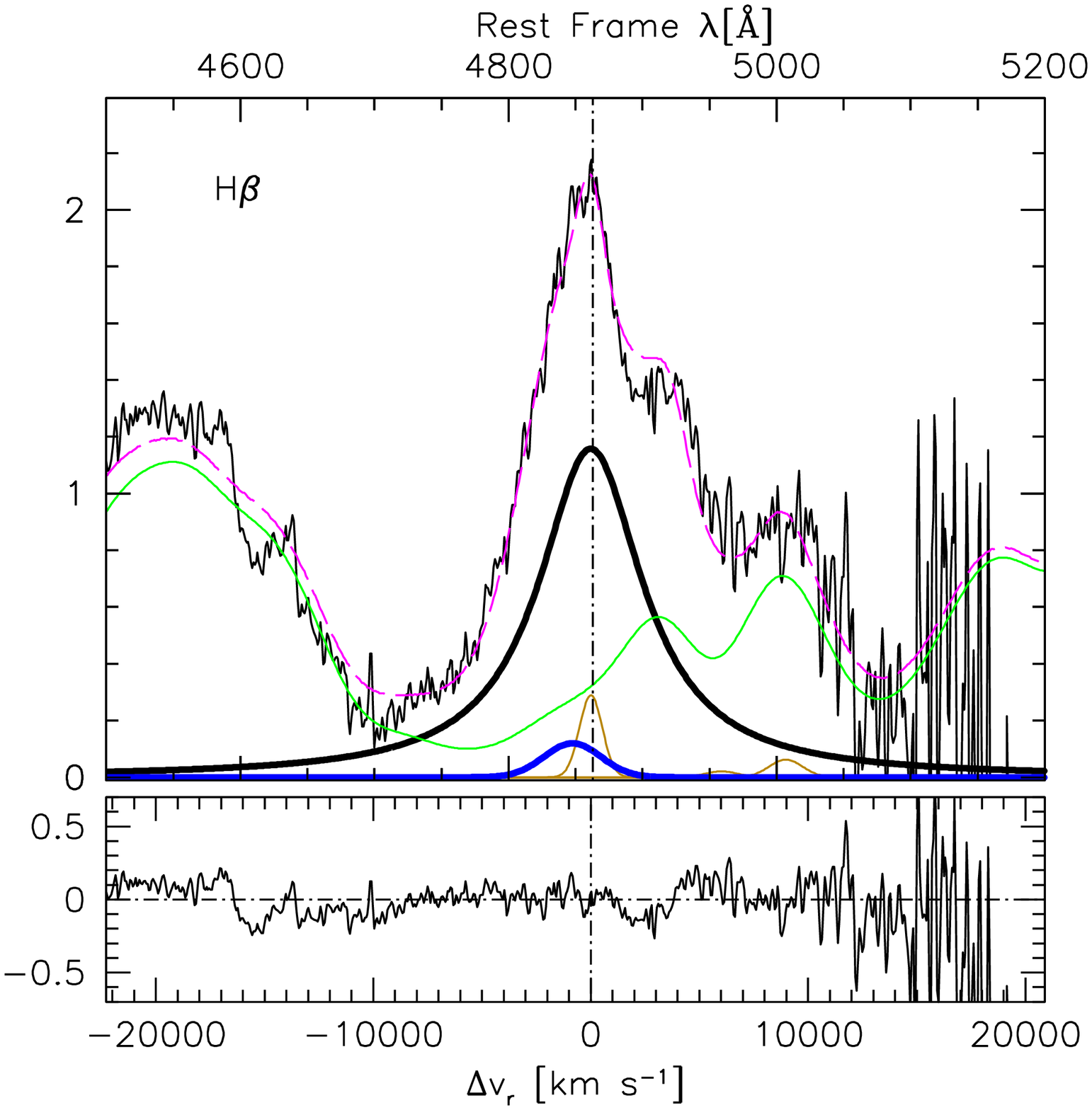}\\
\includegraphics[width=0.34\columnwidth]{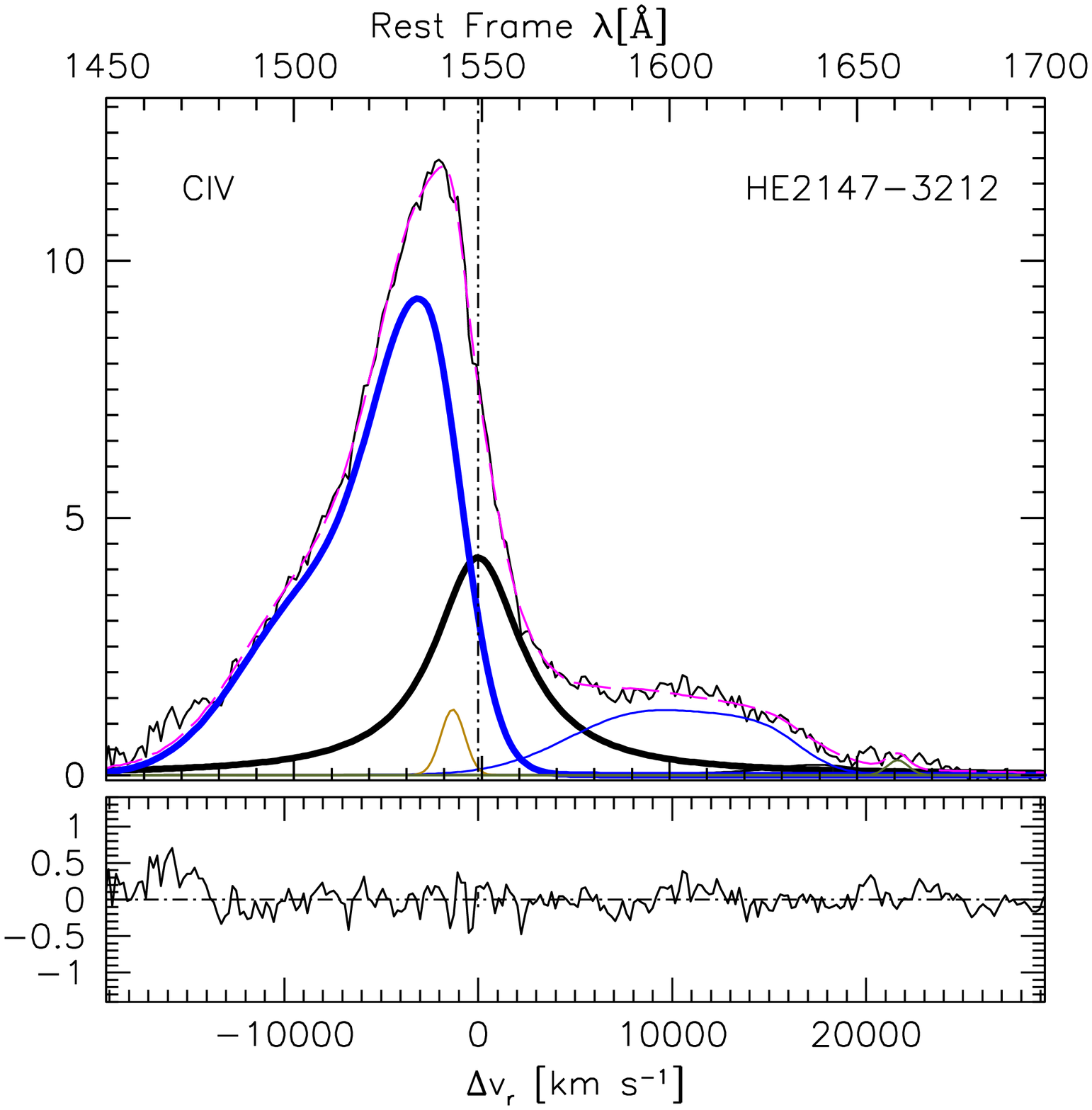}
\includegraphics[width=0.34\columnwidth]{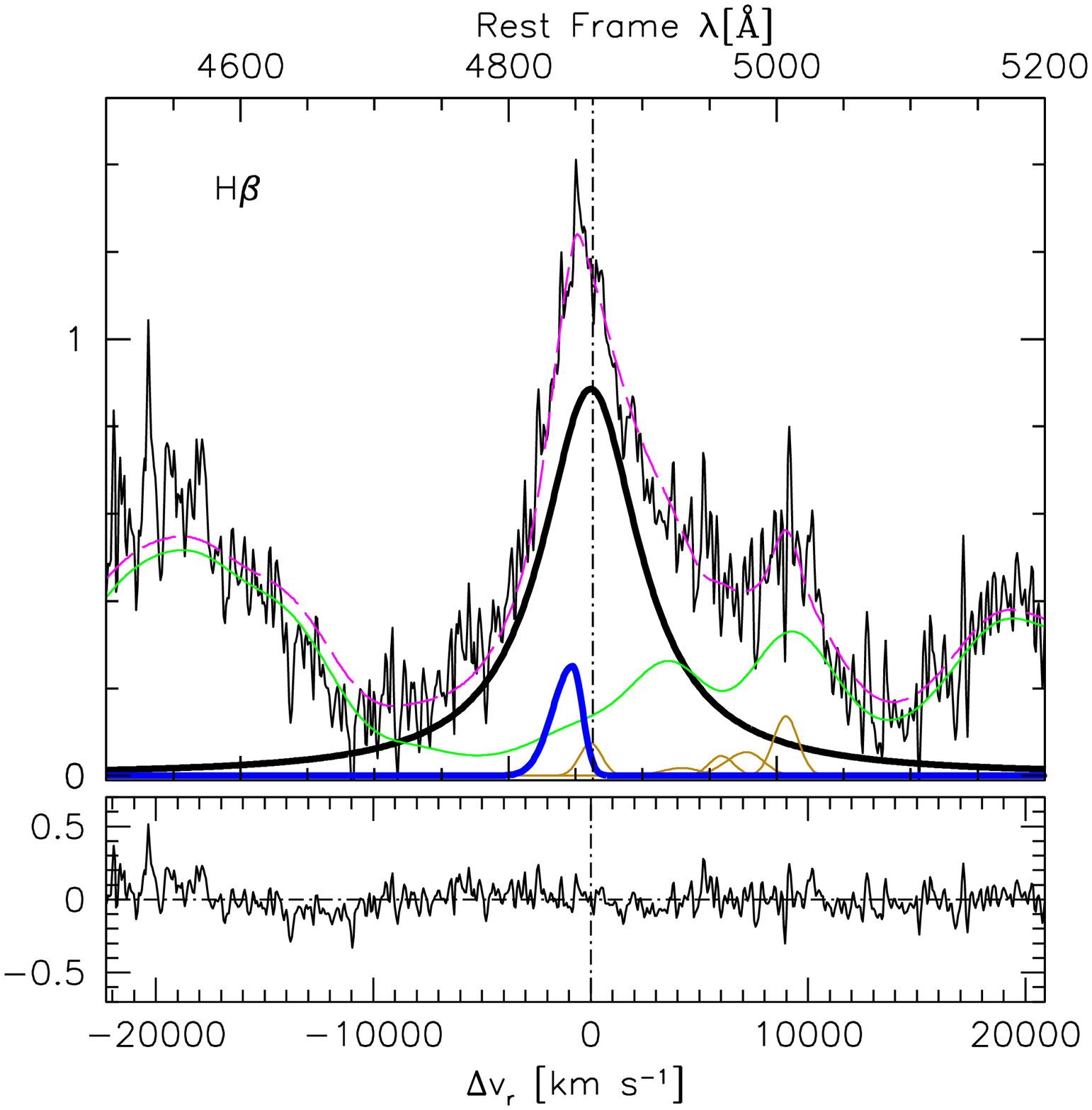}
\includegraphics[width=0.34\columnwidth]{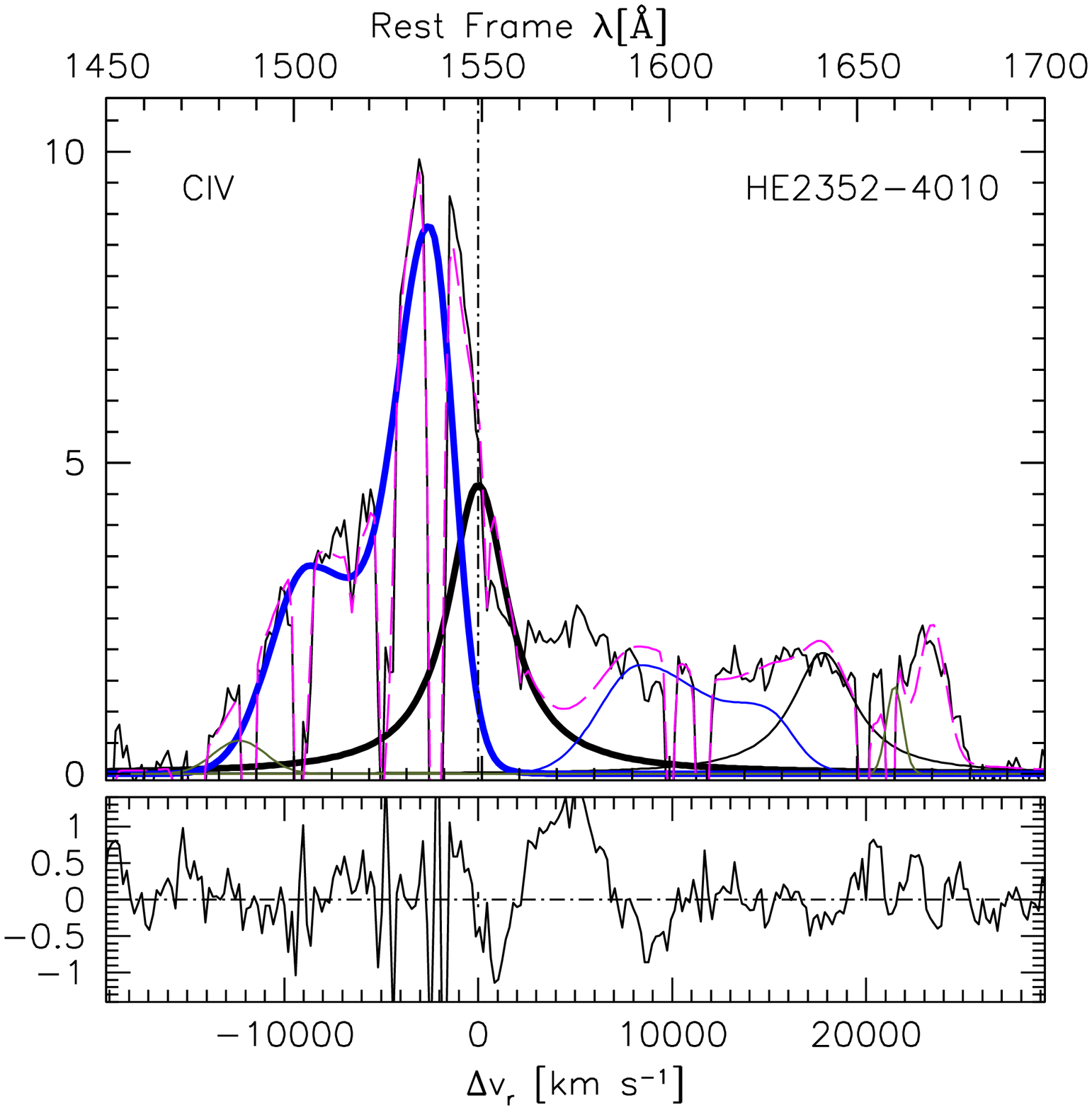}
\includegraphics[width=0.34\columnwidth]{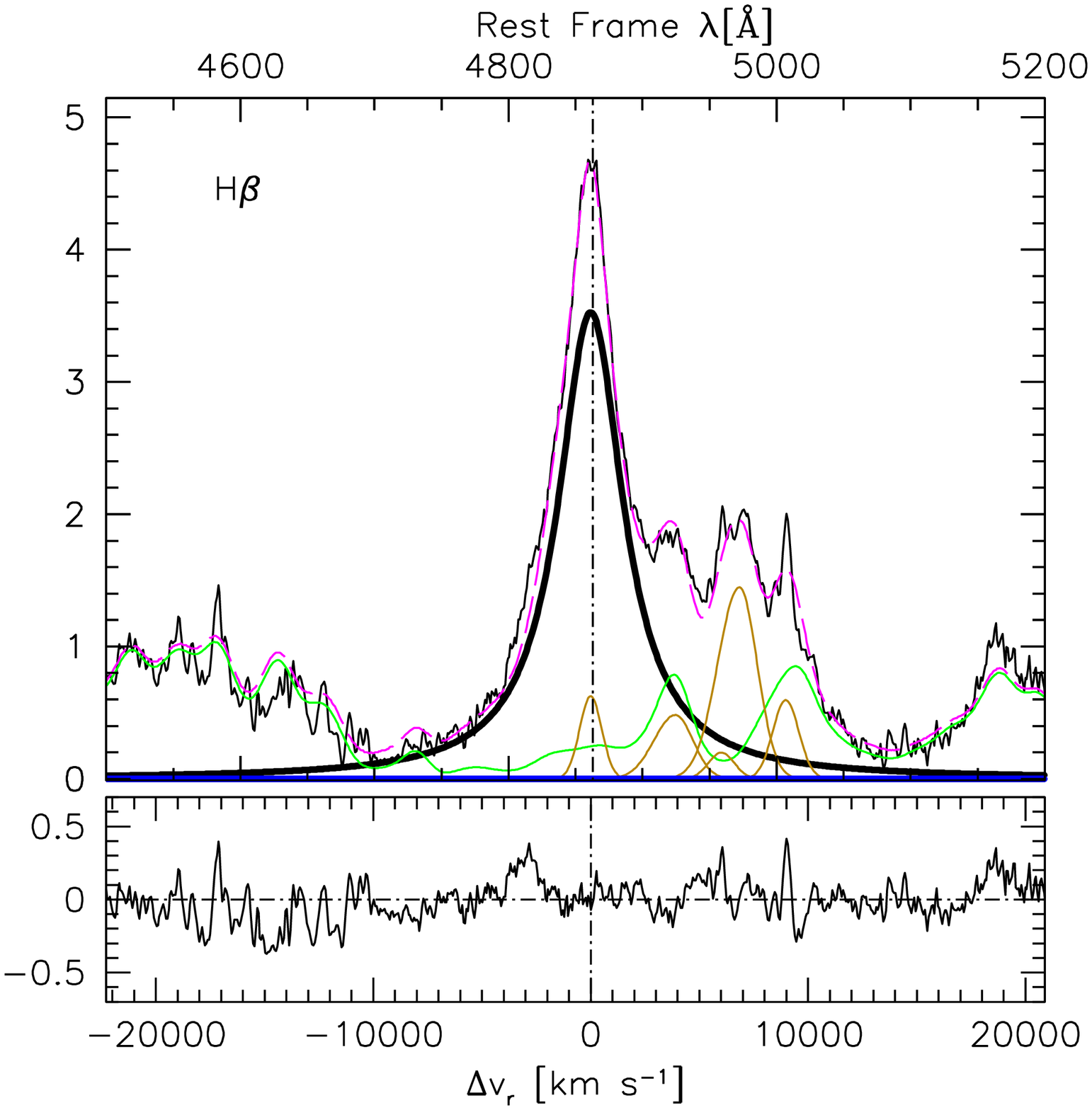}
\caption{Results of the {\tt specfit} analysis on the spectra of Population A sources in the HE sample,  in the \civonly\ and \hb\ spectral range (adjacent left and right panels),  after continuum subtraction. Horizontal scale is rest frame wavelength in \AA\  or radial velocity shift from \civonly\ rest wavelength (left) or  \hb\ (right) marked by dot dashed lines; vertical scale is specific flux in units of $10^{-15}$ \ergss cm$^{-2}$ \AA$^{-1}$.  The panels show the emission line components used in the fit:  \feii\ emission (green), BC (thick black), and BLUE (thick blue), narrow lines (orange). The \oiiiuv\ line at $\approx$ 1660\ \AA\ is shown in grey.  Thin blue and black lines trace the BLUE and BC for \heiiuv. The lower panels show (observed - {\tt specfit} model) residuals.\label{fig:specfita}}
\end{figure*}

\begin{figure*}[htp]
\centering
\includegraphics[width=0.34\columnwidth]{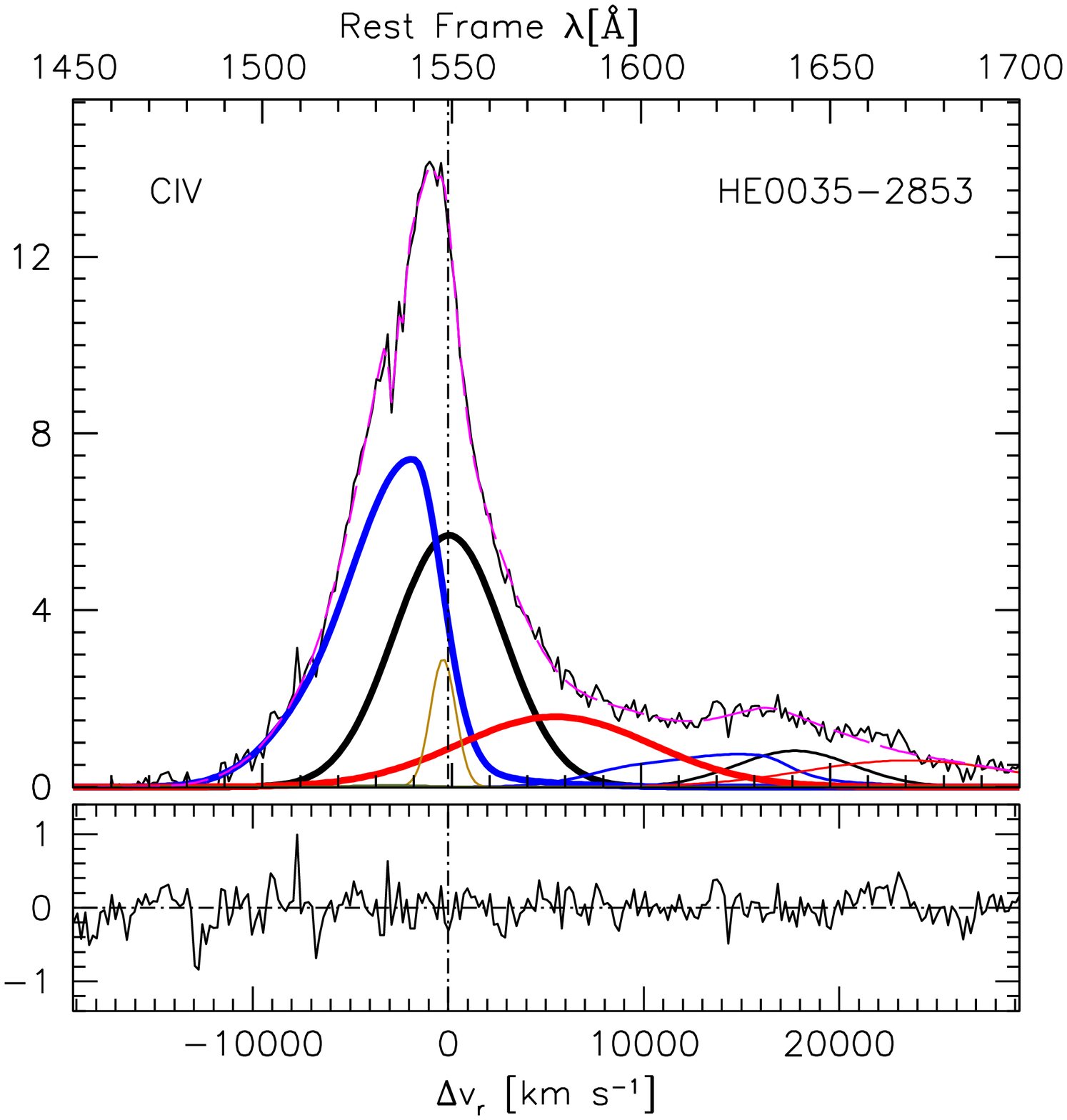}
\includegraphics[width=0.34\columnwidth]{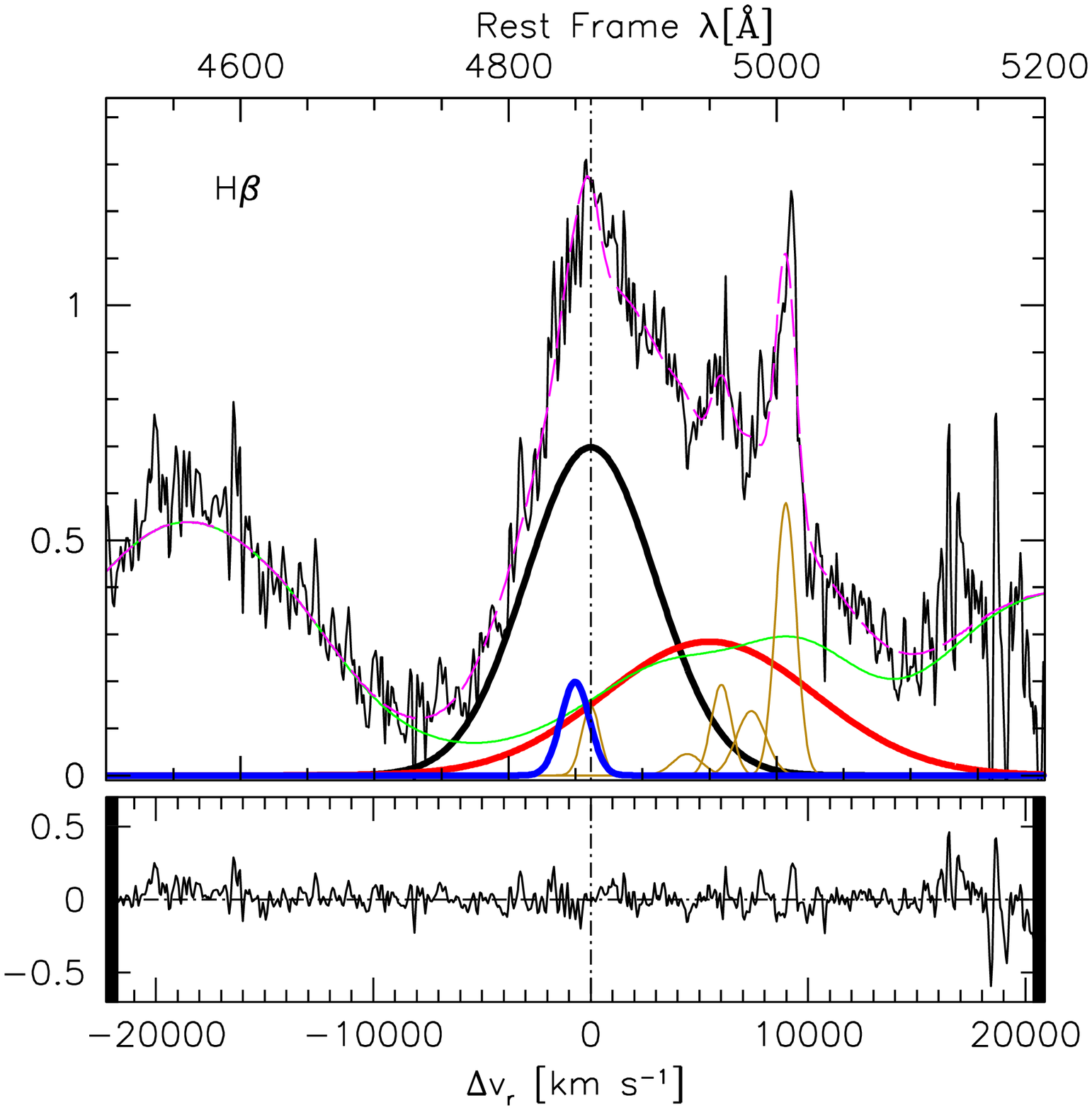}
\includegraphics[width=0.34\columnwidth]{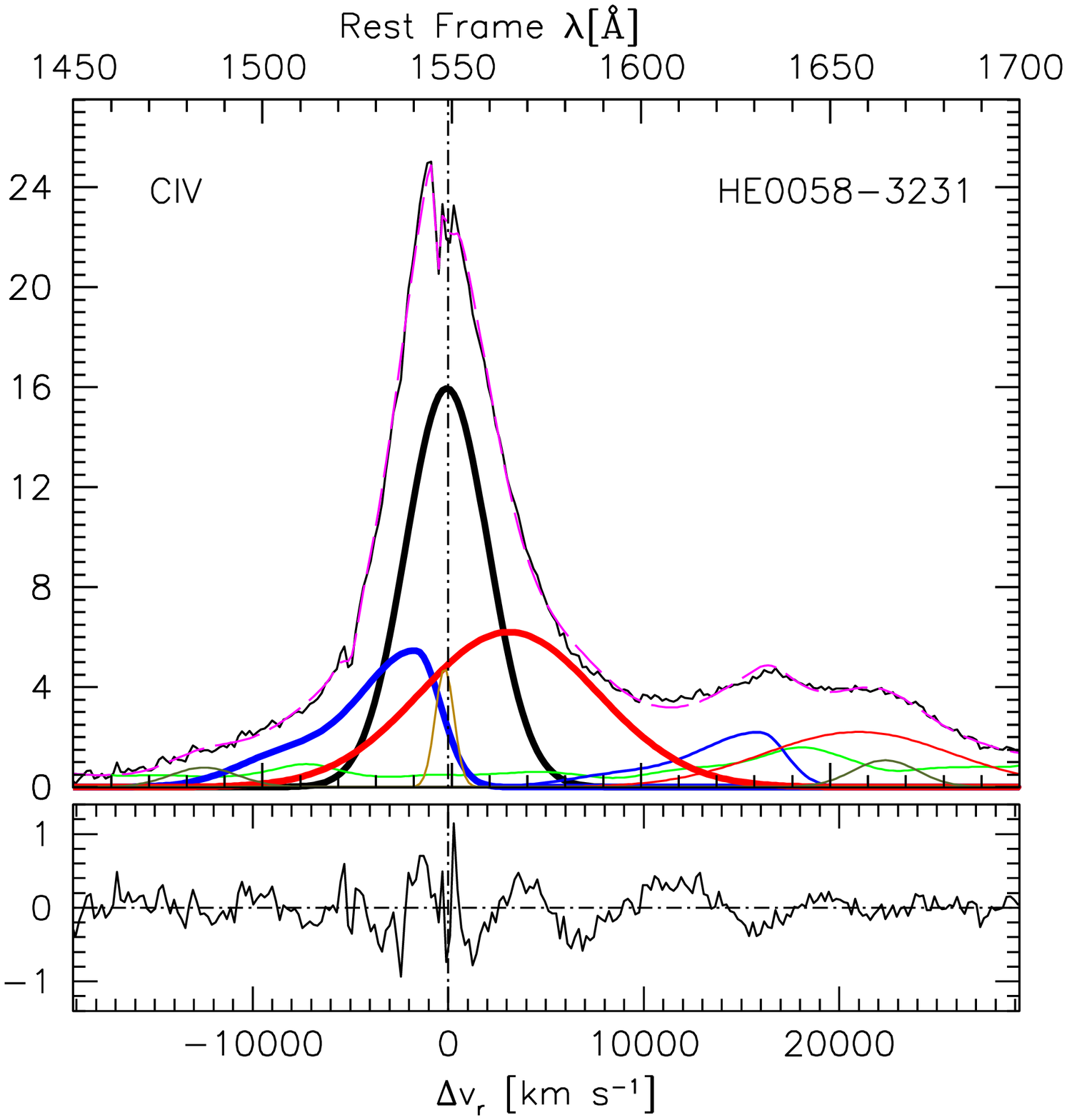}
\includegraphics[width=0.34\columnwidth]{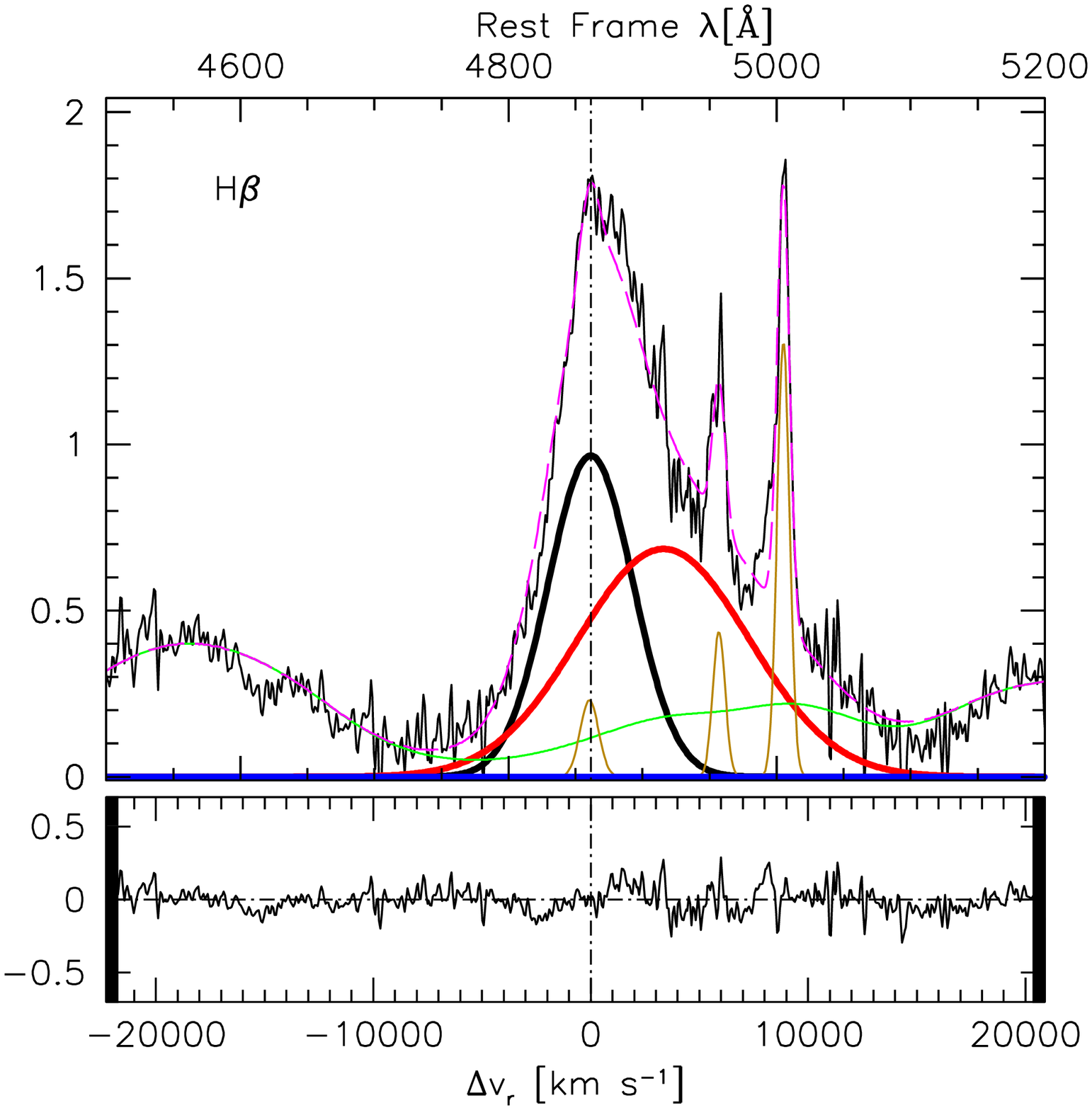}\\ 
\includegraphics[width=0.34\columnwidth]{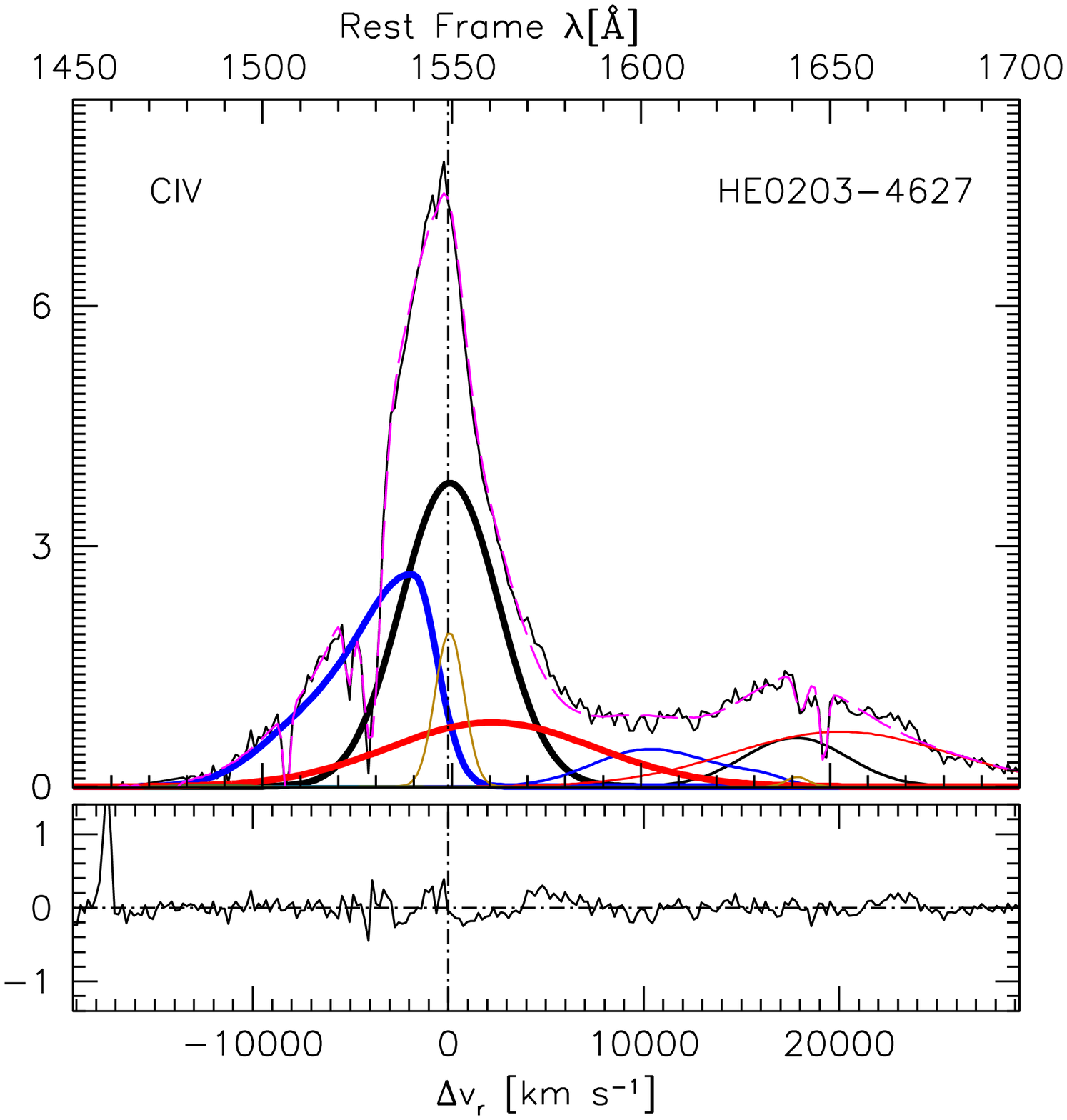}
\includegraphics[width=0.34\columnwidth]{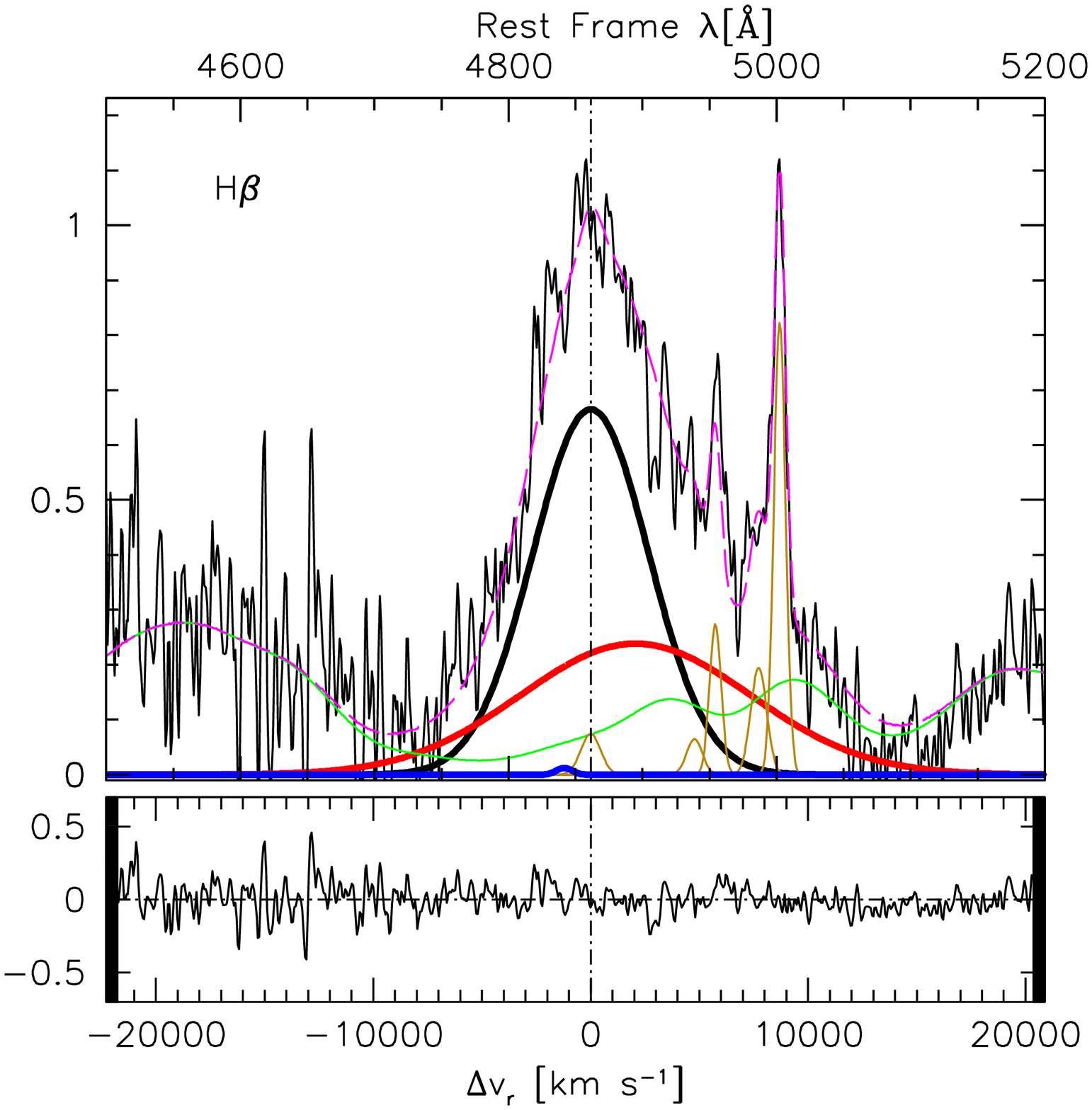}   
\includegraphics[width=0.34\columnwidth]{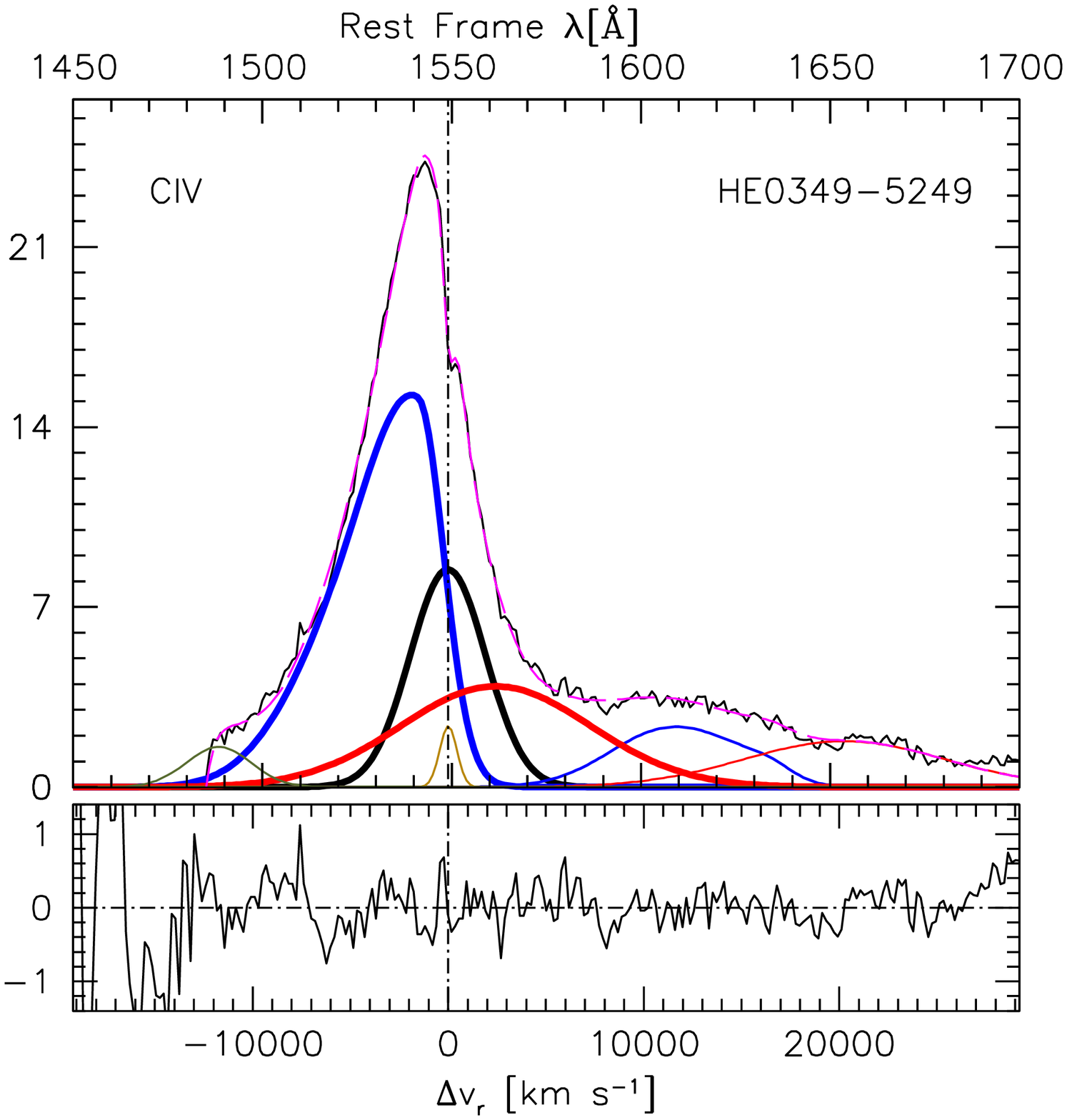}
\includegraphics[width=0.34\columnwidth]{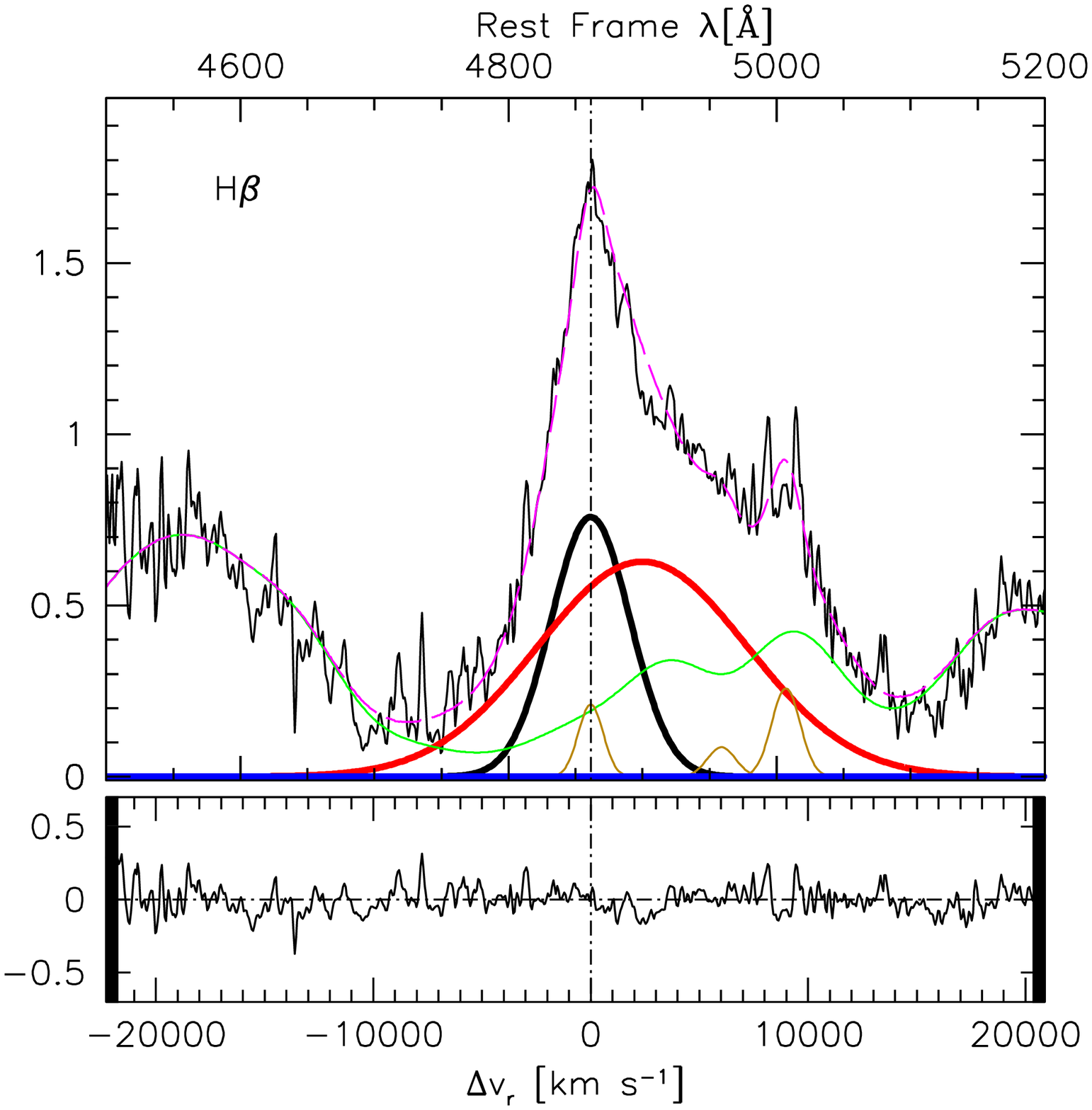}\\ 
\includegraphics[width=0.34\columnwidth]{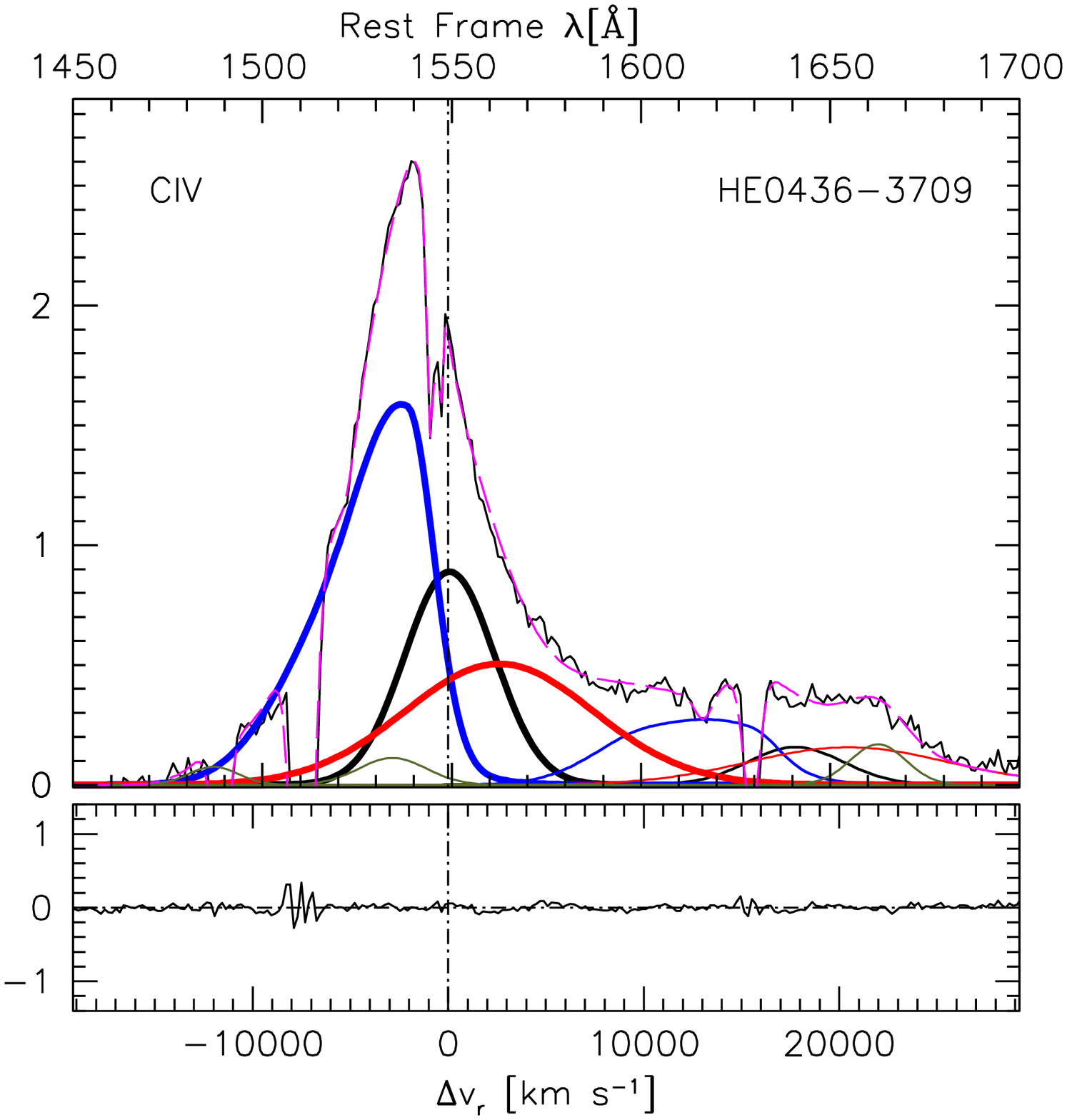}
\includegraphics[width=0.34\columnwidth]{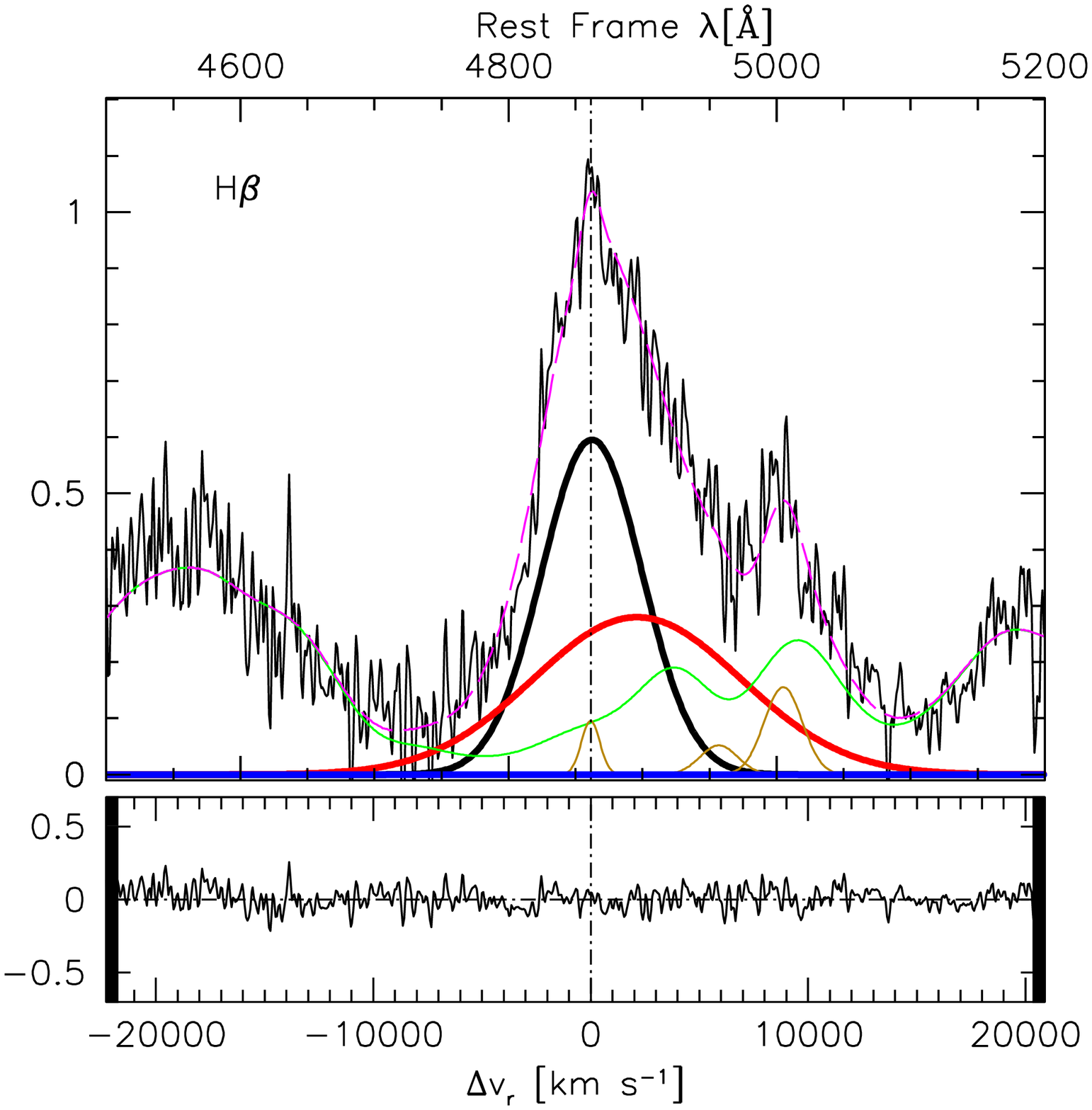} 
\includegraphics[width=0.34\columnwidth]{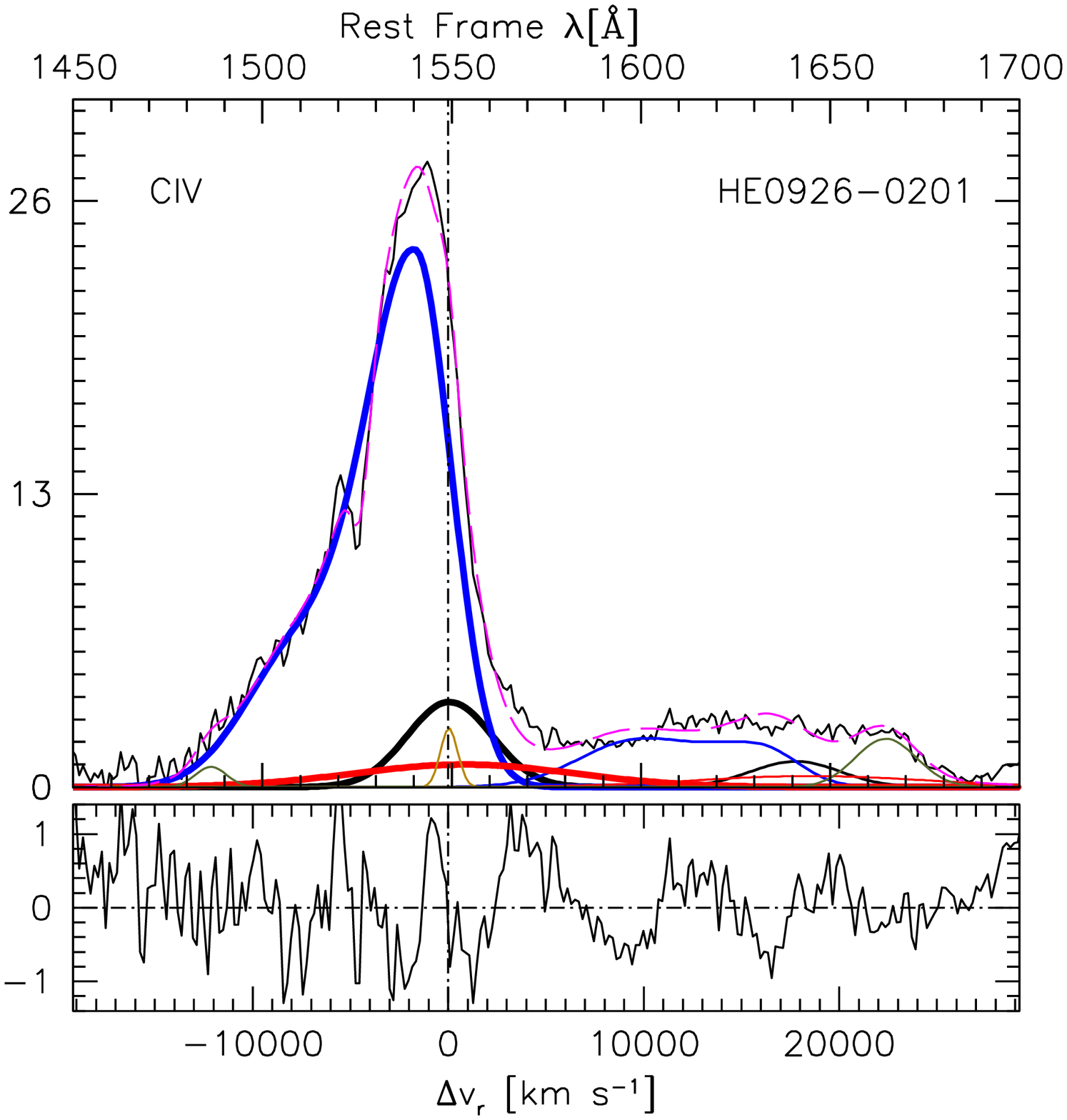}
\includegraphics[width=0.34\columnwidth]{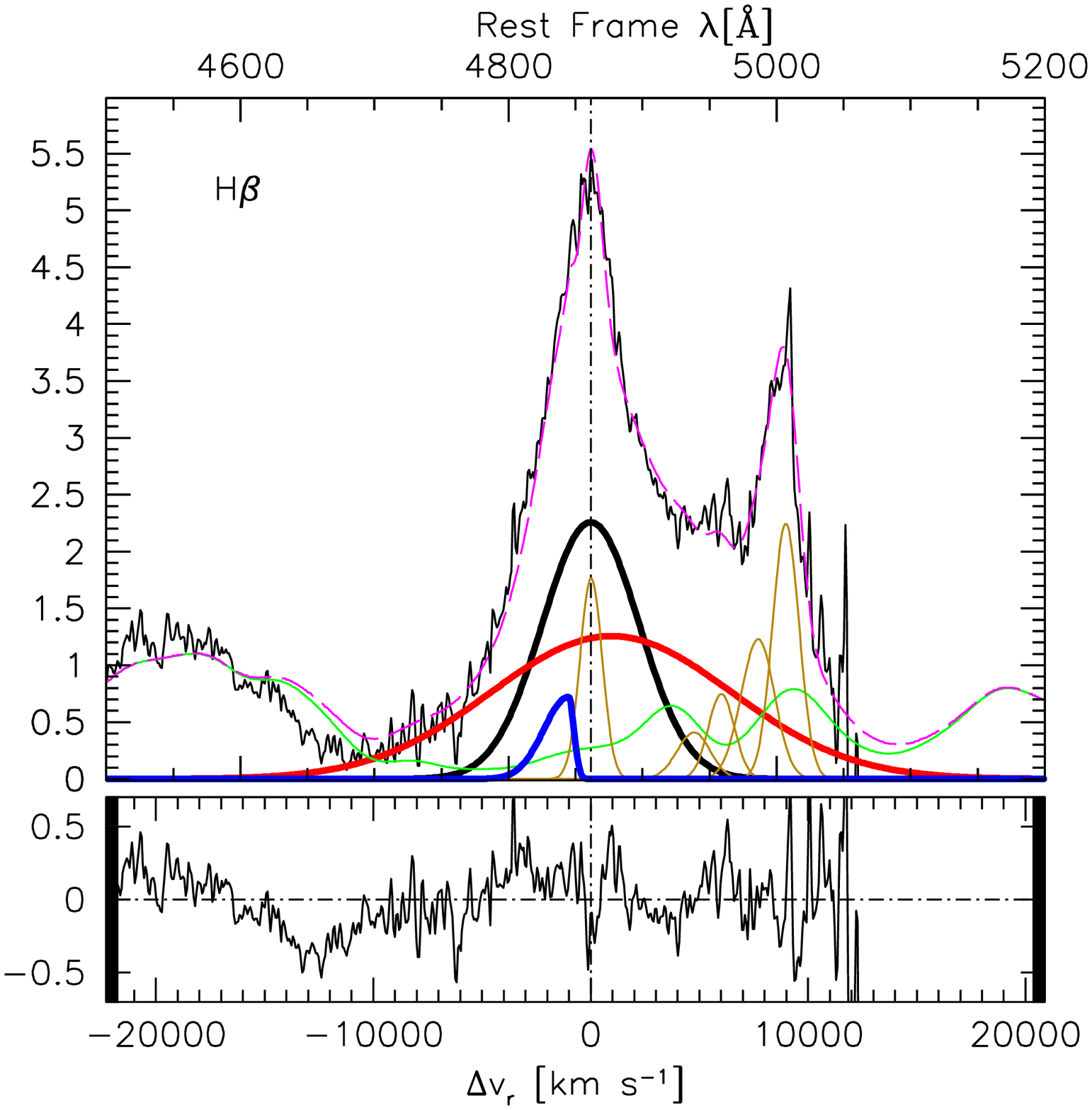}\\
\includegraphics[width=0.34\columnwidth]{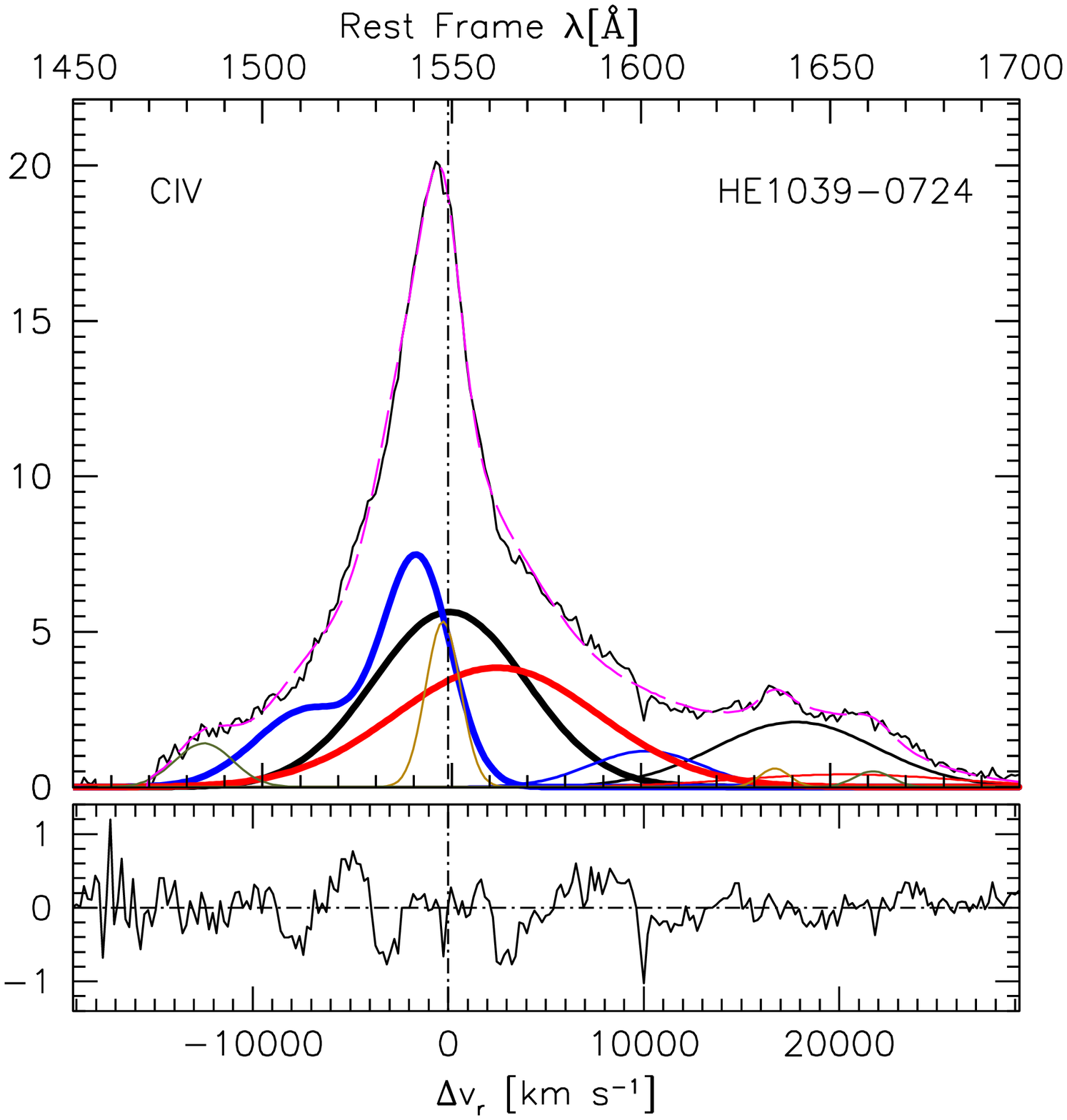}
\includegraphics[width=0.34\columnwidth]{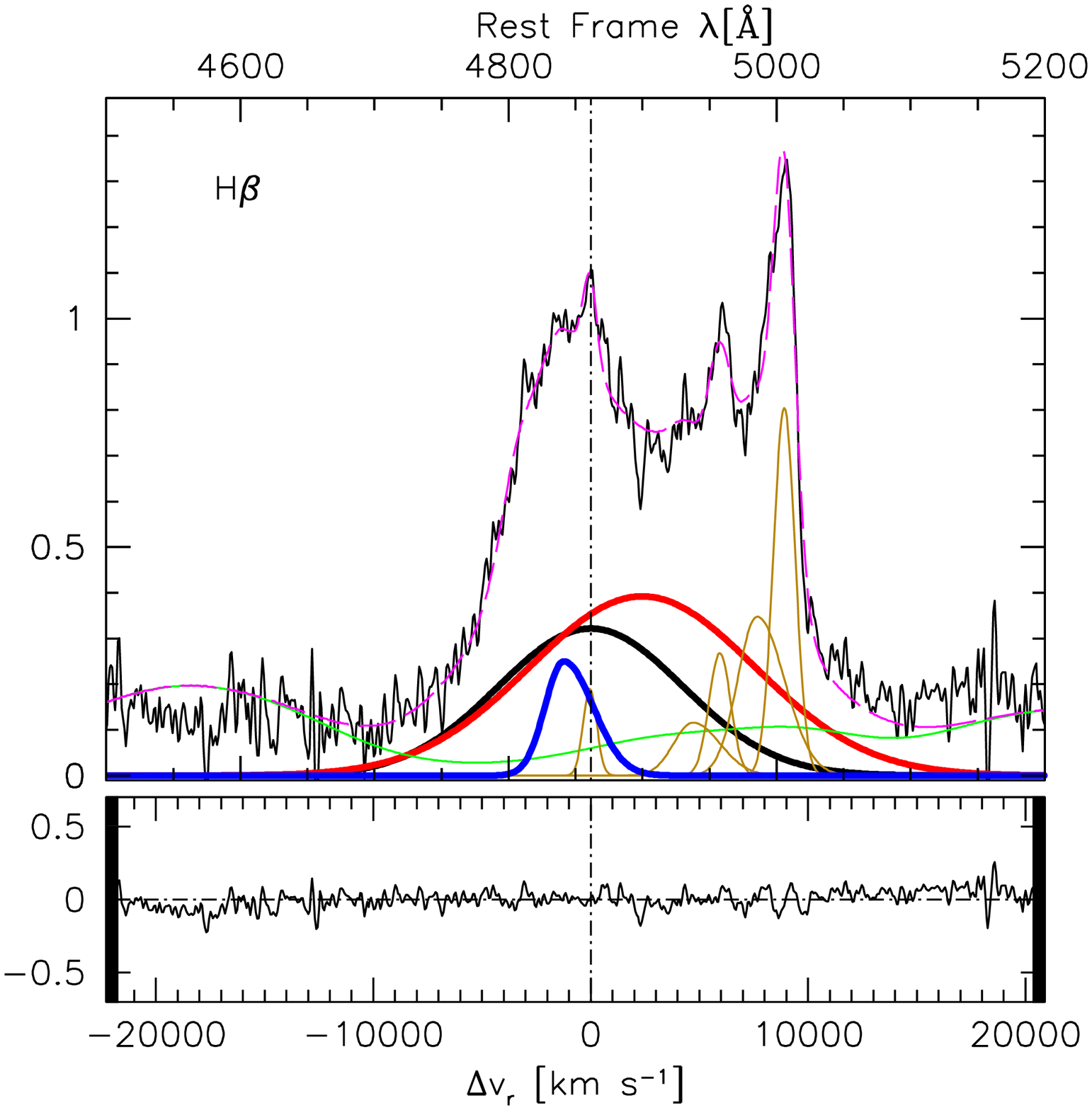} 
\includegraphics[width=0.34\columnwidth]{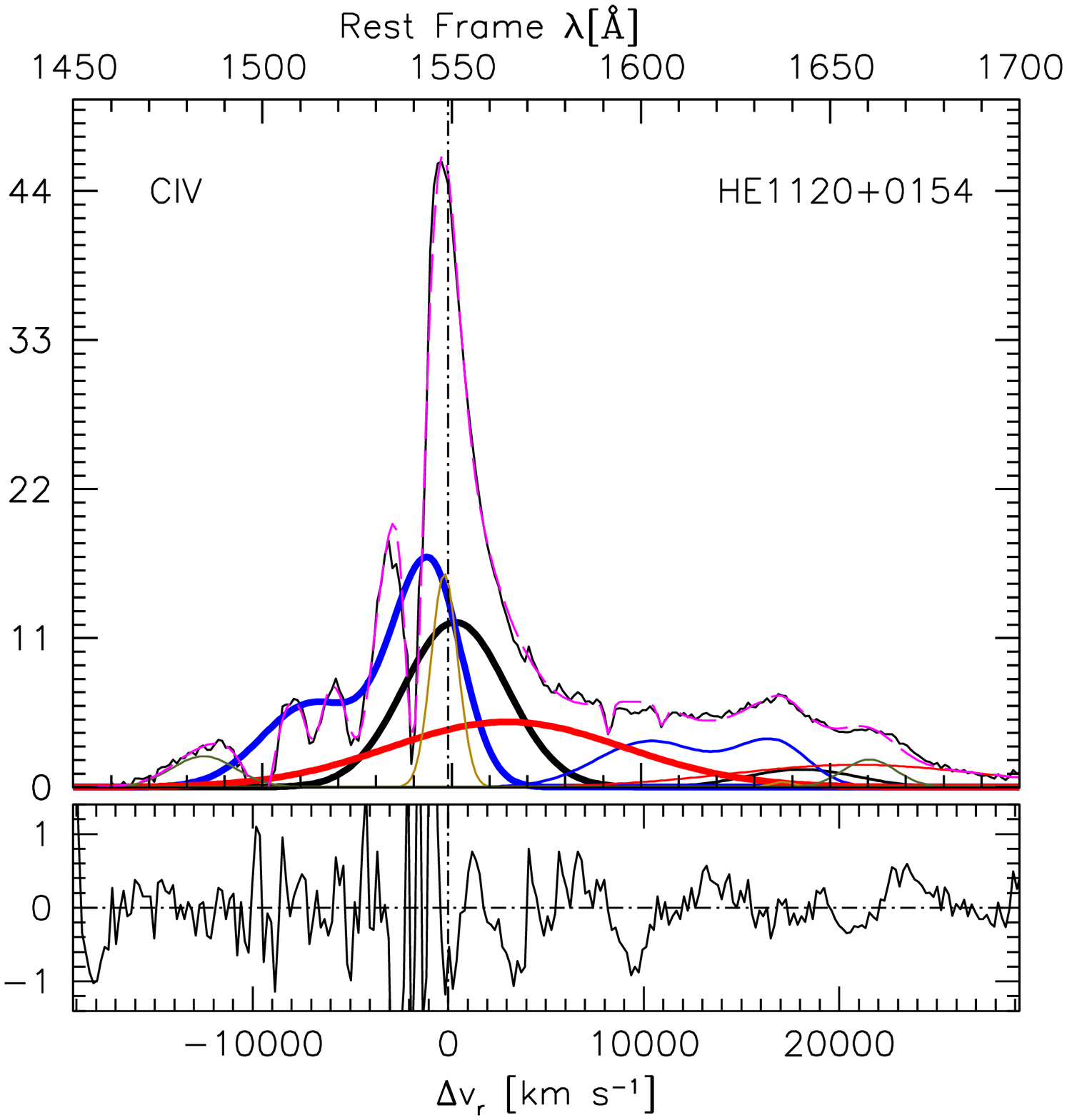}
\includegraphics[width=0.34\columnwidth]{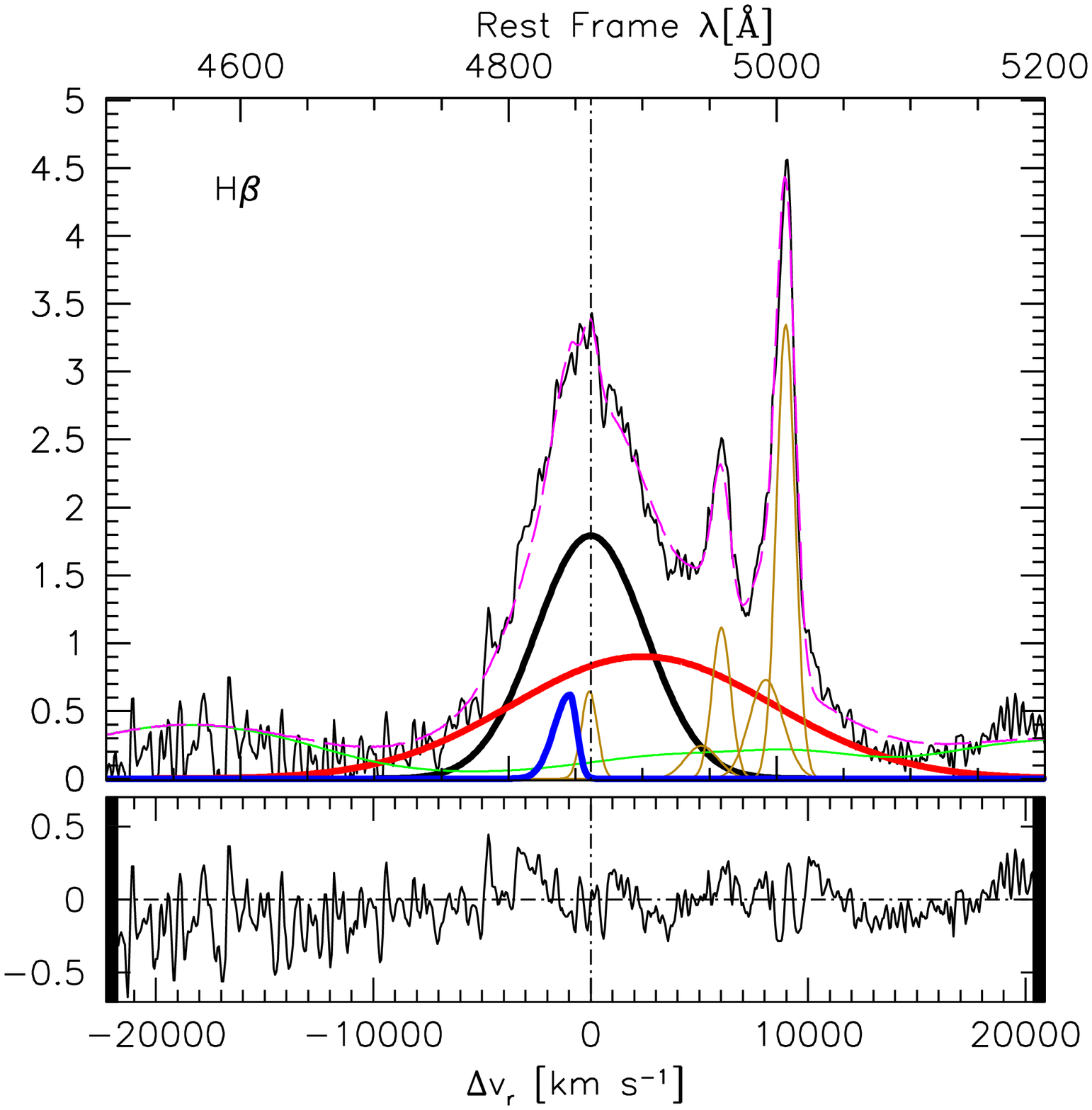}\\ 
\includegraphics[width=0.34\columnwidth]{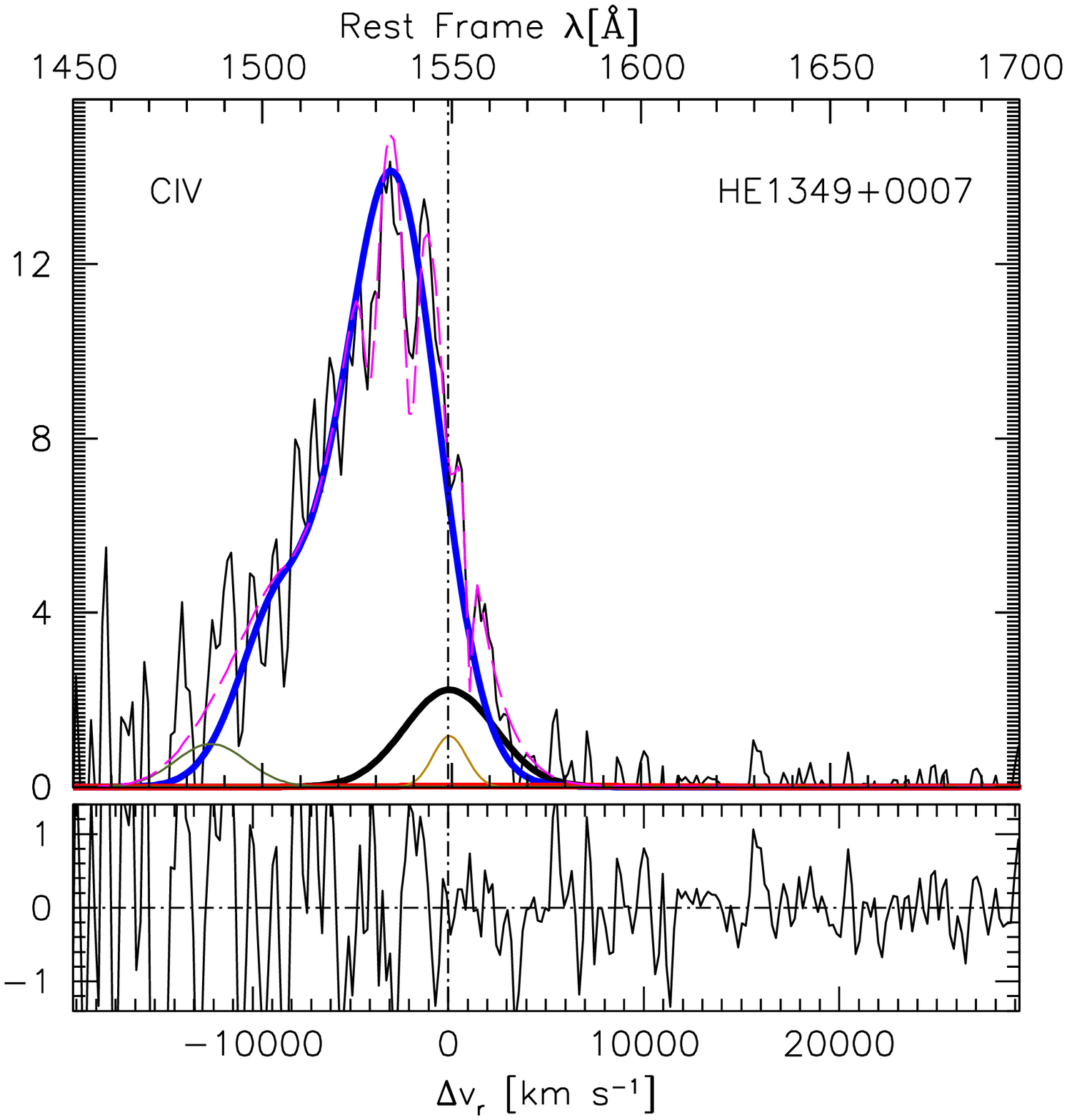}
\includegraphics[width=0.34\columnwidth]{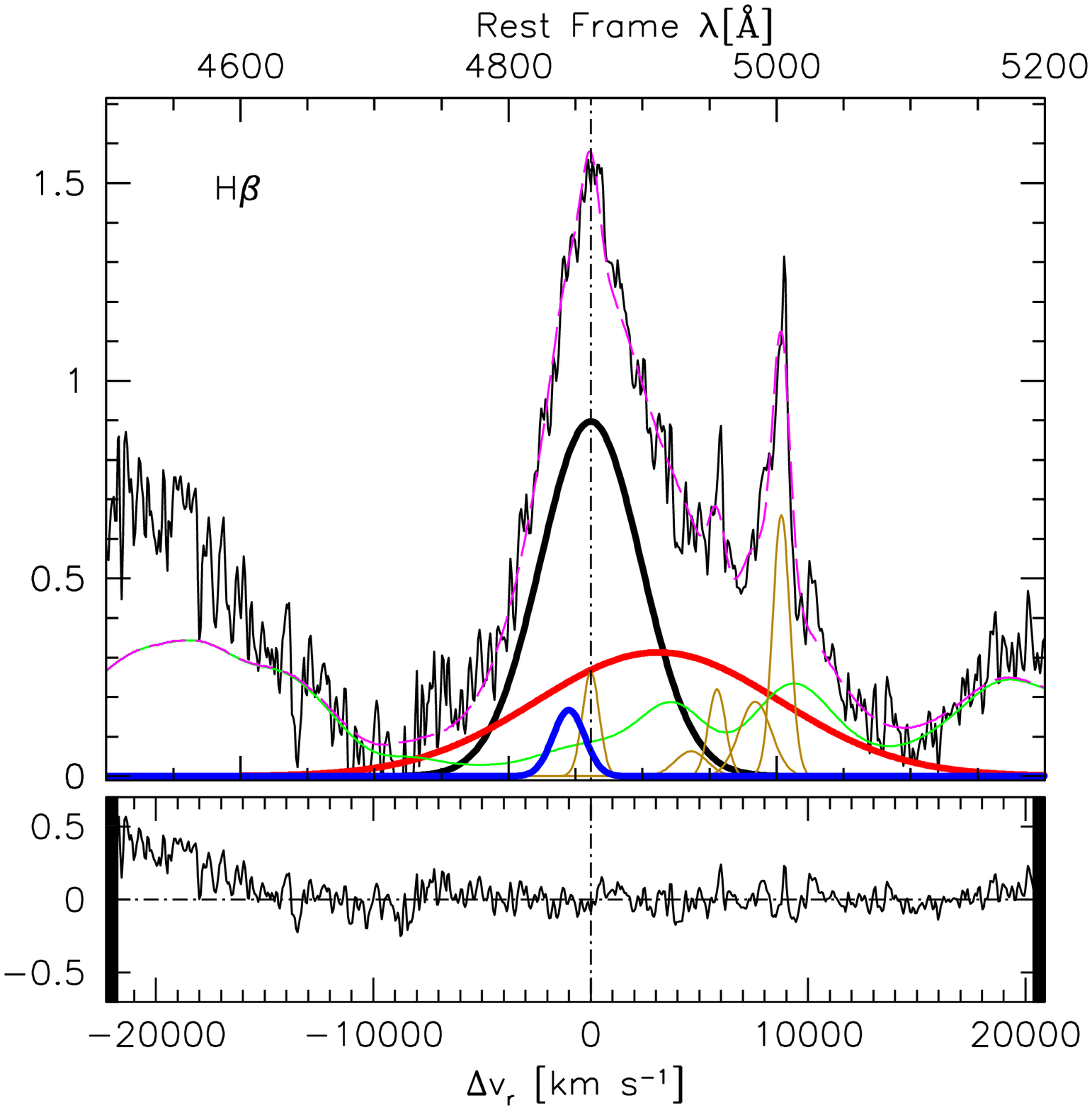}
\includegraphics[width=0.34\columnwidth]{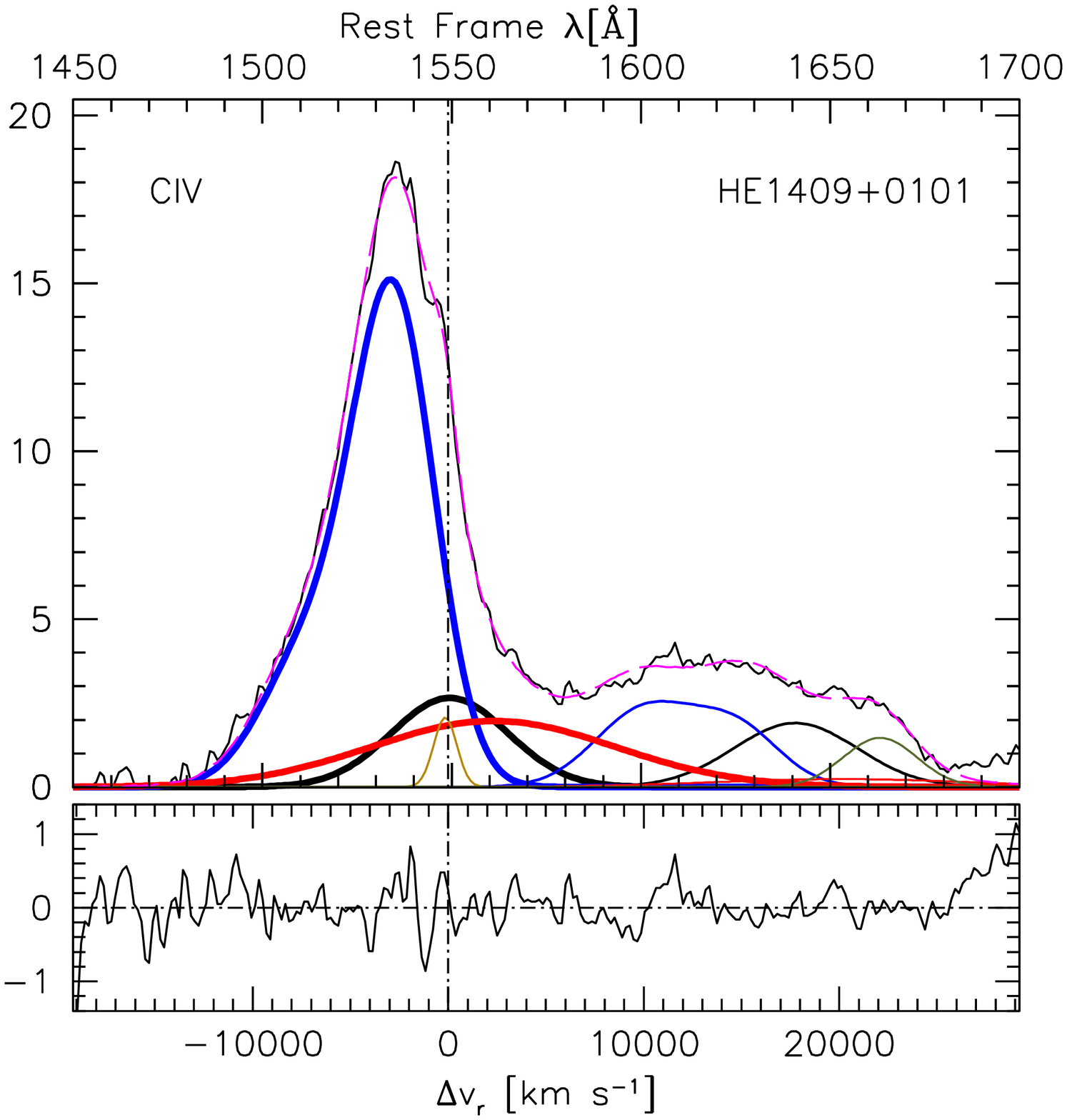}
\includegraphics[width=0.34\columnwidth]{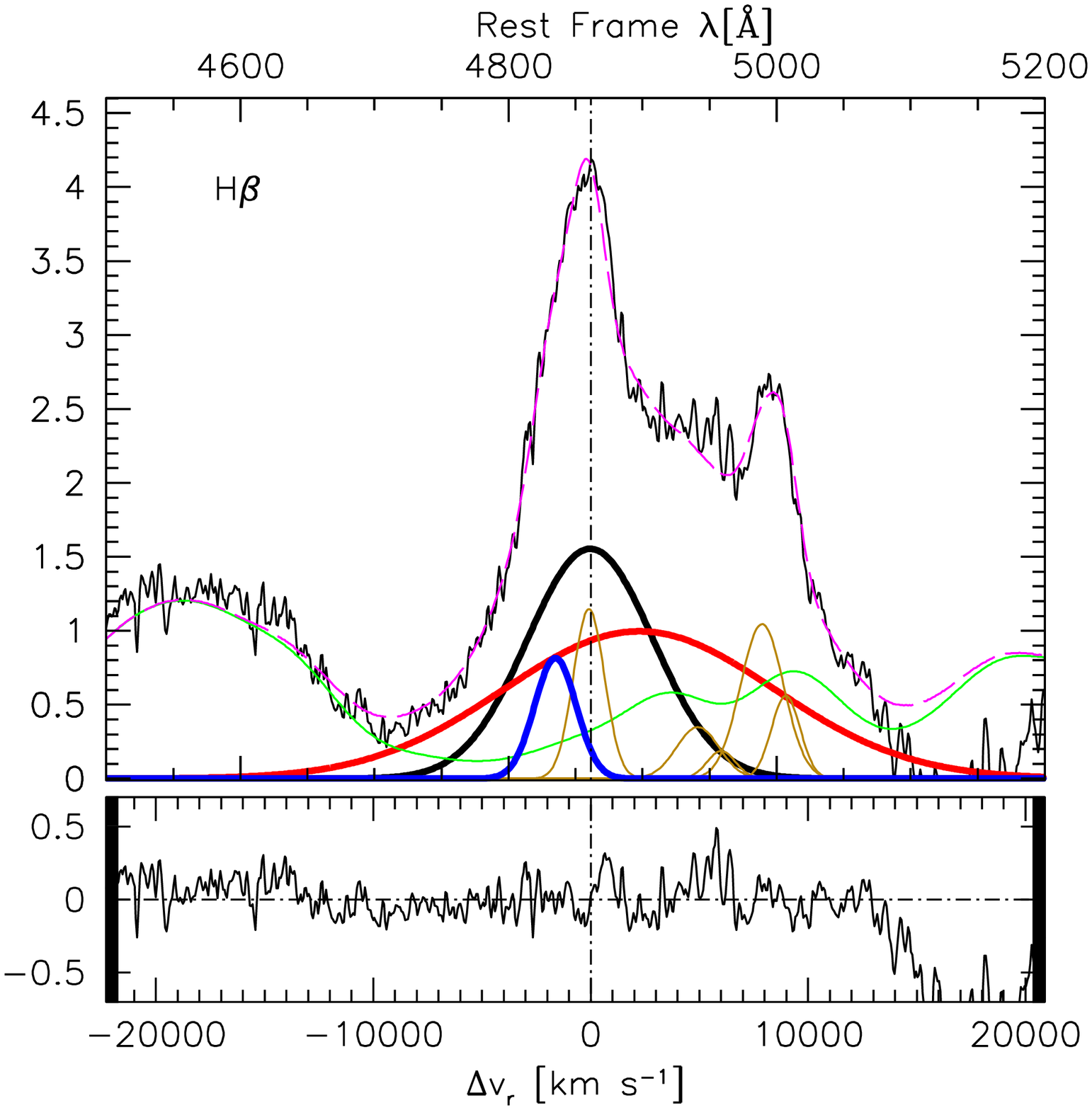}\\
\includegraphics[width=0.34\columnwidth]{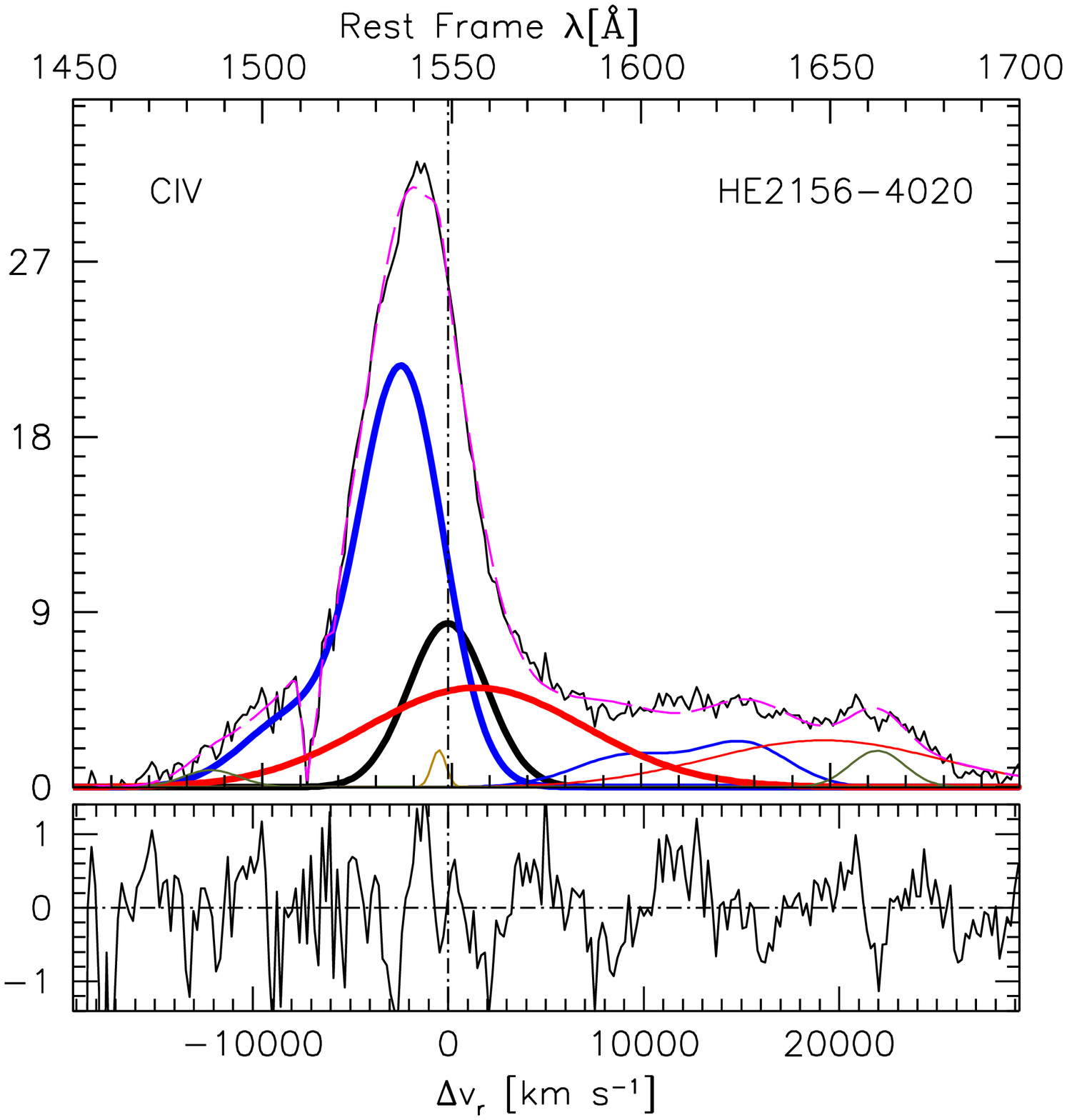}
\includegraphics[width=0.34\columnwidth]{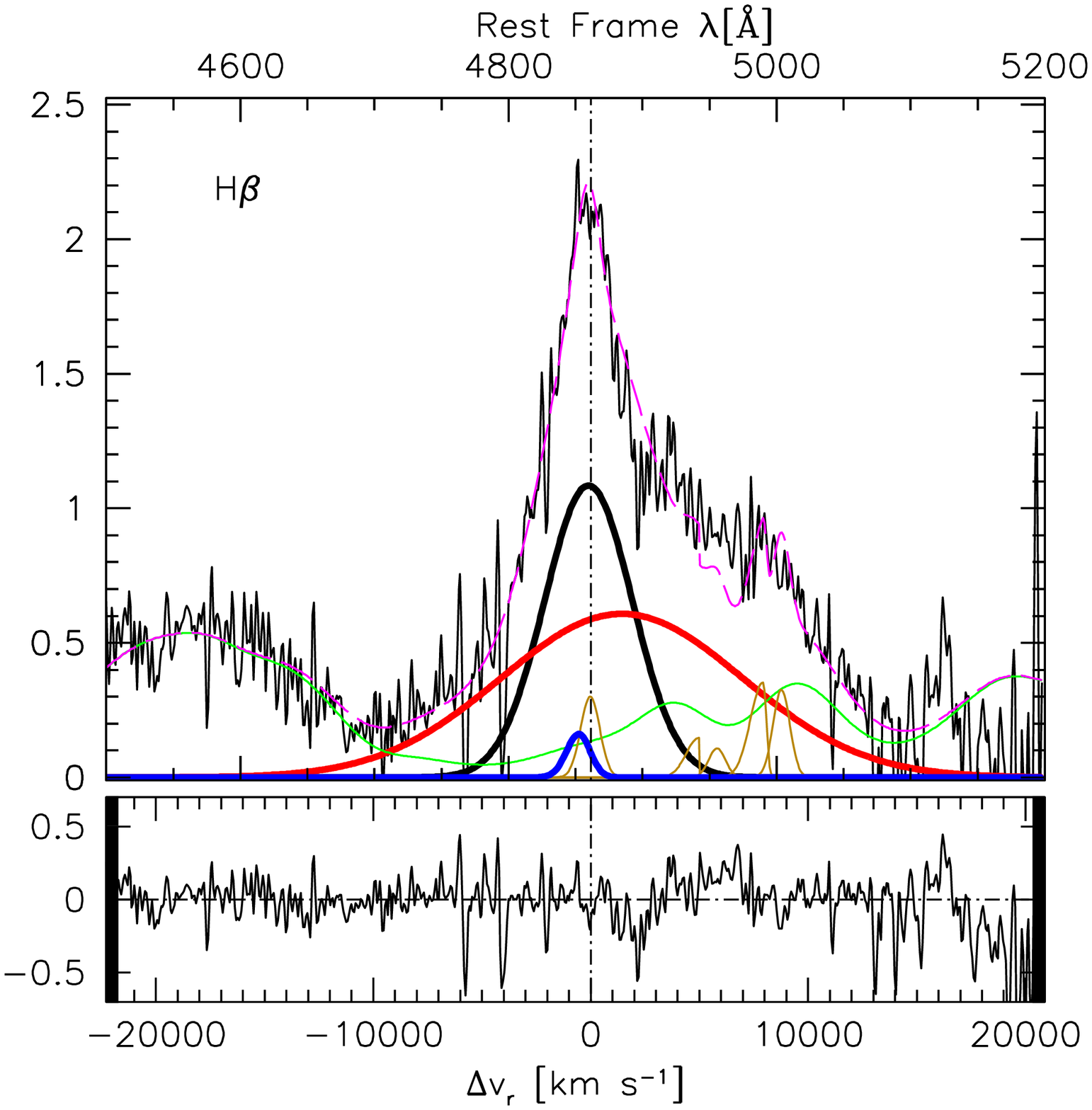}
\includegraphics[width=0.34\columnwidth]{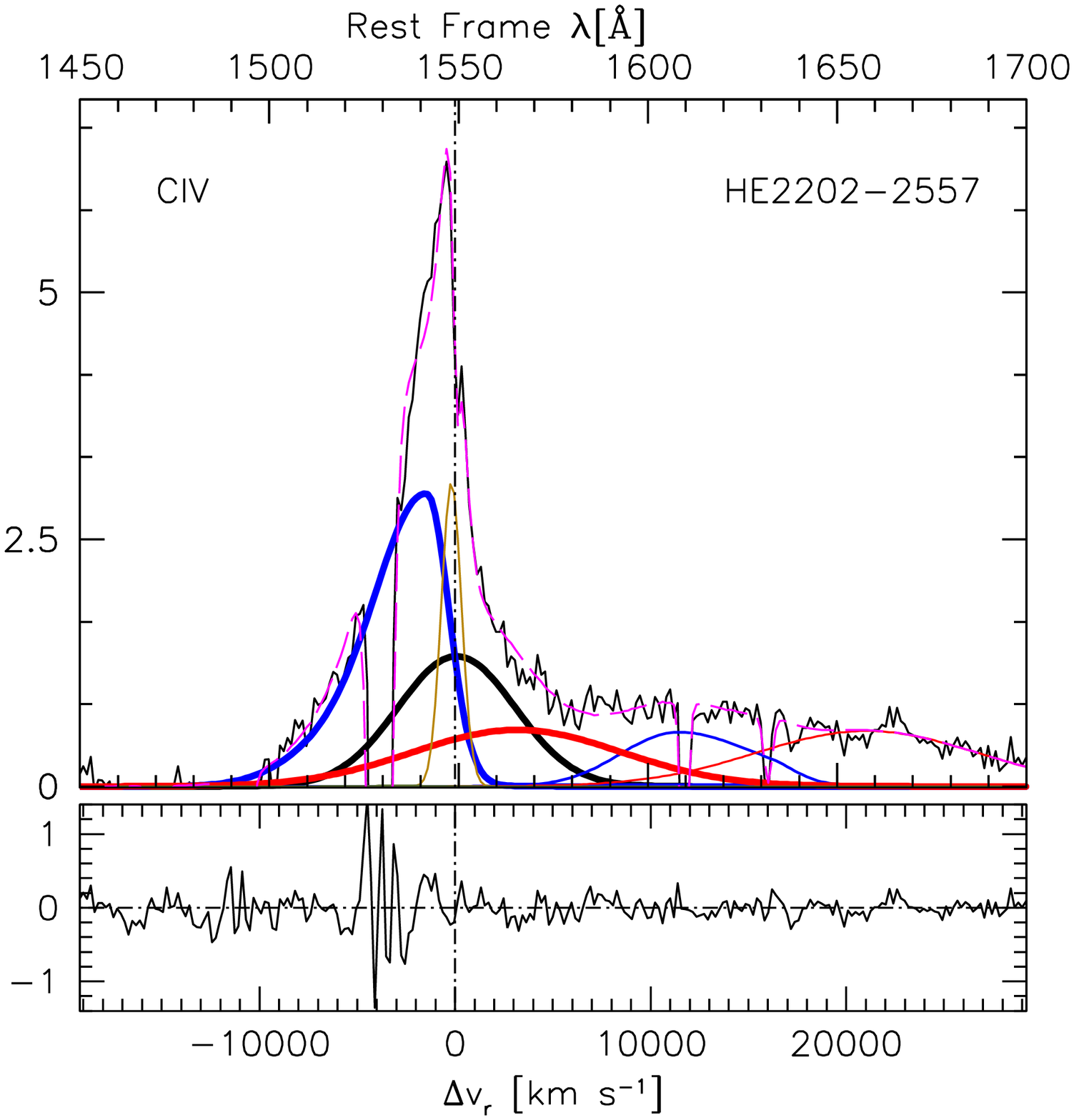}
\includegraphics[width=0.34\columnwidth]{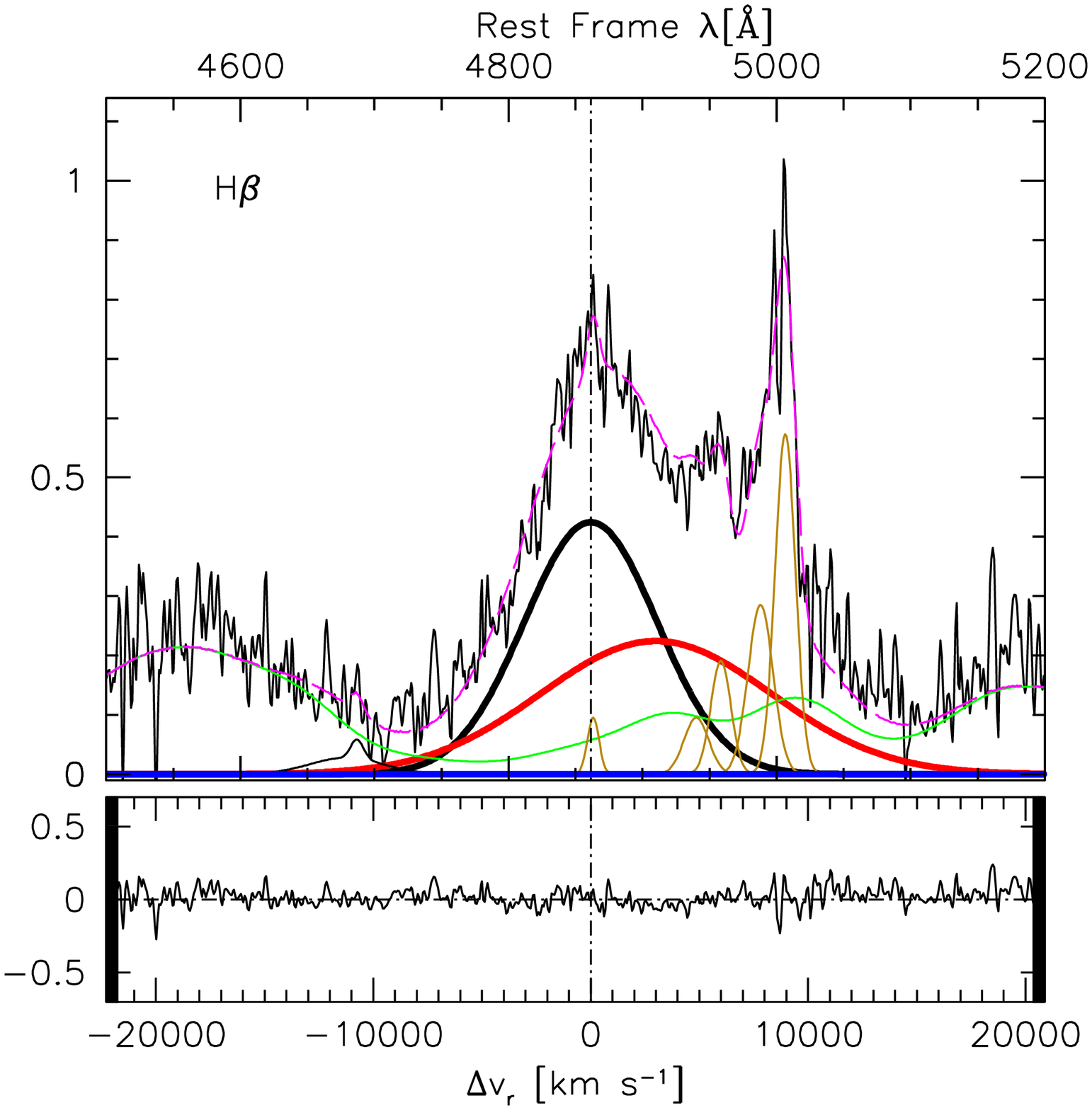}\\
\includegraphics[width=0.34\columnwidth]{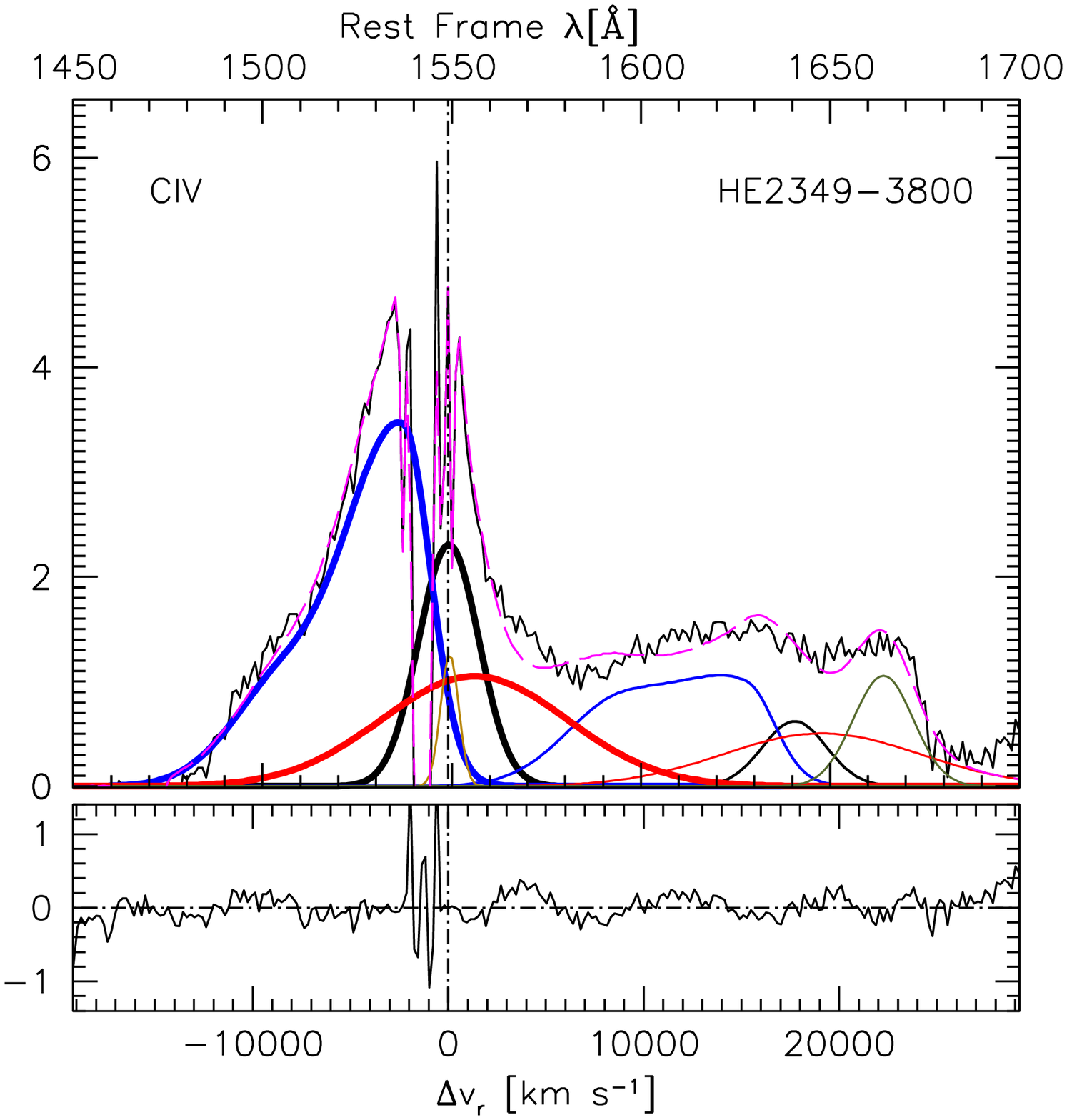}
\includegraphics[width=0.34\columnwidth]{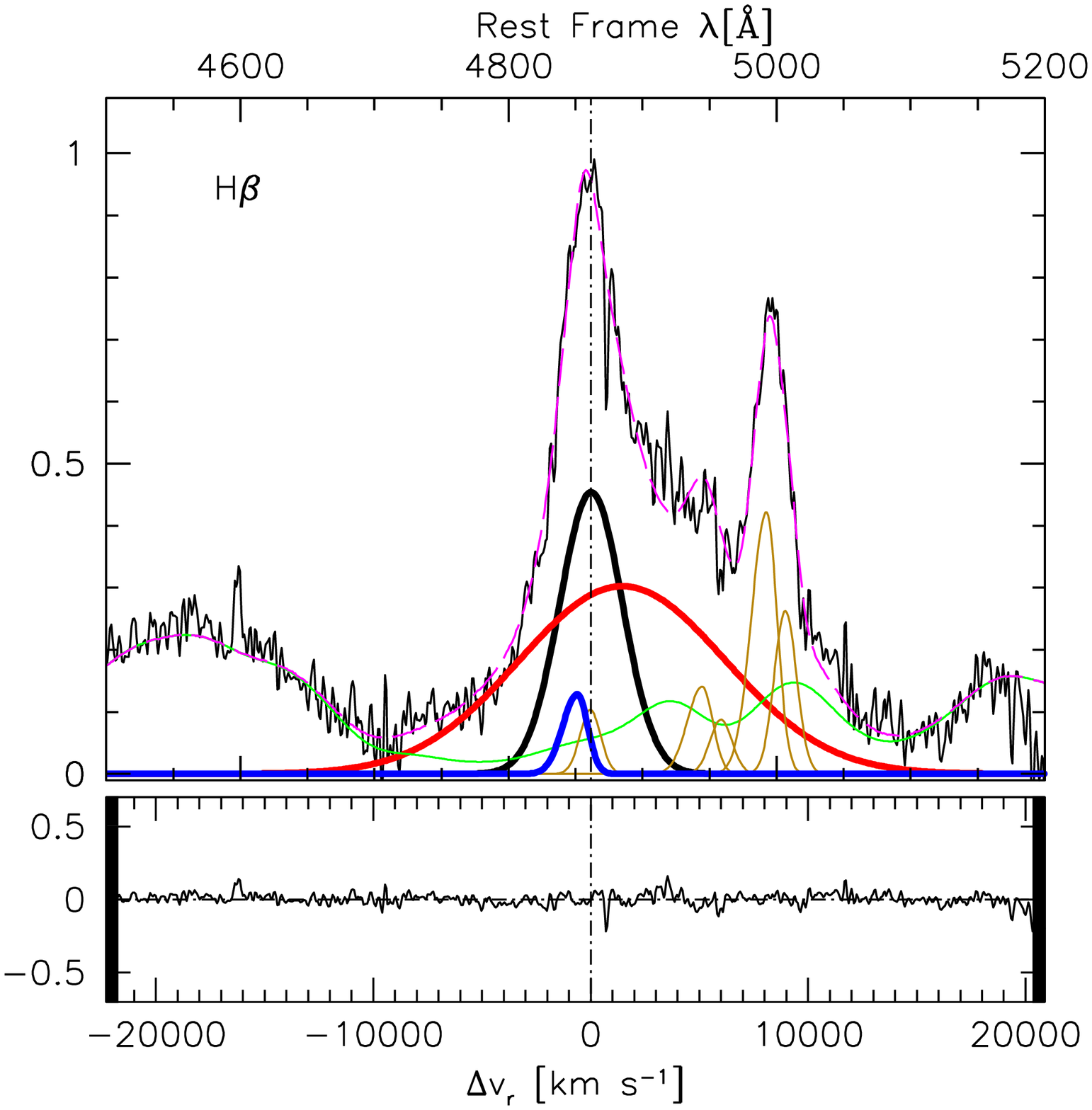} 
\includegraphics[width=0.34\columnwidth]{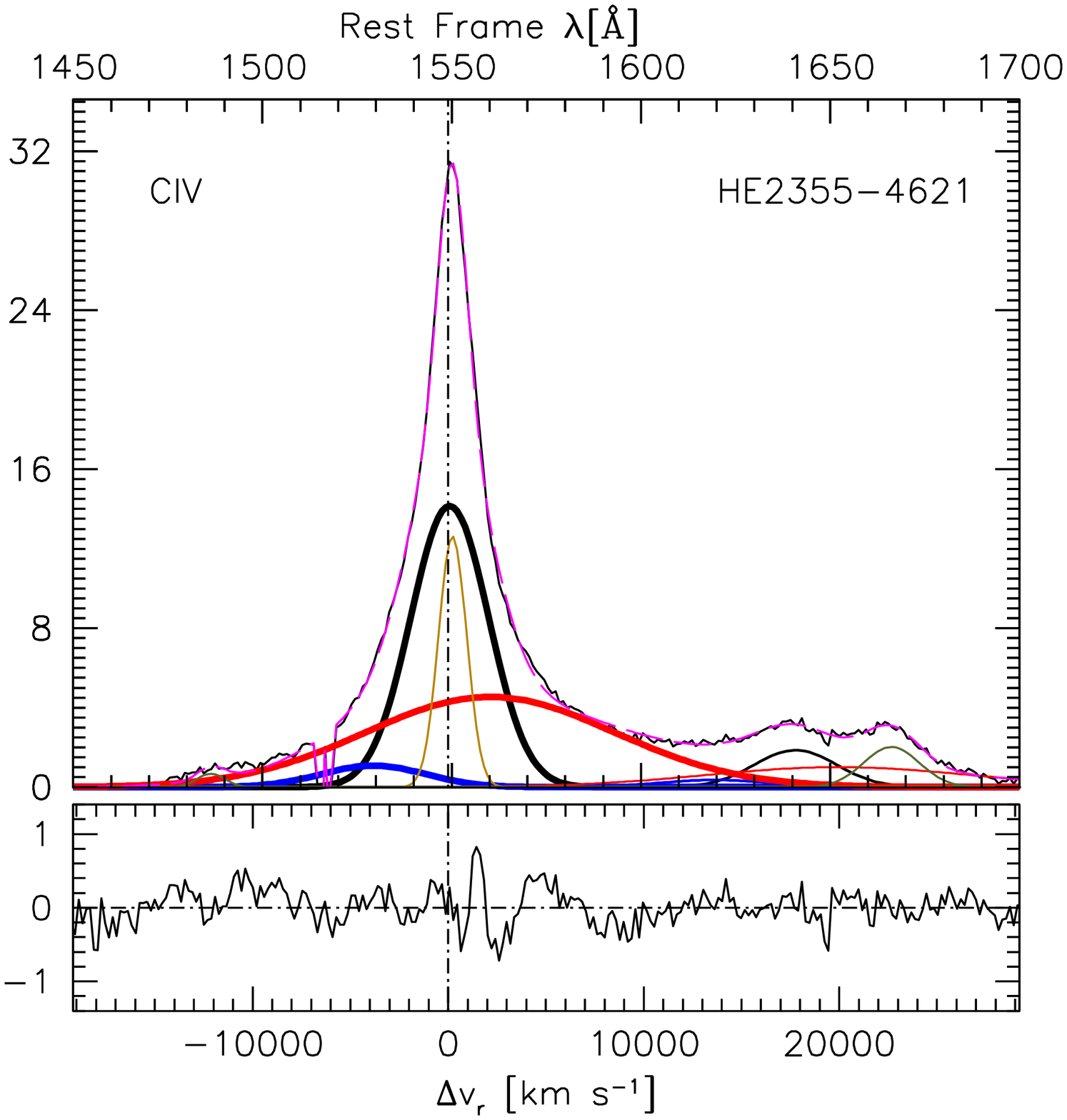}
\includegraphics[width=0.34\columnwidth]{plotHE2349hb.eps}
\caption{Same as for the previous Fig. \ref{fig:specfita}, for Pop. B sources, with the addition of the VBC (in red).}
\label{fig:specfitb}
\end{figure*}

The last four columns of Table \ref{tab:civhb}  report  the rest frame specific continuum flux $f_{\lambda}$\  at 5100 \AA\ as well as the flux and equivalent width of \hb\ and  \rfe. As for the UV range, uncertainties on $f_{\lambda}$\ are derived from the magnitude uncertainties used in the rescaling, and are at 1$\sigma$\ confidence level.  The measurements on the \hb\ full line profile are   reported in Table \ref{tab:hbprof}. We note in passing that there is only one source with an apparently discordant \hb\ FWHM with respect to its population assignment, \object{HE2349-3800}. In this case the FWHM measurement is not stable, as the profile is the addition of a narrower BC and a much broader VBC (Fig. \ref{fig:specfitb}). The asymmetry index AI $\approx$ 0.3, one of the largest in our sample, places this object among Pop. B sources. Table \ref{tab:specfithb} reports the results of the specfit analysis for  \hb, organized in the same way of \civonly.  The new {\tt specfit } \hb\ profile analysis yields results that are consistent with the previous work (M09) on the same objects. VBC shifts are  $\sim 2000 $ \kms\ to the red, and the VBC contribution is large, above 1/2 of the total line flux, consistently with the decomposition made by M09 on composite spectra. The large contribution of the VBC makes the BC FWHM especially uncertain.

\begin{table}
\setlength{\tabcolsep}{2.5pt}
\begin{center}\scriptsize
\caption{Measurements on the \hb\ full line profile   \label{tab:hbprof}}
\begin{tabular}{lcccrrr}\hline\hline\noalign{\vskip 0.05cm}
\multicolumn{1}{l}{Identification}  &\multicolumn{1}{c}{FWHM}& \multicolumn{1}{c}{A.I.} &\multicolumn{1}{c}{Kurt.}     &\multicolumn{1}{c}{$c(\frac{1}{4})$} & \multicolumn{1}{c}{$c(\frac{1}{2})$} \\  
&   [\kms]    & &  & [\kms]   & [\kms]\\ 
\hline\noalign{\vskip 0.05cm}
\multicolumn{6}{c}{Population A}&\\[0.05cm] \hline\noalign{\vskip 0.05cm}
HE0043-2300	&	4150$\pm$\ 300	&	0.00	$\pm$\ 0.10	&	0.33	$\pm$\ 	0.06	&	200	$\pm$\ 	700	&	200	$\pm$\ 	300	\\
HE0109-3518	&	4450$\pm$\ 330	&	0.00	$\pm$\ 0.10	&	0.32	$\pm$\ 	0.06	&	20	$\pm$\ 	760	&	20	$\pm$\ 	330	\\
HE0122-3759 	&	3440$\pm$\ 260	&	-0.01$\pm$\ 0.09	&	0.34	$\pm$\ 	0.05	&	-60	$\pm$\ 	500	&	-100	$\pm$\ 	260	\\
HE0205-3756	&	4680$\pm$\ 350	&	-0.06$\pm$\ 0.10	&	0.33	$\pm$\ 	0.05	&	-70	$\pm$\ 	760	&	50	$\pm$\ 	350	\\
HE0248-3628	&	4480$\pm$\ 320	&	0.00	$\pm$\ 0.10	&	0.33	$\pm$\ 	0.06	&	-40	$\pm$\ 	760	&	-40	$\pm$\ 	320	\\
HE0251-5550	&	5050$\pm$\ 360	&	0.00	$\pm$\ 0.10	&	0.33	$\pm$\ 	0.06	&	-50	$\pm$\ 	850	&	-50	$\pm$\ 	360	\\
HE0359-3959	&	5050$\pm$\ 350	&	-0.10$\pm$\ 0.09	&	0.33	$\pm$\ 	0.04	&	-500	$\pm$\ 	590	&	-610	$\pm$\ 	350	\\
HE0507-3236 	&	3880$\pm$\ 280	&	0.00	$\pm$\ 0.10	&	0.33	$\pm$\ 	0.06	&	-100	$\pm$\ 	650	&	-100	$\pm$\ 	280	\\
HE0512-3329 	&	4080$\pm$\ 250	&	0.02	$\pm$\ 0.09	&	0.38	$\pm$\ 	0.06	&	-20	$\pm$\ 	570	&	-180	$\pm$\ 	250	\\
HE0940-1050	&	3420$\pm$\ 260	&	-0.06$\pm$\ 0.09	&	0.33	$\pm$\ 	0.05	&	-190	$\pm$\ 	470	&	-210	$\pm$\ 	260	\\
HE1104-1805	&	4650$\pm$\ 330	&	0.00	$\pm$\ 0.10	&	0.33	$\pm$\ 	0.06	&	-80	$\pm$\ 	780	&	-80	$\pm$\ 	330	\\
HE1347-2457	&	5500$\pm$\ 390	&	0.03	$\pm$\ 0.10	&	0.34	$\pm$\ 	0.06	&	10	$\pm$\ 	940	&	-70	$\pm$\ 	390	\\
HE2147-3212	&	4680$\pm$\ 340	&	0.17	$\pm$\ 0.12	&	0.31	$\pm$\ 	0.06	&	-30	$\pm$\ 	860	&	-160	$\pm$\ 	340	\\
HE2352-4010 	&	3640$\pm$\ 260	&	0.00	$\pm$\ 0.10	&	0.33	$\pm$\ 	0.06	&	30	$\pm$\ 	610	&	30	$\pm$\ 	260	\\[0.05cm]
 \hline\noalign{\vskip 0.05cm}
\multicolumn{6}{c}{Population B}&\\[0.05cm] \hline\noalign{\vskip 0.05cm}

HE0035-2853	&	7290$\pm$\ 590	&	0.36	$\pm$\ 0.07	&	0.30	$\pm$\ 	0.05	&	1910	$\pm$\ 	1320	&	800	$\pm$\ 	590	\\
HE0058-3231	&	6720$\pm$\ 640	&	0.31	$\pm$\ 0.06	&	0.34	$\pm$\ 	0.04	&	2210	$\pm$\ 	770	&	1110	$\pm$\ 	640	\\
HE0203-4627	&	7310$\pm$\ 420	&	0.09	$\pm$\ 0.08	&	0.41	$\pm$\ 	0.05	&	620	$\pm$\ 	870	&	300	$\pm$\ 	420	\\
HE0349-5249	&	6530$\pm$\ 650	&	0.28	$\pm$\ 0.08	&	0.31	$\pm$\ 	0.04	&	1900	$\pm$\ 	970	&	710	$\pm$\ 	650	\\
HE0436-3709	&	6640$\pm$\ 410	&	0.13	$\pm$\ 0.09	&	0.39	$\pm$\ 	0.05	&	920	$\pm$\ 	950	&	420	$\pm$\ 	410	\\
HE0926-0201 	&	6250$\pm$\ 400	&	0.29	$\pm$\ 0.10	&	0.37	$\pm$\ 	0.06	&	410	$\pm$\ 	1060	&	30	$\pm$\ 	400	\\
HE1039-0724	&	9380$\pm$\ 690	&	0.26	$\pm$\ 0.07	&	0.33	$\pm$\ 	0.04	&	1390	$\pm$\ 	960	&	1170	$\pm$\ 	690	\\
HE1120+0154	&	6520$\pm$\ 480	&	0.34	$\pm$\ 0.09	&	0.33	$\pm$\ 	0.06	&	950	$\pm$\ 	1330	&	330	$\pm$\ 	480	\\
HE1349+0007	&	6090$\pm$\ 390	&	0.22	$\pm$\ 0.10	&	0.38	$\pm$\ 	0.07	&	810	$\pm$\ 	1280	&	260	$\pm$\ 	390	\\
HE1409+0101	&	7620$\pm$\ 530	&	0.34	$\pm$\ 0.08	&	0.34	$\pm$\ 	0.05	&	1040	$\pm$\ 	1290	&	290	$\pm$\ 	530	\\
HE2156-4020	&	5850$\pm$\ 430	&	0.18	$\pm$\ 0.11	&	0.32	$\pm$\ 	0.06	&	740	$\pm$\ 	1290	&	140	$\pm$\ 	430	\\
HE2202-2557 	&	8770$\pm$\ 530	&	0.14	$\pm$\ 0.08	&	0.41	$\pm$\ 	0.05	&	1330	$\pm$\ 	1010	&	720	$\pm$\ 	530	\\
HE2349-3800 	&	4390$\pm$\ 350	&	0.30	$\pm$\ 0.12	&	0.28	$\pm$\ 	0.06	&	1110	$\pm$\ 	1220	&	200	$\pm$\ 	350	\\
HE2355-4621 	&	7100$\pm$\ 450	&	0.14	$\pm$\ 0.11	&	0.37	$\pm$\ 	0.06	&	990	$\pm$\ 	1300	&	350	$\pm$\ 	450	\\[0.05cm] \hline
\end{tabular}
\end{center}
\end{table}

\begin{table*}
\setlength{\tabcolsep}{1pt}
\begin{center}\scriptsize
\caption{Results of {\tt specfit} analysis  on \hb\   \label{tab:specfithb}}
\begin{tabular}{lcccccccccccrrrr}\hline\hline\noalign{\vskip 0.05cm}
\multicolumn{1}{l}{Identification}  &\multicolumn{3}{c}{BLUE}&& \multicolumn{3}{c}{BC} &&\multicolumn{3}{c}{VBC}  &&\multicolumn{3}{c}{NC} \\[0.05cm] \cline{2-4}\cline{6-8}\cline{10-12}\cline{14-16}\noalign{\vskip 0.05cm}
& $I/I_\mathrm{tot}$ & FWHM & Shift  && $I/I_\mathrm{tot}$ & FWHM &Shift && $I/I_\mathrm{tot}$ & FWHM & Shift  && $I/I_\mathrm{tot}$ & FWHM & Shift \\ 
& & \multicolumn{2}{c}{$\rm [km~s^{-1}]$} &&   &\multicolumn{2}{c}{$\rm [km~s^{-1}]$} & && \multicolumn{2}{c}{$\rm [km~s^{-1}]$}& && \multicolumn{2}{c}{$\rm [km~s^{-1}]$}\\ 
\hline\noalign{\vskip 0.08cm}
&\multicolumn{14}{c}{Population A}&\\[0.05cm] \hline\noalign{\vskip 0.08cm}
HE0043-2300	&	0.00	&	\ldots	&	\ldots	&	&	0.96	&	4150	&	180	&	&	\ldots	&	\ldots	&	\ldots	&	&	0.039	&	1290	&	-10	\\
HE0109-3518	&	0.00	&	\ldots	&	\ldots	&	&	0.96	&	4570	&	0	&	&	\ldots	&	\ldots	&	\ldots	&	&	0.037	&	860	&	-20	\\
HE0122-3759 	&	0.03	&	1450	&	-1980	&	&	0.96	&	3250	&	0	&	&	\ldots	&	\ldots	&	\ldots	&	&	0.005	&	390	&	-20	\\
HE0205-3756	&	0.05	&	\ldots	&	-2640	&	&	0.93	&	4430	&	230	&	&	\ldots	&	\ldots	&	\ldots	&	&	0.017	&	420	&	-20	\\
HE0248-3628	&	0.00	&	\ldots	&	\ldots	&	&	0.98	&	4480	&	-20	&	&	\ldots	&	\ldots	&	\ldots	&	&	0.025	&	1000	&	-30	\\
HE0251-5550	&	0.00	&	\ldots	&	\ldots	&	&	0.98	&	5050	&	-40	&	&	\ldots	&	\ldots	&	\ldots	&	&	0.023	&	890	&	10	\\
HE0359-3959	&	0.09	&	2560	&	-2770	&	&	0.88	&	3960	&	-20	&	&	\ldots	&	\ldots	&	\ldots	&	&	0.033	&	890	&	-80	\\
HE0507-3236 	&	0.00	&	\ldots	&	\ldots	&	&	0.95	&	3880	&	-130	&	&	\ldots	&	\ldots	&	\ldots	&	&	0.054	&	1300	&	90	\\
HE0512-3329 	&	0.04	&	1730	&	-1680	&	&	0.90	&	3700	&	0	&	&	\ldots	&	\ldots	&	\ldots	&	&	0.057	&	1060	&	-10	\\
HE0940-1050	&	0.05	&	1970	&	-1900	&	&	0.95	&	2990	&	0	&	&	\ldots	&	\ldots	&	\ldots	&	&	0.000	&	\ldots	&	\ldots	\\
HE1104-1805 	&	0.00	&	\ldots	&	\ldots	&	&	1.00	&	4650	&	0	&	&	\ldots	&	\ldots	&	\ldots	&	&	0.000	&	\ldots	&	\ldots	\\
HE1347-2457	&	0.03	&	3090	&	-840	&	&	0.93	&	5800	&	0	&	&	\ldots	&	\ldots	&	\ldots	&	&	0.034	&	1240	&	10	\\
HE2147-3212	&	0.05	&	2180	&	-830	&	&	0.94	&	5450	&	0	&	&	\ldots	&	\ldots	&	\ldots	&	&	0.012	&	1200	&	0	\\
HE2352-4010 	&	0.08	&	3210	&	\ldots	&	&	0.88	&	3640	&	0	&	&	\ldots	&	\ldots	&	\ldots	&	&	0.035	&	1200	&	0	\\[0.05cm]
\hline\noalign{\vskip 0.05cm}
&\multicolumn{14}{c}{Population B}&\\[0.05cm] \hline\noalign{\vskip 0.05cm}
HE0035-2853	&	0.04	&	1560	&	-710	&	&	0.57	&	6980	&	0	&	&	0.38	&	 	11240	&	5470	&	&	0.017	&	990	&	0	\\
HE0058-3231	&	0.00	&	\ldots	&	\ldots	&	&	0.40	&	4580	&	0	&	&	0.58	&		9150	&	3350	&	&	0.021	&	1000	&	-60	\\
HE0203-4627	&	0.00	&	\ldots	&	\ldots	&	&	0.58	&	6270	&	0	&	&	0.41	&	 	12390	&	2060	&	&	0.011	&	1070	&	0	\\
HE0349-5249	&	0.00	&	\ldots	&	\ldots	&	&	0.30	&	4270	&	0	&	&	0.67	&	 	11250	&	2380	&	&	0.027	&	1350	&	0	\\
HE0436-3709	&	0.00	&	\ldots	&	\ldots	&	&	0.50	&	5420	&	50	&	&	0.49	&	 	11190	&	2140	&	&	0.015	&	990	&	0	\\
HE0926-0201 	&	0.04	&	2680	&	-1020	&	&	0.37	&	5230	&	0	&	&	0.52	&	 	13010	&	950	&	&	0.067	&	1200	&	20	\\
HE1039-0724	&	0.04	&	1420	&	-1170	&	&	0.54	&	8880	&	0	&	&	0.39	&	 	12170	&	2430	&	&	0.019	&	810	&	-100	\\
HE1120+0154	&	0.03	&	1720	&	-960	&	&	0.43	&	5920	&	0	&	&	0.52	&	 	14040	&	2400	&	&	0.022	&	850	&	-60	\\
HE1349+0007	&	0.03	&	1720	&	-1010	&	&	0.51	&	5450	&	0	&	&	0.43	&	 	13070	&	3110	&	&	0.025	&	910	&	0	\\
HE1409+0101	&	0.06	&	2270	&	-1620	&	&	0.37	&	6810	&	-20	&	&	0.50	&	 	14560	&	2230	&	&	0.067	&	1690	&	-80	\\
HE2156-4020	&	0.02	&	1340	&	-540	&	&	0.38	&	4760	&	-100	&	&	0.58	&	 	13050	&	1440	&	&	0.023	&	1030	&	-20	\\
HE2202-2557 	&	0.00	&	\ldots	&	\ldots	&	&	0.52	&	7170	&	0	&	&	0.47	&	 	12340	&	3000	&	&	0.010	&	630	&	120	\\
HE2349-3800 	&	0.03	&	1580	&	-620	&	&	0.30	&	3390	&	20	&	&	0.65	&	 	11090	&	1450	&	&	0.022	&	1110	&	-20	\\
HE2355-4621 	&	0.00	&	\ldots	&	\ldots	&	&	0.45	&	5660	&	40	&	&	0.52	&	13900	&	2150	&	&	0.027	&	1200	&	-80	\\[0.05cm]  
\hline
\end{tabular}
\end{center}
\end{table*}

\begin{table*}\scriptsize
\setlength{\tabcolsep}{2pt}
\centering
\caption{Two sample comparison   \label{tab:2samp}}
\begin{tabular}{lccccccccccccccc}\hline\hline\noalign{\vskip 0.05cm}
 \multicolumn{1}{l}{Parameter }  & \multicolumn{2}{c}{A LOWZ  } &&\multicolumn{2}{c}{B LOWZ  }  &&\multicolumn{2}{c}{A HE } &&\multicolumn{2}{c}{B HE  } &&  \multicolumn{1}{c}{A HE vs LOWZ}  &&\multicolumn{1}{c}{B HE vs LOWZ}  \\[0.05cm]  
 \cline{2-3} \cline{5-6} \cline{8-9}  \cline{11-12} \cline{14-14} \cline{16-16}\noalign{\vskip 0.05cm}
\multicolumn{1}{l}{}  & \multicolumn{2}{c}{mean $\pm  \sigma$} &&\multicolumn{2}{c}{mean $\pm  \sigma$}  &&\multicolumn{2}{c}{mean $\pm  \sigma$} &&\multicolumn{2}{c}{mean $\pm  \sigma$} &&   $P_\mathrm{KS}$  &&    $P_\mathrm{KS}$\\ [0.05cm] 
\hline\noalign{\vskip 0.05cm}
$z$	&	0.17	&	0.14	&&	0.28	&	0.23	 	&&	2.02	&	0.52	&&	1.67	&	0.35	&&	3.0E-10	&&	1.2E-09	\\	
$L$	&	45.42&	0.79	&&	45.76	&	1.00	 	&&	47.69	&	0.36	&&	47.48	&	0.22	&&	9.3E-10	&&	1.7E-08	\\
\lledd\ &	-0.56	&	0.43	&&	-1.27	&	0.51	 	&&	-0.02	&	0.22	&&	-0.51	&	0.20	&&	3.3E-05	&&	5.8E-06	\\
\\
FWHM(\civonly)&	4321	&	1375	&&	5266	&	1487	&&	6479&	1722&&	6454&	661	&&	2.8E-04	&&	3.5E-04	\\
\cqp\ (\civonly)	&-1090	&	1245	&&	-322	&	927	&&	-3431	&	1586&&	-1656	&	1269	&&	1.3E-04	&&	2.5E-05	\\
\cmp\ (\civonly)	&-865	&	964	&&	-250	&	557	&&	-2742	&	1516	&&	-1559	&	920	&&	2.0E-04	&&	5.0E-04	\\
AI(civ)	&	-0.12	&	0.14	&&	-0.04	&	0.14		&&	-0.19	&	0.09	&&	-0.03	&	0.13	&&	3.3E-01	&&	3.7E-01	\\
curt(\civonly)	&	0.36	&	0.08	&&	0.33	&	0.05		&&	0.37	&	0.07	&&	0.34	&	0.03	&&		7.8E-01	&&	4.9E-01	\\
$W$(\civonly)	&	55	&	36	&&	119	&	84	&&	25	&	13	&&	35	&	13	&&		1.9E-03	&&	7.3E-06	\\	
\\ 
$R_\mathrm{FeII}$	&	0.66	&	0.48	&&	0.27	&	0.21 &&	0.62	&	0.32	&&	0.42	&	0.17	&&	4.2E-01	&&	2.1E-02	\\	
FWHM(\hb)		&	2366	&	850	&&	6390	&	2513 &&	4368	&	630	&&	6890	&	1221	&&	6.8E-08	&&	7.7E-02 \\	
\cqp\ (\hb)$^\mathrm{a}$	&	59	&	472	&&	1225	&	1463	&&	-63	&	153	&&	1166	&	526	&&	8.3E-02	&&	2.1E-01	\\	
\cmp\ (\hb)$^\mathrm{a}$	&	1	&	309	&&	497	&	886	&&	-93	&	182	&&	488	&	355	&&	2.4E-01	&&	6.1E-01	\\
AI(\hb)$^\mathrm{a}$	&	0.01	&	0.08	&&	0.11	&	0.20	&&	0.00	&	0.06	&&	0.24	&	0.09	&&	8.3E-02	&&	 	3.0E-02	\\
curt(\hb)$^\mathrm{a}$&	0.33		&	0.04	&&	0.37	&	0.09	&&	0.33	&	0.02	&&	0.35	&	0.04	&&	1.4E-01	&&	 1.1E-01	\\
$W$(\hb)$^\mathrm{a}$	&	88	&	34	&&	108	&	33	&&	68	&	29	&&	76	 &	16	&&	1.4E-01	&&	8.6E-04	\\[0.05cm] \hline\noalign{\vskip 0.05cm}
 \multicolumn{16}{l}{$^\mathrm{a}$ Restricted to the sample for which \hb\ and \civonly\ are both available (\S \ref{fos}). }\\
\end{tabular}

\end{table*}

\subsection{HE  and LOWZ sample comparison }
\label{helowz} 
 
The meaning of the \civ\ properties in the HE sample can be properly evaluated if compared to a low-$L$\ sample. 
The main statistical results from the inter-sample LOWZ-HE comparison  are summarized in Table \ref{tab:2samp} where parameter  averages and sample standard deviations  are reported for Pop. A and B, along with the statistical significance 
of the Kolmogorov-Smirnov  test, $P_\mathrm{KS}$,  of the null hypothesis that the HE and LOWZ samples are undistinguishable.

The first two rows list the results for $z$ and UV luminosity $\lambda L_{\lambda}$\  at 1450 \AA\  which are obviously different at  a high significance level. The third parameter listed is the Eddington ratio. {The \lledd\ has been computed from the 5100 \AA\ luminosity, assuming a bolometric correction factor 10 and $L_\mathrm{Edd} \approx 1.5\times10^{38}\cdot(M_\mathrm{BH}/M_{\odot})$ \ergss. The black hole mass \mbh\ was estimated using the same 5100 \AA\ luminosity, the FWHM \hbbc\ with the scaling law of \citet{vestergaardpeterson06}.   Black hole masses in the HE sample are $\sim $ several {$10^{9} $ \msol}, close to the largest \mbh\ values observed in active quasars \citep[][and references therein]{sulenticetal06,marzianisulentic12} and  spent ones in the local Universe \citep[such as Messier 87, ][]{walshetal13}.   A \lledd\ systematic difference between LOWZ and HE is present  for both Pop. A Pop. B. Indeed, several of the HE Pop. B sources are close to the boundary between Pop. A and B set from the analysis of low-$z$ quasars.  This is, at least in part, the effect of a redshift-dependent  \lledd\ bias in a flux limited survey \citep{sulenticetal14}. In addition, a systematic increase of \lledd\ with z is also expected \citep{trakhtenbrotnetzer12}. The sample difference in \lledd\ is important for interpreting  any luminosity difference, and makes it  advisable to consider, in addition to Pop. A and B separately, also the full A+B sample (\S \ref{corr}). 

The second group of rows provides  \civonly\  line parameters: FWHM, \cmp, \cqp, AI, kurtosis and rest frame EW.   The third groups provides first the 4DE1 parameters \rfe\ and FWHM \hb, and then the profile \hb\ parameters reported as for \civonly.  

We note in passing that the \civonly\ equivalent width is significantly lower for both Pop. A and B, reflecting the well known anticorrelation between equivalent width of HILs and luminosity (the Baldwin effect e.g., \citealt[][]{baldwinetal78,bianetal12}). The \civonly\ EW is also strongly affected by \lledd\ \citep{bachevetal04,baskinlaor04,shemmerlieber15}, but the  issue that will be discussed in a follow-up paper. 

The 4DE1 parameter  \rfe\ (third row)  is also significantly higher than at low-$z$, but only for B sources, probably reflecting the increase of \rfe\ as a function of Eddington ratio \citep{marzianietal01,sunshen15}. A similar effect is expected to operate for W(\hb) which is systematically higher in Pop. B than in Pop. A sample at low $L$: Table \ref{tab:2samp} indicate a significant decrease in the HE sample, $\approx 76$\ \AA\ with respect to $\approx 100$\ at low $L$.

A main result concerning the \civonly\ line profile is that the  LOWZ and HE \civonly\ centroids and FWHM  are significantly different at a confidence level $\gtrsim 4\sigma$\  for both Pop. A and B (4th to 6th rows in Table \ref{tab:2samp}). While at low-$L$ a strong effect of the blueshifted excess is restricted mainly to extreme Population A sources, at high $L$ the effect of the blue shifted excess is significant  also for Pop.  B sources as shown by the profile decomposition of Fig. \ref{fig:specfitb}.  In the case of the most extreme blueshifts, the kurtosis value can be $\gtrsim 0.4$\ (Table \ref{tab:civprof}), although sample differences between LOWZ and HE Pop. A and B are never significant (Table \ref{tab:2samp}). 
On average, {\em profile shape} (AI and kurt) parameters are preserved for both \civonly\ and \hb. 
 
The \civonly\ FWHM is correlated with  \civonly\ \cmp\ in Pop. A sources (see Figure \ref{fig:windfwhm}; c.f. \citealt{coatmanetal16}):
\begin{equation}
 c(\frac{1}{2}) \approx (-0.62 \pm 0.07) \cdot \mathrm{FWHM} + (1630 \pm 380)\, \mathrm{km \ s}^{-1}
\end{equation}
with a significance for the null hypothesis of $P \approx 10^{-8}$.  The correlation can be straightforwardly explained by the assumption of a blueshifted component  of increasing prominence summed to an unshifted symmetric component  represented through a scaled \hb\ profile, as actually done with the {\tt specfit} analysis (\S \ref{immediate}). The blueshifted excess resolved in radial velocity fully motivates the modelization of the \civonly\ profile as due to the sum of BLUE and BC, and accounts for the \civonly\ FWHM  being larger than the FWHM \hb\ by almost a factor $\approx 2$ in Pop. A sources (Tables \ref{tab:civprof} and \ref{tab:hbprof}).  

\begin{figure}
\centering
\includegraphics[width=0.9\columnwidth]{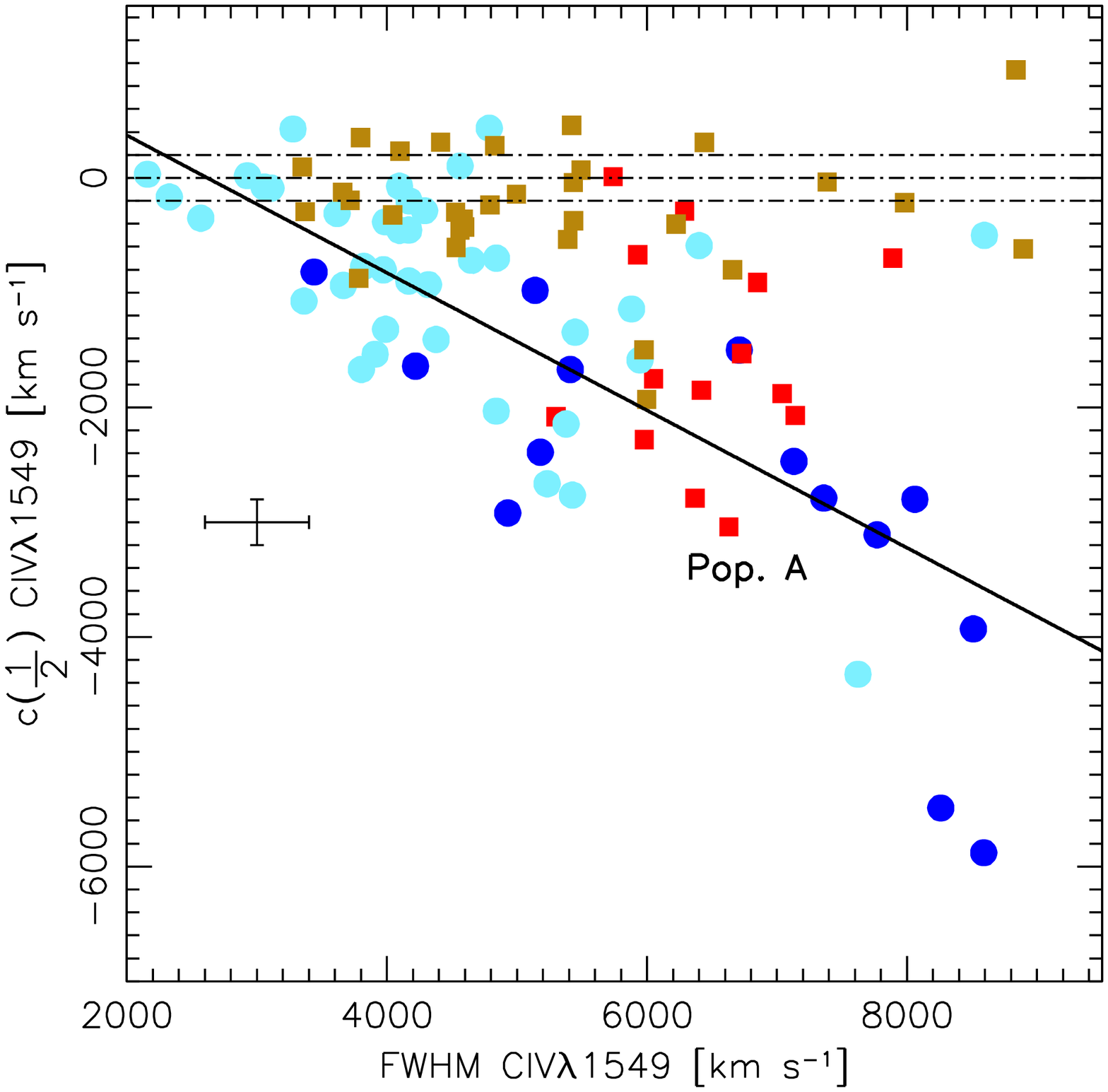}
\caption{Correlation between \civonly\ FWHM and \civonly\ \cmp. Pop. A sources are indicated by filled circles and Pop. B  sources by filled squares. HE sources are colored blue (Pop. A) and red (Pop. B),  and  LOWZ sources in pale blue (Pop. A) and golden  (Pop. B). The lsq best fit line (black) refers to Pop. A sources. The dot-dashed lines at $\pm$200 \kms\ mark the \civonly\ \cmp\ range consistent with no centroid shift. Error bars refer to 1$\sigma$\ level uncertainty.}
\label{fig:windfwhm}
\end{figure}

The large \civonly\ blueshifts suggest that in Pop. A sources the \civonly\ emission is dominated by BLUE:  BLUE is $\lesssim \frac{1}{2}$\ of the \civonly\ total flux for   5 Pop. A  objects  and $\gtrsim 60$\%\ for 9 of them. On the converse, 11 Pop. B sources show I(BLUE) $\lesssim \frac{1}{2}$\ the total \civonly\ emission I(\civonly), and only 3 Pop. B sources have I(BLUE)/I(\civonly)  $\gtrsim 60$\%\  (Table \ref{tab:specfitciv}). The scheme in \citet{marzianietal16} indicates that the lower shifts in Pop. B can be ascribed to the combined effect of BLUE plus a   BC+VBC profile resembling the one of \hb. 

\section{Discussion}
\label{disc}
\subsection{CIV in EXTREME QUASARS in 4DE1}

The basic picture emerging from the HE sample involves a high prevalence of large \civonly\ shifts, in several cases as large as several thousands \kms\ with respect to the adopted rest frame, and in two cases exceeding 5000 \kms. Twenty-one out of  24 HE sources  show \cmp\ \civonly\ blueshifts larger than 1000 \kms\ (see Table \ref{tab:civprof}). This result is consistent with past finding \citep[e.g.,][]{richardsetal11} as well as with more recent works   which show that quasars at the highest luminosities ($\log L \gtrsim 47 $\ [\ergss])  display large \civonly\ blueshifts \citep[e.g.][]{coatmanetal16,vietri17}. 

A two-population distinction was introduced by several works, on the basis of the \civ\ blueshifts \citep[][]{richardsetal11} and of the ratio FWHM to velocity dispersion \citep{collinetal06,kollatschnyzetzl11}. Population A and B quasars as defined in \citet{sulenticetal00a} provide an effective  way to discriminate extremes in quasar properties. Figure \ref{fig:composites} shows normalized median composite \civonly\ profiles for the 14 Pop A  (blue) and 14 Pop B (red) sources where the vertical line represents rest frame wavelength of \civonly, derived from the matching \hb spectra. 
The median composite profiles presented in Fig. \ref{fig:composites} show that Pop. A sources have  a stronger signature of a blue component indicative of a wind or outflow. Median composite profiles are a good illustration of the individual blueshift measures presented in Table \ref{tab:civprof}, where the high luminosity Pop A profiles show blueshifts with a median of 80\% of the flux on the blue side of the rest frame. The fraction of blue component flux   among Pop B sources is 60\%, also consistent with the one of the composite.  In the comparison LOWZ sample  only extreme Pop A bins (A3, A4) show similar percentage of blueshift flux.

The distributions of the individual values as well as the averages and medians of profile measurements allow a first order comparison of  Pop A-B consistency in the three samples: one of lower L/higher $z$\ (GTC) and the other of  higher L/higher z (HE), in addition to LOWZ. 

\begin{figure}[htp!]
\centering
\includegraphics[width=0.98\columnwidth,angle=0]{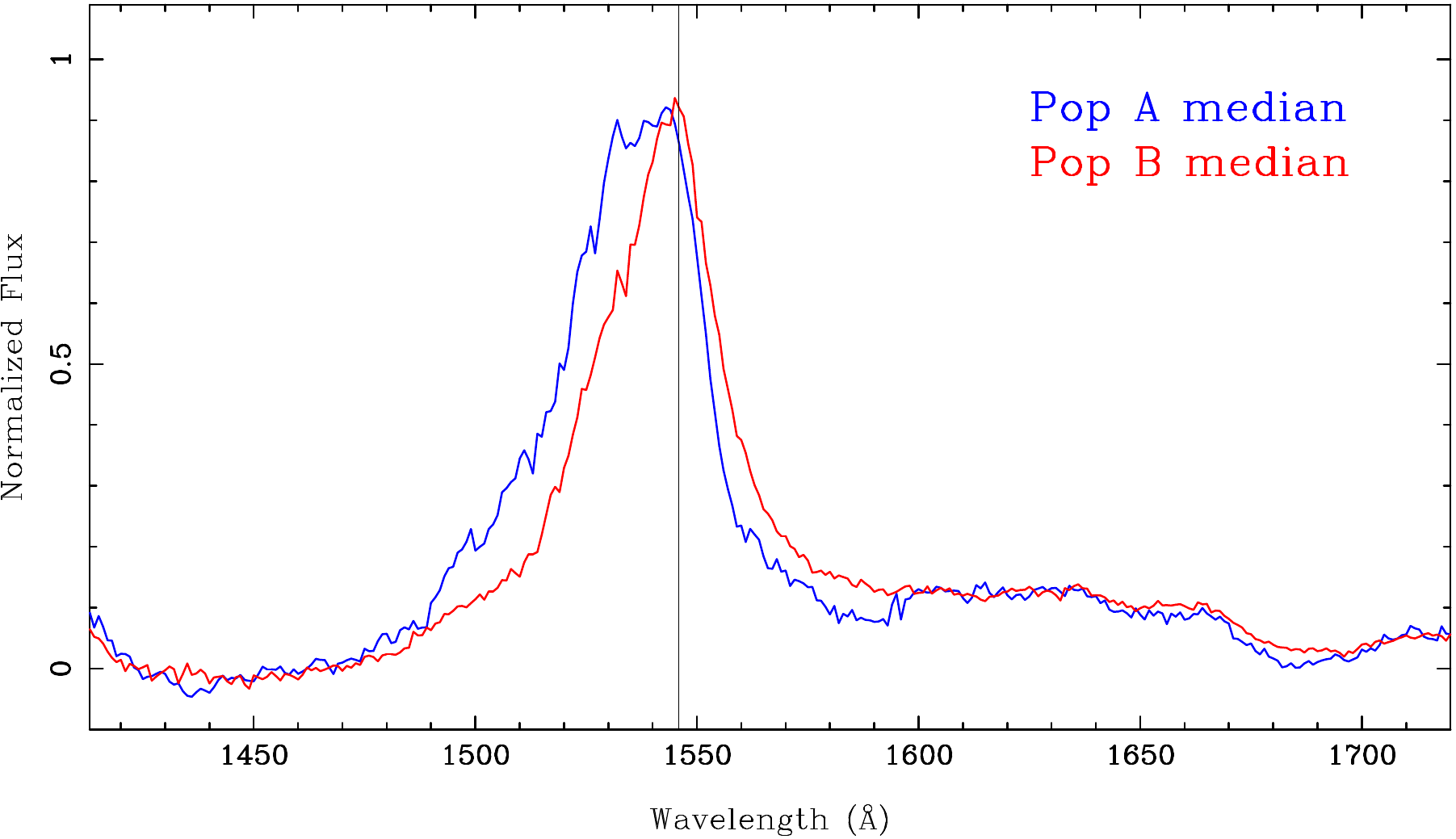}
\caption{Median composite \civonly profiles for the 14 Population A (blue line)
 and 14 Pop. B (red) sources in the HE sample. Vertical line marks the source rest frame derived from the matching \hb spectra.}
\label{fig:composites}
\end{figure}

Our LOWZ RQ sample showed Pop A sources with a median \civonly\  \cmp\  $\approx -650$ \kms\  while Pop B showed ${\rm -230}$\kms\ (averages presented in Table  \ref{tab:2samp} yield the same trend). A K-S test indicates a significant difference between the distribution of \cmp\ values of LOWZ Pop. A and B at a $4 \sigma$\ confidence level. The value for low z Pop. B is close to the typical uncertainty of individual measurements although  a ranked-sign test  \citep[][]{dodge08} provides a $Z$-score $\approx 2.3$, for the null hypothesis that the value is not different from 0. In other words, Pop. B sources show modest blueshifts, often within the instrumental uncertainties. Hence our previous inferences that only higher \lledd\, Pop A quasars show a \civonly blueshift \citep{sulenticetal07}. RL sources, which are predominantly Pop. B quasars, are overrepresented  in the Hubble archive that provided low z \civonly measures  \citep{willsetal93,marzianietal96}. The median Pop B \civonly shift decreases  to -40\kms if the RL quasars are included. Since the majority of HE (and GTC)  quasars are RQ we exclude RLs from comparisons. The GTC sample showed median shifts of -400 and -10\kms   for Pop A and B 
respectively (significantly  different at a $\approx 4 \sigma$ \ confidence level), consistent with LOWZ.  In the LOWZ and GTC samples Pop B  spectra show no significant blue flux excess with a large excess for extreme Pop. A sources \citep{sulenticetal14}. The HE sample shows median \civonly shifts of -2600 and -1800\kms for Pop A and B respectively:  \civonly is dominated by blueshifted emission  although a Pop A-B difference persists in this high z high L sample (Table \ref{tab:civprof}). The  difference between HE and  LOWZ is highly significant for both Pop. A and B,   as shown by   K-S tests whose results are reported in Table \ref{tab:2samp}: the $P_\mathrm{KS}$ corresponds to a significance level of $\gtrsim 3.5 \sigma$, for both \civonly\ \cmp\ and \cqp.

Irrespective of whether one interprets the Pop. A and B designations  as evidence for two distinct quasar populations (e.g. one capable of radio  hyperactivity and one not) or simply a measure of extremes along a 4DE1 
sequence, a search for \civonly FWHM and profile shift sample differences  in the Pop. A-B context gives us more sensitivity to changes due to  redshift, luminosity or Eddington ratio. Figure \ref{fig:histogram-shiftCIV} shows a  histogram of \civonly shifts in the HE sample. We see that the majority of Pop. A HE sources show large blueshifts  and a tendency towards smaller shifts  in Pop B thus preserving profile differences established at low z. However the difference  between A and B appears to be decreasing in extreme sources. There may be a selection bias in the HE sample towards sources with higher accretion efficiency but a much  larger sample will be needed to prove it (as discussed in \S \ref{corr}). In the HE sample we find only one Pop B  source without a blueshift and a few Pop A sources with shifts similar to I Zw 1, a A3 prototype  extreme blueshift source at low z, with \cmp\ = -1100\kms. 

\begin{figure}[htp!]
\centering
\includegraphics[width=0.9\columnwidth,angle=0]{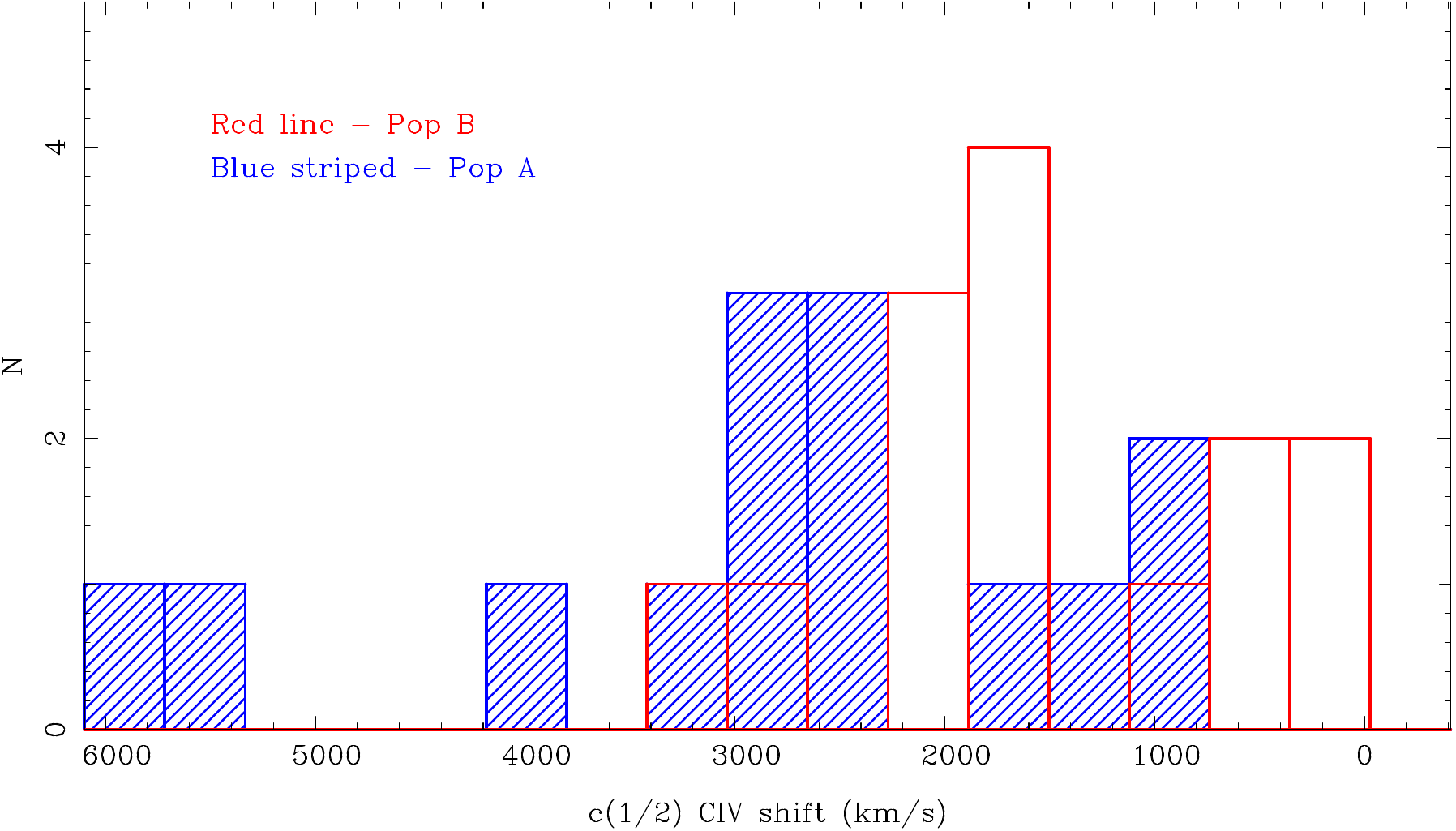}
\caption{Distribution of \civonly profile shifts, \cmp, for Pop. A (blue filled) and Pop. B (red line) 
in our extremely luminous HE quasars}
\label{fig:histogram-shiftCIV}
\end{figure}

Full profile FWHM \civonly measures for Pop A and B sources yield medians of 6900 and 6400 \kms respectively (averages are $\approx 6500$ \kms for both populations, as reported in Table \ref{tab:2samp}). These values are  higher than LOWZ medians  (with 4130 and 4910\kms for Pop A and B respectively) and GTC ($\approx$ 5200 \kms\ and 5800 \kms) estimates. }{  The LOWZ and GTC FWHM \civonly  distributions differ from the HE sample with statistical significance $\gtrsim  4 \sigma$. The  \civonly\  FWHM in the HE sample is higher for both Population A and B due to the growing strength of  the wind/outflow component shown in \S \ref{helowz} -- as \civonly blueshift increases,  FWHM \civonly is likely to increase as well (Figure \ref{fig:windfwhm}; see also Figure 7 in \citealt{coatmanetal16}). This non-virial broadening  associated with the wind \citep[e.g.,][]{richardsetal11} lead to overestimation of FWHM \civonly based black hole masses \citep[e.g.,][and references therein]{netzeretal07,sulenticetal07,parketal13,mejia-restrepoetal16,coatmanetal16a}. The amplitude of the effect depends on the location of 4DE1 sequence, being stronger for Pop. A \citep{sulenticetal07,brothertonetal15}.

\subsection{The Eigenvector 1 optical plane at low- and high-$L$}
\label{lum}

\begin{figure}
\centering
\includegraphics[width=1.\columnwidth]{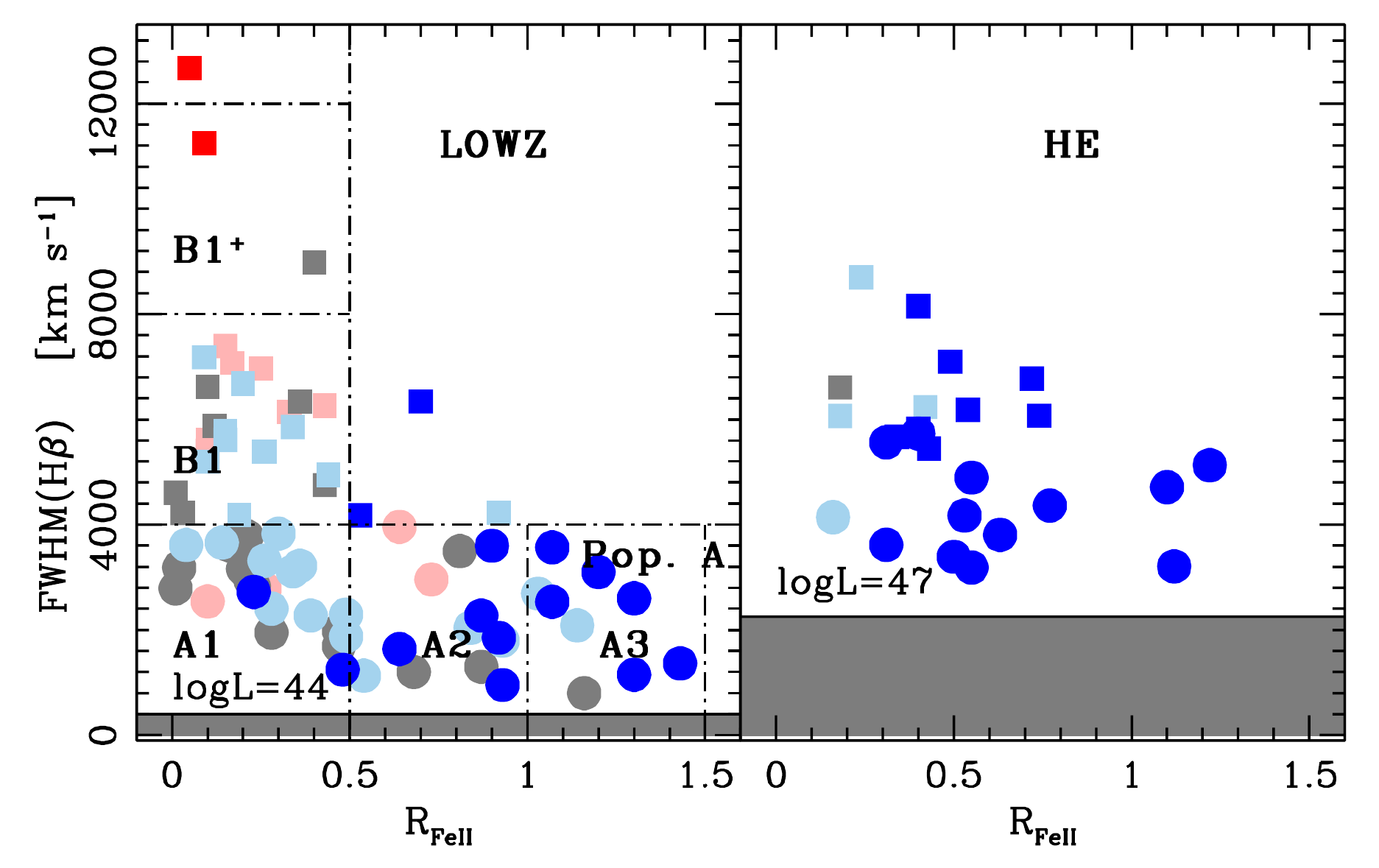}
\caption{Amplitude of \civonly \cmp\, in the 4DE1 optical plane (FWHM \hbbc vs. \rfe). Population A and B sources are marked as circles and squares respectively. Colors represent shift amplitudes. Blue: large \civonly   blueshifts (\cmp>\!-1000\kms), aquamarine: significant  blueshifts (\cmp=\,-200\! to\! -1000\kms); grey: 
no significant shift (|\cmp|<200\kms). Red and pink, large and significant \cmp\, redshift, respectively. 
Left: 4DE1 plane for the RQ LOWZ sample. Right: 4DE1 plane for HE sample. The horizontal filled lines 
mark the minimum FWHM\hb for virialized systems at $\log L=44\ {\rm and} \ 47$ \citep[no sources 
may exist in the shaded area,][]{marzianietal09}.}
\label{fig:4DE1withCIV}
\end{figure}

Figure \ref{fig:4DE1withCIV} incorporates \civonly\ into the 4DE1 optical plane using color to represent the strength of the \civonly blueshift. The LOWZ sample (left panel) illustrates  the strong preference for \civonly blueshifts but only in Population A sources and especially sources with higher \lledd\ in bins A2 and A3. Large blueshifts are rare in LOWZ Pop. 
B. The extreme HE sources presented in this paper (right panel Fig. \ref{fig:4DE1withCIV}) show a larger fraction of \civonly blueshifts due to the blue excess in the \civonly profiles of most sources. The FWHM(\hb) of HE sample shift upwards as expected if \hb\ arises in a  structure  where profile width is dominated by virial motions \citep[\S \ref{pop}; ][a prime candidate is the accretion disk]{marzianietal09}. 

The advantage of the present work is to introduce accurate rest frame estimates and S/N high enough  to decompose the \civonly\ line profile in a symmetric virialized  component and a blue shifted excess at high $L$. Clear trends then emerge by joining low- and high-$L$\ sources  which are not even guessed at in low-$z$\ samples. The question is now whether the change in the HE sample is driven by L or \lledd.

\subsection{Is it L or \lledd\ that drives the \civonly outflow?}
\label{corr}

Evidence  at low $z$\ suggests that \cmp\, \civonly is an Eigenvector 1 parameter with \lledd\  likely to be  the driver of the wind/outflow. Source luminosity, L, has long been known to be an Eigenvector 2  parameter \citep{borosongreen92} and therefore not strongly related to any of the 4DE1 parameters which, in low-$L$ samples have been shown to depend more strongly on \lledd\ than on $L$ {\bf \citep{marzianietal03b,sunshen15}}. In other words, luminosity effects are not strong in type-1, and, for a moderate luminosity range (1-2 dex), the diversity of quasars may be {\bf mainly accounted for by \lledd, along with other parameters not immediately related to $L$, such as the viewing angle \citep[e.g.,][]{marzianietal01}}. The addition of \civonly measures for extreme HE sources, coupled with our low $z$\   sample, makes it possible to search for correlations over the entire range of  \lledd\ and over a range of more than 5 dex in quasar luminosity. 

\begin{figure*}[htp!]
\centering
\includegraphics[width=1.8\columnwidth]{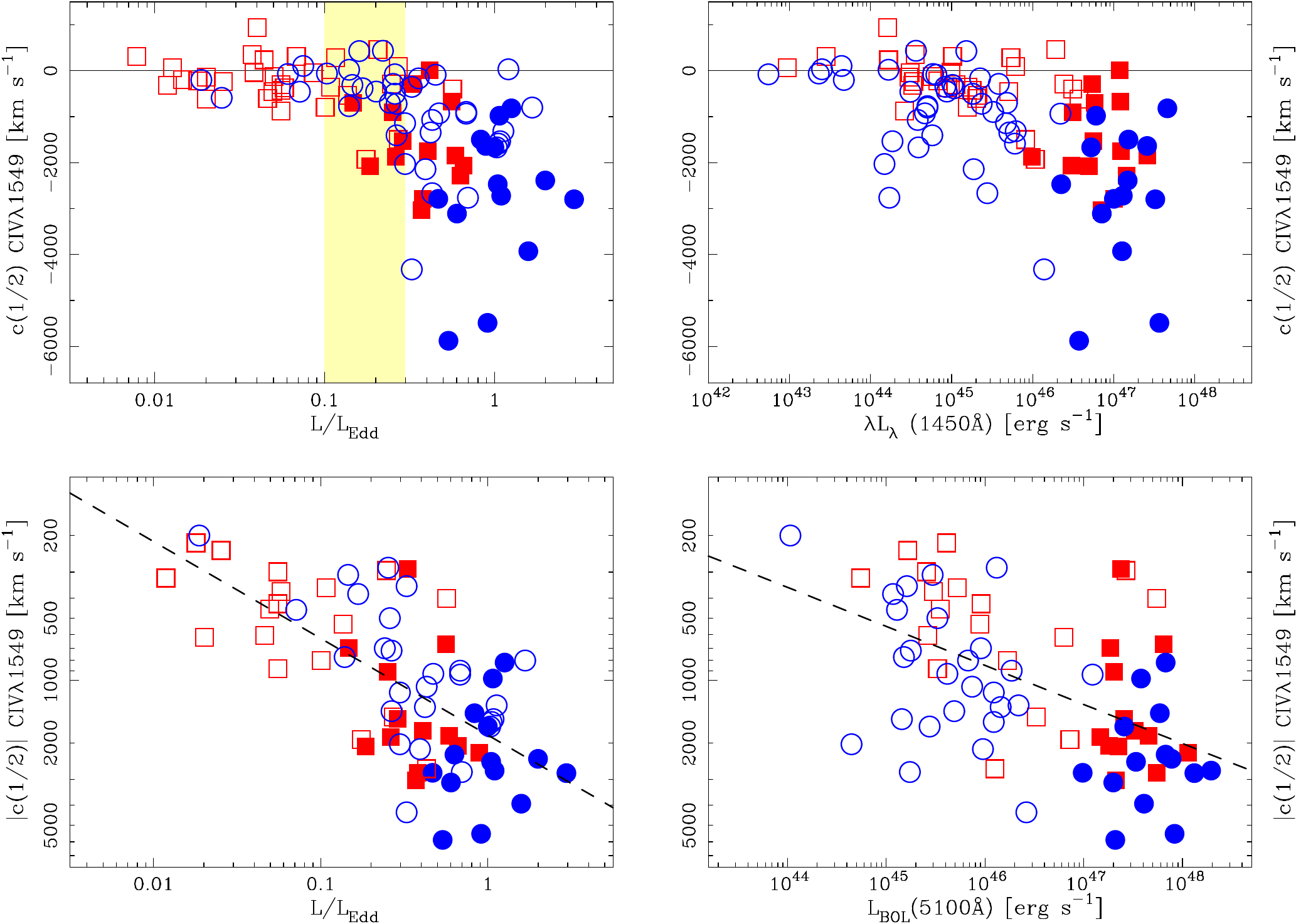}
\caption{Dependence of \civonly \cmp,  on \lledd\  (left panels),  on  $\lambda L_{\lambda}(1450\AA)$  (right upper), and 
on $\log {L}$\ {  derived from the continuum at 5100 \AA\ }(lower right). Population  A sources are indicated by blue circles and Pop. B  sources by red squares. Filled symbols represent HE sources and empty ones correspond to low z sources. The pale yellow band corresponds to \lledd=0.2$\pm$0.1. The dashed lines in bottom frames trace unweighted least squares}
\label{fig:llledd}
\end{figure*}

Figure \ref{fig:llledd} compares the correlations of \cmp\  with L and \lledd\, for 54 Pop A and 45 Pop B quasars in HE and LOWZ samples. We find weak correlation  between L and \cmp\  with sources  showing large and similar \civonly shifts over 4 of the 5 dex range in source luminosity: excluding the four extreme blueshifts  with $\log L \gtrsim 47$ [erg s$^{-1}$], blueshifts between 1000 and 4000 \kms\ are observed over 4dex in luminosity, as  shown in Fig. \ref{fig:llledd} (upper right). The weak trend is  consistent with previous results based on SDSS that found a dependence on luminosity with a large scatter \citep{richardsetal11,shenetal16}. The left panels of Fig. \ref{fig:llledd} show a clearer trend between \lledd\ and \cmp\ . The largests blueshifts ($>$ 1000\kms) are observed for \lledd\ $\gtrsim\ $\  0.2$\pm$0.1. In fact the observed distribution of points in the blueshift-\lledd\, diagram supports the suggestion of a discontinuity between \lledd\, and \cmp\  -- a critical value of \lledd, above which high amplitude winds are possible. If this interpretation is correct, then the Population A-B hypothesis would argue that the largest blueshifts are a Pop A phenomenon and only arise in sources with \lledd$\gtrsim$0.2. Figure \ref{fig:llledd} epitomizes a challenging aspect of the \civonly\ blueshift interpretation: on the one hand, largest shifts are observed only at high $L$, at least in our samples; on the other hand, \cmp\ $\lesssim$ -1000\kms\ are not observed at low \lledd. This is not to say that outflows do not occur at low \lledd, but that the outflowing component BLUE is not dominating the line broadening. The threshold at \lledd\  $\approx$ 0.2$\pm$0.1 would provide a necessary condition for such dominance to occur, with large negative values of \cmp\ possible also at modest $L$. In other words, high \lledd\ appears the condition {\em sine qua non } that makes possible outflows dominating the HIL profiles.

The lower panels of Fig. \ref{fig:llledd} centre the attention in the 72 sources with larger blueshifts ($>$200 \kms). The logarithm  of $|$\cmp$|$ is shown as a function of \lledd\,(left) and L (right) for 42 Pop. A and 30 Pop B sources in the LOWZ and HE samples. The correlation results for the larger blueshift  sources  are reported in Table \ref{tab:correlation} were we list the correlation coefficients, computed following Pearson's  definition and the probability of a spurious correlation $P$.   Linear fit parameters (intercept $a$\ and slope $b$),  obtained with  unweighted least squares (LSQ) and with the orthogonal method \citep{pressetal92} are reported in the next columns. The orthogonal method is known to provide a robust estimator of fit parameters if they are not normally distributed.  The values of the Pearson's correlation coefficient $r$\     suggest   significant correlations  in all cases, at a confidence level $> 3 \sigma$. The values of $r$\  and fitting parameters are listed for Pop. A and B separately first.  We then join population A and B to improve the statistics  since we do not notice relevant differences between the two populations. A significant  correlation  (slope $b \approx 0.45$ or $\approx 0.58$\ for LSQ and orthogonal method, respectively) between blueshift \civonly\,and \lledd\, is found.  Sources of the joined  Pop. A and B sample (last row of Table \ref{tab:correlation})  show an increase in blueshift amplitude with \lledd\ and a Pearson's correlation coefficient of $\approx$0.625, corresponding to a confidence level of $\approx$ 5.2 $\sigma$\ and a probability $P \sim 10^{-7}$\ for the null  hypothesis of uncorrelated data. A weaker (  slope $b \approx 0.15$ for both LSQ and orthogonal method) but still significant dependence  of \cmp\, on bolometric $L$\ is also found (Fig. \ref{fig:llledd}\  bottom  right). The luminosity trend becomes appreciable over a very wide range in luminosity (4dex), when both the LOWZ and HE sample are put together,  with   Pearson's correlation coefficient of $ \approx0.465$ implying a confidence level $\approx 3.9 \sigma $\ (Table \ref{tab:correlation}). 

\begin{table*}[htp!]
\setlength{\tabcolsep}{3pt}\scriptsize
\centering
\caption{Results of correlation analysis \cmp\ vs $\log L $\ and \lledd$^\mathrm{a}$\   \label{tab:correlation} }
\begin{tabular}{lccccccccccccc}\hline\hline\noalign{\vskip 0.05cm}
Sample  & Var.  &   $r$    &  $P$\ & \multicolumn{2}{c}{LSQ$^\mathrm{b}$}  && \multicolumn{2}{c}{Orthogonal$^\mathrm{b}$} && $r$\ (partial)$^\mathrm{c}$ \\[0.05cm] \cline{5-6} \cline{8-9}\noalign{\vskip 0.05cm}
              &     &       &  &              $a$  & $b$ & &     $a$  & $b$   &&   \\[0.05cm]  
\hline\noalign{\vskip 0.07cm}
 A & $L$   & 0.472 & 2.5E-03 &   -3.867  $\pm$  2.095    &   0.151     $\pm$ 0.045 && 
-3.121   $\pm$   1.721    &  0.135     $\pm$ 0.037 && 0.252 $\pm$ 0.152  \\
 B  & $L$ &  0.617  & 7.2E-04 &  -8.443  $\pm$   2.361  &   0.244  $\pm$   0.051 
 &&    -9.569   $\pm$  2.468  &   0.268 $\pm$    0.053 && 0.268 $\pm$\  0.179 \\
  A+B & $L$     & 0.465 & 9.0E-05 &-4.818   $\pm$    1.870 &    0.169  $\pm$     0.040   
  &&      -4.120  $\pm$   1.906   &   0.154 $\pm$    0.041 && 0.262 $\pm$ 0.116   \\
  A & \lledd  & 0.537 & 5.9E-04 &    3.269  $\pm$    0.049  &   0.527  $\pm$    0.102  
  &&   3.346         $\pm$ 0.078    &   0.749       $\pm$    0.208 && \ldots   \\
 B & \lledd   & 0.622  & 6.6E-04  &      3.294   $\pm$ 0.118  &   0.458 $\pm$    0.096 
 &&   3.435          $\pm$  0.122   & 0.619        $\pm$  0.113  && \ldots  \\
  A+B & \lledd   & 0.625 & 1.4E-07 &  3.265 $\pm$    0.045   &  0.451  $\pm$   0.048  
   &&     3.334      $\pm$    0.055&   0.584     $\pm$    0.084 && \ldots  \\[0.07cm]
  \hline\noalign{\vskip 0.07cm}
\multicolumn{14}{l}{$^\mathrm{a}${$\log$| \cmp| in \kms, for \cmp $ \le -200$ \kms, yielding 42 and 30 Pop. A and B sources. $\log L = \log \lambda L_{\lambda}$(1540 \AA) +BC, }}\\
\multicolumn{14}{l}{BC =  $\log(3.5)$, $\log \lambda L_{\lambda}$(1540 \AA) gives same slope $b$\ and intercept $a' = a - 0.544$.}\\
\multicolumn{14}{l}{$^\mathrm{b}${Least square fit and orthogonal method parameters computed with {\tt SLOPES} \citep{feigelsonetal92}.}}\\
\multicolumn{14}{l}{$^\mathrm{c}$\ \lledd\ hidden variable.}\\
\\ 
\end{tabular}
\end{table*}

The higher frequency of large blueshifts at high luminosity may reflect a bias in \lledd\, distribution of the observable quasars if large shifts are dependent on \lledd. The partial correlation coefficient \citep[e.g.,][]{walljenkins12}  between \cmp\ and $L_{1450}$\ with \lledd\ as an hidden variable is presented in the last column of Table \ref{tab:correlation}, along with its 1$\sigma$ uncertainty. The partial correlation reveals that, in our sample,  the \lledd\ bias is not strong enough to account for the \cmp\ -- $L$\ correlation: the A+B partial correlation coefficient of \cmp\ -- $L$\ is  significant (at slightly more than $2 \sigma$\ confidence level) if the \lledd\ is considered as the third variable affecting \cmp\ (Table \ref{tab:correlation}). The residual luminosity dependence is rather evident from our data, as \cmp\ blueshifts larger than 2000 \kms\ are not found in the LOWZ sample for  $\lambda L_{\lambda}(1450\AA) \lesssim 10^{44}$ \ergss. It is unlikely that a large fraction of high-amplitude blueshifts would have gone undetected at   $10^{44}$ \ergss\ $\lesssim L_{BOL} \lesssim 10^{47}$ \ergss.

\subsection{Inferences on the outflowing component  dynamical conditions}
\label{dynamics}
  
The main results of the present investigation involve (1) the detection of extreme \civonly\ blueshifts, (2) the dependence of \cmp\ on \lledd, and (3) the weaker dependence on luminosity that becomes appreciable  over a wide range in luminosity (i.e.,  large blueshifts in Pop. B at high $L$\   even if their  \lledd\ is $\lesssim$ 0.3).  The threshold  above which  large \civonly\ blueshifts ($\gtrsim 1000$ \kms) are observed may be associated with a change in accretion disk structure: at \lledd $ \gtrsim$0.2, a geometrically and optically thick accretion (``slim'') disk  may form in an advection-dominated accretion flow \citep[ADAF; e.g.,][and references therein]{abramowiczetal88,abramowiczstaub14}. In this case, outflows may be confined at the edge of a funnel \citep{sadowskietal14}, and geometry is expected to favor large shift amplitudes.  Estimates of the radiative efficiency suggest that ``slim'' disk may be very frequent at high-luminosity and high-$z$\ where highly accreting quasars are preferentially selected \citep{netzertrakhtenbrot14,trakhtenbrotetal17}.  The slim configuration may develop in the innermost part of the disk, while further out the disk is expected to retain a radiatively-efficient, geometrically thin regime \citep[e.g.,][and references therein]{franketal02}. The interplay between wind and the BLR in general and the disk structure at high \lledd\  is a complex problem that it is  just beginning to be faced \citep[e.g.,][and last paragraph of \S \ref{phys}]{wangetal14a}  and that goes beyond the scope of the present paper. In the following, we mainly restrict ourselves to basic considerations on the outflow acceleration mechanism.

Wind models suitable for accretion disks around supermassive black holes can be broadly grouped into two major classes: models based on magnetohydrodynamic centrifugal acceleration  \citep{emmeringetal92,everett05,elitzuretal14}, and radiation-driven models \citep[][]{murrayetal95,elvis00,everett05,proga07a,risalitielvis10}. Both acceleration mechanisms predict that the outflow terminal velocity is strongly dependent  on Eddington ratio: for magneto-centrifugal acceleration, the terminal velocity scales as $v_\mathrm{t, mag} \propto L^{\frac{1}{3}}$ \lledd. This case implies  the conservation of the angular momentum of the gas originally in the Keplerian accretion disk. Predicted line profiles are fairly symmetric and often show a prominent core.The second class can more easily account for fully blueshifted profiles as observed in Pop. A sources because  radiation-driven winds are expected to give rise to a predominantly radial outflow.  Results (1) and (2) also indicate a predominant role of radiative forces, and   condition (3) is expected in the case of a radiation-driven outflow.    

When radiation forces dominate (which may be the case of a radiation driven wind, or of discrete clouds), the outflow terminal velocity can be written as $  v_\mathrm{t} \propto v_\mathrm{K}(r) \sqrt{{\cal M} \frac{L}{L_\mathrm{Edd}}} \propto   M_{BH}^{\frac{1}{2}} L^{-\frac{\alpha}{2}} \sqrt{{\cal M} \frac{L}{L_\mathrm{Edd}}}$, if\ $r \propto L^{\alpha}$\ \citep[e.g.,][]{laorbrandt02,netzermarziani10}.  
${\cal M}$ is the force multiplier, and $v_\mathrm{K}$\ the Keplerian velocity at the wind launching radius. The slope of the \lledd\ ratio dependence (Table \ref{tab:correlation}) is in agreement with the predictions of a dominant role of radiation forces.  
A concomitant  dependence on luminosity occurs  for increasing black hole mass  if the term $v_\mathrm{K}$\ or $v_\mathrm{K}\sqrt{\cal M}$\ depend  on luminosity. For example, $ v_\mathrm{K}\sqrt{\cal M} \propto L^{k}$ with $k \approx 0.15$\ is possible   if the force multiplier scales roughly with the Eddington ratio.

\subsection{Physical conditions of the line emitting gas}
\label{phys}

The increase in \civonly\ FWHM along with \civonly\ shift argues against a smaller value of the emissivity weighted distance for \civonly\ in a virial velocity field as the {\em dominant} broadening effect for \civonly. If this were the case, the profile would remain roughly similar to the one of \hb.  

The balance between radiative and gravitational forces is not  the same for the whole BLR. Specifically, if we consider the HIL and LIL emitting BLR as two distinct sub-regions, radiation forces may dominate in the high-ionization sub-region, while gravitational ones in the denser and thicker low-ionization sub-region \citep{ferlandetal09,marzianietal10,negreteetal12}.  The I(\civonly$_\mathrm{BC}$)/I(\hbbc) intensity ratio values (which can be deduced from the data reported in Tables \ref{tab:civhb}, \ref{tab:specfitciv} and \ref{tab:specfithb}) imply low ionization for the gas emitting the virialized BC, and confirm much higher ionization for  emission associated with BLUE. The median values for the BC intensity ratios  I(\civonly)/I(\hb) are $\approx 1.4$\ and $\approx 2.2$, and for BLUE I(\civonly BLUE)/I(\hb\ BLUE) $\approx$ 34 and 60 (Pop. A and B, respectively, Table  \ref{tab:specfitciv} and \ref{tab:specfithb}).  The overall scenario envisaged at low-$z$  is basically confirmed for the high-$z$ quasars. The low BC I(\civonly)/I(\hb)   along with several other diagnostic flux ratios such as I(\lya/I(\civonly), I(\aliii/I(\siiii), I(\mgii)/I(\lya), I(\feii)/I(\hb), I(\lya)/I(\hb)  suggested that the BC is being emitted by a dense region of low-ionization (\nh\ $\sim 10^{11} - 10^{12} $\ cm$^{-3}$, ionization parameter $\sim10^{-2} - 10^{-2.5}$) while BLUE requires lower density and higher ionization (\nh\ $\lesssim 10^{10} $\ cm$^{-3}$, ionization parameter $\sim10^{-1}$, \citealt{marzianietal10}).   The  I(\caii)/I(\hbbc)\ intensity ratio measured by    \citet{martinez-aldamaetal15}\ for several sources of the HE sample also indicates the existence of a region of low ionization and high density as the emitter of the BC, and sets a robust {  lower} limit on the density of the line emitting gas, $\sim 10^{10.5} $ cm$^{-3}$ \citep[as  found by][]{matsuokaetal07}.  The inferences on the physical conditions of the two emitting region add further support to the idea that  the emitting gas resolved in radial velocity -- i.e., emitting BC and BLUE --    is also spatially segregated \citep{collinsouffrinetal88,elvis00}. 

Figure \ref{fig:bluebc} shows the relation between \cmp\ and the prominence of BLUE with respect to total \civonly\ emission (upper panel), and the I(\civonly$_\mathrm{BC}$)/I(\hbbc)\ intensity ratio (lower panel).  The BLUE prominence and \cmp\ are almost perfectly anti-correlated (Pearson's $r \approx -0.90$), with intensity ratio I(\civonly BLUE)/I(\civonly) $\approx (-0.15  \pm  0.014)$ \cmp $+0.21 \pm 0.04$, which is not surprising given the definition of BLUE. More interestingly,  \cmp\ and the intensity ratio of the {\em broad component} of \civonly\ and \hb\ are also correlated ($r \approx 0.65$): I(\civbc)/I(\hbbc)$ \approx (0.615 \pm 0.14)$ \cmp\ $  + (3.12 \pm 0.36)$. As the shifts grow larger, the BC of \civonly\ diminishes dramatically. The equivalent width W(\civonly) is anti-correlated with \cmp\   and the \cmp\ amplitude is correlated with \rfe\ \citep{marzianietal16}: the most extreme shifts are associated with the weakest emission lines, and strongest \feii\ emitters. In other words, the largest blueshifts are associated with a very low ionization degree:  \cmp\ is highly correlated with \rfe: $r\ \approx -0.82$, corresponding to a significance above $4 \sigma$\ for 28 objects. In addition minimum  I(\civonly$_\mathrm{BC}$)/I(\hbbc)\ and the BLUE/\civ\ are also associated with the strongest \feii\ emission: BLUE/\civ\ is correlated with \rfe\ with $r \approx 0.63$\ ($3.9 \sigma$);  and  I(\civonly$_\mathrm{BC}$)/I(\hbbc)\ is anti-correlated with $r \approx -0.36$\ ($\approx 2 \sigma$). This is why it is so difficult to identify a correction to the FWHM of \civonly\ that is directly connected to \cmp:  the \cmp\ values are large when blue dominates the \civonly\ emission, and the \civonly\ BC is concomitantly very faint, especially in Pop. A. When the \civonly\ shift is small, however, it might be possible to recover a \civonly-based virial estimator. The issue will be the subject of an eventual paper.

\begin{figure}[htp!]
\centering
\includegraphics[width=0.9\columnwidth]{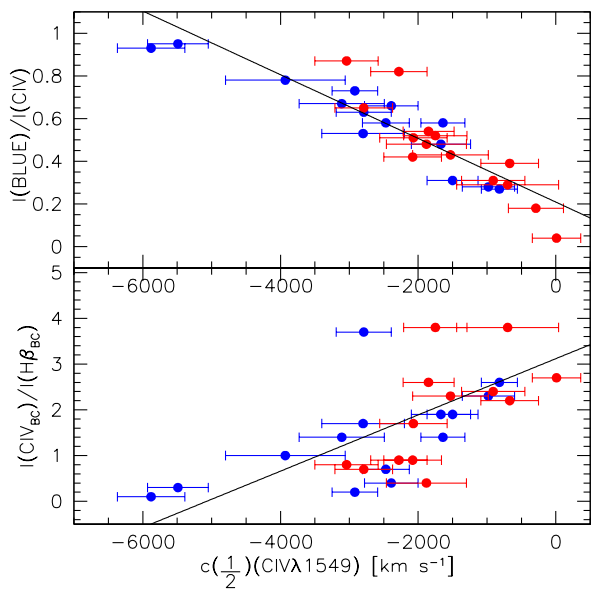}
\caption{Relation between \cmp\ \civonly\ and intensity ratio between the BLUE component and the full \civonly\ line (upper panel), and intensity ratio of the broad components of \civonly\ and \hb\ (lower panel) for the sources of the HE sample. The horizontal scale is in units of \kms. Population A sources are indicated by blue symbols and Pop. B by red ones.}
\label{fig:bluebc}
\end{figure}

Extreme Pop. A (\rfe $\gtrsim 1.0$, in the HE sample: {HE0122-3759}, {HE0359-3959}, {HE1347-2457}) all have I(\civbc)/I(\hbbc) $\lesssim 1$.   The  ratio I(\civbc)/I(\hbbc) for  {HE0359-3959} is just $\approx 0.1$, since the \civonly\ is almost fully blue shifted. In general, density and ionization   parameter appear to reach   extreme values in correspondence of sources with \rfe\ $\gtrsim 1$, $\log$ \nh\ $\sim10^{12}$ \cm3, and $\log U \lesssim -2.5$\ \citep{negreteetal12}. These sources (called xA by \citealt{marzianisulentic14}) show such very  low equivalent width of emission lines along the  4DE1 main sequence. Several (but not all; in our sample {HE0359-3959} and the borderline object \object{HE2352-4010}) meet the definition of weak-lined quasars (WLQs, \citealt{diamond-stanicetal09}, W(\civonly) $\le$ 10 \AA). The WLQs in our sample show high-amplitude blueshifts: --5900 (HE0359)  and --2390 (HE2352) and also very low I(\civbc)/I(\hbbc)\ $\lesssim 0.5$.  }{  WLQs as a class are also characterized by high-amplitude \civ\ blueshifts \citep{plotkinetal15,shemmerlieber15}. WLQs properties are, in general,  consistent with  the properties of xA sources. It is interesting to note that WLQ are present also at low-$L$. By joining the \citet[][]{sulenticetal07} and the \citet{plotkinetal15} data, \citet{marzianietal16} showed that 9 out of 11 sources are consistent with the xA location in the 4DE1 optical plane FWHM(\hb) vs. \rfe.  In turn, xA sources show low W \civonly \ $\lesssim$ 30 \AA\  could perhaps provide a more appropriate physical limit for isolating the weakest \civ\ emitters: WLQs are extreme cases of xA sources but apparently not fundamentally different from the other xA sources, as they show prominent HIL blueshifts along with strong \feii\ emission.

The result of the present paper yield some constraints  on the structure of the BLUE and BC emitting regions in Pop. A.  The     LIL-BLR may be exposed to a low-level ionizing continuum,  due to scatter or self-absorption by the high-ionization wind (not necessarily the same gas emitting the \civonly\ line, \citealt[e.g.,][]{leighlyetal07}). An alternative is that the LIL-BLR may be shadowed from the continuum by  the optically and geometrically thick  structure as provided by an ADAF.  At the same time the 1900 blend   of HE  sources (described in the follow-up paper), shows a rather high I(\aliii)/I(\siiii) ratio,  implying high density following \citet{negreteetal12}.  This result suggests that LIL emission may occur from a remnant   emitting regions  made of the densest gas, while much lower-density gas is being ablated away and dispersed into the quasar circumnuclear regions.

\section{Conclusions}

Comparison of low (\hb) and high (\civonly) excitation  broad emission line profiles in high and low redshift quasars offers a method to search for quasar  evolution in redshift, luminosity and source Eddington ratio. However there  are three desiderata in order to make such comparisons: 1) an accurate measure of  the local rest frame in each source, 2) spectral S/N must be high enough to allow  decomposition of both \hb and \civonly, and 3) a context in which one can interpret  quasar diversity (i.e. Population A and B discrimination). This explains why there have been few studies of this kind \citep[e.g.,][]{espeyetal89,marzianietal96,sulenticetal07,coatmanetal16}. Low redshift comparisons  show that the two lines are very different in Population A and B quasars.
This paper combines  samples of low z (ground based/HST-FOS), high z/low L (GTC sample)  and high z/high L (HE ISAAC/FORS and TNG-LRS sample) quasars which explore the full observed ranges of L and \lledd. \civonly\ measures of extreme quasars show blueshift values among the largest ever observed. Previously published \hb\ spectra provide  accurate rest frame measures, Eddington ratios and Population A-B  classifier while high S/N \civonly\ measures, through accurate line profile decomposition,  add insights about disk structure and especially about outflows or winds 
arising from the disk.  Major results are:

\begin{enumerate}
\item \civonly\, outflows become stronger and more frequent at higher z in sources 
with $\rm{2-4}$\ dex higher luminosity than low z quasars. The emission component 
associated with the outflow becomes a dominant source of line broadening,
and confirms that the full \civonly\ profile is not an  useful virial estimator of black hole mass for most sources.

\item We find that the 4D Eigenvector 1 sequence persists but is less strong among the extreme  HE quasars { as far as \civonly\ blueshifts are concerned: } many Pop B sources show a significant \civonly blueshift. In the optical plane of the 4DE1 parameter space, data points are displaced upwards toward larger FWHM(\hb) values, as expected for the systematic increase of line width with luminosity in virialized systems. 

\item We show evidence that the \civonly outflows are driven by \lledd\ rather than source
luminosity. A clear dependence of \civonly blueshifts on Eddington ratio is observed 
while a weaker trend is observed with source luminosity that becomes appreciable only 
over a wide luminosity range (4 -- 5 dex). Large shifts (\cmp\ $\lesssim -1000$ \kms)\ are apparently possible above a threshold \lledd $\approx 0.2 \pm 0.1$.  

\item {The relations between \cmp\ and \lledd\ and $L$\ are consistent with a radiation-driven outflow, and largest shifts may support the existence of an optically-thick ADAF  in at least part of Pop. A sources.}   
\end{enumerate}

\begin{acknowledgements}
A.d.O., J.W.S. and M.L.M.A. acknowledge financial support from the Spanish Ministry for Economy and 
Competitiveness through grants AYA2013-42227-P and AYA2016-76682-C3-1-P. \  J. P.  acknowledges financial support from the Spanish Ministry for Economy and Competitiveness through grants AYA2013-40609-P and AYA2016-76682-C3-3-P. \ D. D.  and A. N. acknowledge support from grants PAPIIT108716, UNAM, and CONACyT221398.
\end{acknowledgements}
\vfill\eject\eject

\bibliographystyle{aa}

\vfill
\newpage
\pagebreak

\end{document}